\documentclass[%
	reprint,
	amsmath,amssymb,
	aps,
	prb,
]{revtex4-2}

\usepackage{tikz}
\usepackage{ifthen}
\usepackage{tikz-3dplot}
\usepackage{graphicx}
\usepackage{dcolumn}
\usepackage{bm}

\usepackage{mathrsfs}
\usepackage{hyperref}
\hypersetup{
	colorlinks = true,
	urlcolor = blue,
	linkcolor = blue,
	citecolor = green,
	filecolor = magenta,
}

\newcommand{\ud}{\mathrm{d}}
\newcommand{\pd}{\partial}

\newcommand{\sD}{\mathscr{D}}

\newcommand{\sH}{\mathcal{H}}

\newcommand{\cF}{\mathcal{F}}
\newcommand{\cS}{\mathcal{S}}

\newcommand{\x}{x}
\newcommand{\y}{y}
\newcommand{\p}{p}

\begin{document}


\title{Gravitational lensing of gravitational waves: universal characteristics of strongly lensed memory waveforms}
\author{Ruanjing Zhang}
\email{zhangruanjing@126.com}
\affiliation{College of Physics, Henan University of Technology, Zhengzhou 450001, China}
\author{Zhi-Chao Zhao}
\email{zhaozc@cau.edu.cn}
\affiliation{Department of Applied Physics, College of Science, China Agricultural University, Qinghua East Road, Beijing 100083, China}
\author{Shaoqi Hou}
\email{hou.shaoqi@whu.edu.cn}
\affiliation{School of Physics and Technology, Wuhan University, Wuhan, Hubei 430072, China}
\author{Xi-Long Fan}
\email{xilong.fan@whu.edu.cn}
\affiliation{School of Physics and Technology, Wuhan University, Wuhan, Hubei 430072, China}

\author{Kai Liao}
\email{liaokai@whu.edu.cn}
\affiliation{School of Physics and Technology, Wuhan University, Wuhan, Hubei 430072, China}

\author{Zong-Hong Zhu}
\email{zhuzh@whu.edu.cn}
\affiliation{School of Physics and Technology, Wuhan University, Wuhan, Hubei 430072, China}


\date{\today}

\begin{abstract}
	In this work, the strong lensing effect of the memory signal was considered.
	In the geometric optics limit, the lensed memory signal becomes oscillatory, while the unlensed is basically monotonic.
	This is because only the high frequency Fourier modes contribute strongly to the lensed signal.
	Due to the step function like behavior of the unlensed memory waveform, the lensed waveform possesses characteristic morphology that is dependent on the type of the image, but independent of the lens model and the binary system.
	That is, for each type of the lensed image, the lensed memory waveform has an approximate reflection symmetry about a symmetrical axis in the time domain.
	More specifically, for the type I and type III images, the lensed memory signals are nearly odd under the reflection, while the type II signal is roughly even.
	In addition, at the symmetrical axis, the sign of the slope for type I image is different from that for the type III image.
	These universal characteristic features would help determine the type of the lensed image.
	This is particularly because the memory waveform can be well approximated by a suitable step function, which involves just two parameters, the overall amplitude and the time of arrival.
	It is fast and cheap to simulate this approximated waveform.
	Once the type of the lensed image is determined with the approximated memory waveform, one can use the appropriate waveform template for the oscillatory component of the gravitational wave to perform the parameter estimation.
\end{abstract}

\maketitle


\section{Introduction}

As is well known, the detection of the gravitational wave event GW150914 marked a new era of the gravitational wave physics and astrophysics \cite{gw150914}.
This long-awaited triumph confirmed the existence of the gravitational wave, the prediction of Einstein's theory of general relativity made nearly a hundred years ago \cite{Einstein:1916cc,Einstein:1918btx}.
According to this theory, whenever the spacetime is perturbed, the gravitational wave is produced \cite{Wald:1984rg}.
Then, it travels from its source to the detector.
During the trip, it interacts with the spacetime background and itself, since the gravitational interaction is nonlinear.
Interesting effects and phenomena thus occur.
Two of them are the gravitational lensing effect \cite{gravlens1992} and the memory effect \cite{Zeldovich:1974gvh,Braginsky:1986ia,Christodoulou1991,Thorne:1992sdb}.

The gravitational lensing effect happens when the gravitational wave interacts with the spacetime background, in particular, the one produced by celestial objects.
These celestial objects are named lenses.
Like the light \cite{Hou:2019wdg}, the trajectory of the gravitational wave will be bent near the lens.
After passing by the lens, its amplitude is modified, and its phase is shifted.
Since in some situation, the amplitude becomes larger, the lensed gravitational wave signal would be easier to be detected.
It may also happen that two lensed gravitational wave signals exist at the detector simultaneously, interfering with each other and forming the beat pattern \cite{Hou:2019dcm}.

The lensing effect takes place if the linear interaction is considered, while the memory effect is related to the nonlinear nature of the gravitation.
It is the phenomenon that the spacetime metric is permanently changed after the gravitational wave has passed by.
It manifests as the everlasting contraction or stretch of the arms of the interferometer.
Earlier studies of this effect were mainly based on the post-Minkowskian or the post-Newtonian approximations \cite{Zeldovich:1974gvh,Braginsky:1986ia,Christodoulou1991,Thorne:1992sdb,Favata:2008ti,Favata:2008yd,Favata:2009ii,Favata:2011qi}.
The idea of solving the Einstein's equation using the double-null foliation formalism of the spacetime was applied to analyze this phenomenon quite rigorously \cite{Christodoulou1991}.
Recent works heavily relied on the Bondi-Sachs formalism \cite{Bondi:1962px,Sachs:1962wk,Madler:2016xju}, and revealed the rich connection between the memory effect and the asymptotic symmetry of the isolated system \cite{Sachs1962asgr,Strominger2014bms,Strominger:2014pwa}.
It was also found out that the memory waveform can be obtained by manipulating the flux-balance laws associated with the asymptotic symmetries \cite{Flanagan:2015pxa}.

One usually calculates the memory waveform assuming there is no gravitational lens between the binary system and the detector.
One thus finds out that in many cases, the memory signal is almost a monotonically increasing function of time \cite{Favata:2010zu,Lasky:2016knh}.
This signal starts at the zero value at the infinite past, increases very slowly during the inspiral phase of the binary evolution, surges in the merger and the ringdown phases, and finally, plateaus.
This feature occurs when the inclination angle $\iota$, the angle between the line-of-sight and the angular momentum of the binary, is nearly $\pi/2$.
Once $\iota$ deviates from this value further and further on either side, the memory signal gradually ceases to be monotonic, and becomes more and more oscillatory \cite{Hou:2024rgo}.
When $\iota$ is very close to 0 or $\pi$, the memory signal is largely oscillating.
But no matter what the value of $\iota$ is, the final memory magnitude is never zero.
Moreover, although the memory signal possesses more interesting structures when $\iota$ is far away from $\pi/2$, the overall magnitude is smaller, making it more difficult to be detected.
Therefore, we would mainly consider the memory signal with $\iota$ is close to $\pi/2$, and treat the memory effect as a low frequency phenomenon.
This is distinctively different from the waveforms of the quadrupole radiation \cite{mtw} and other parts of the gravitational wave that can usually be computed using the post-Newtonian method \cite{Poisson2014}.
We would call these later waveforms oscillatory in the following.

It was fairly reasonable to ignore the possibility of the lensing of the gravitational wave.
All of the confirmed gravitational wave events by LIGO-Virgo-KAGRA (LVK) collaboration have sources quite near to the earth with the largest redshift $\sim1.18$ \cite{LIGOScientific:2018mvr,Abbott:2020niy,LIGOScientific:2021usb,KAGRA:2021vkt}.
The probability for at least one of them being lensed is expected to be low.
Indeed, several works found no strong evidence of the lensing effect \cite{LIGOScientific:2021izm,LIGOScientific:2023bwz}.
In future, more advanced interferometers will be able to probe gravitational wave events occurring at larger redshifts, and the probabilities of lensing effect are appreciable \cite{Sereno:2010dr,Piorkowska:2013eww,Ding:2015uha,Piorkowska-Kurpas:2020mst,Hou:2020mpr,Gao:2021sxw,Yang:2021viz,Lin:2023ccz}.
At the same time, the memory effect might also be detected in future.
Although the ground-based interferometers will have difficulties in detecting the memory effect \cite{Hubner:2019sly,Grant:2022bla}, the space-borne interferometers would have a good chance to observe several to even hundreds of strong enough memory signals per year \cite{Zhao:2021zlr,Sun:2022pvh,Gasparotto:2023fcg,Inchauspe:2024ibs,Hou:2024rgo}.
So it would be interesting to study how the memory waveform is lensed.

In this work, the strong lensing effect of the memory waveform will be considered.
For the light, the strong lensing occurs when the lens is massive enough so that the images of the source can be separated viewed by a distant observer.
For the gravitational wave, the strong lensing effect usually refers to the case when the wavelength is much smaller than the curvature radius of the lens.
It is well-known that (the oscillatory component of) the gravitational wave would travel along the null geodesic from the binary system to the detector, passing by a lens.
For a generic lensing system, it is expected that the gravitational wave will reach the earth through multiple paths at different times.
Each path corresponds to an ``image" of the binary system, and the signals ``produced" by different ``images'' will be magnified differently.
Similarly, the memory signal, as another part of the gravitational wave, also shares these properties.
This is because although the memory signal is mainly of low frequencies, a heavy enough lens would have a sufficiently large curvature radius $\mathcal R$ such that at least the high frequency Fourier modes of the memory signal can be considered to be strongly lensed, \textcolor{black}{as long as the frequency of the mode is much larger than $1/\mathcal R$.}

Since the matched filtering method, employed by the gravitational wave interferometers, relies on the precise waveform of the signal, the lensed memory waveform is to be computed.
For the oscillatory component of the gravitational wave, the strongly lensed waveform in the frequency domain can be obtained simply by multiplying the unlensed one by the amplitude magnification factor and a phase shift due to the time delay.
In addition, a third factor shall be considered, which is related to the extreme points of the Fermat potential \cite{Nakamura1999wo,Takahashi:2003ix}.
The Fermat potential is the time it takes for the photon or the gravitational wave to travel from the source to the observer along a fictitious null trajectory \cite{gravlens1992}.
Taking the variational principle of the Fermat potential determines the actual path, and the positions of the ``images''.
So these ``images'' might be the local minima, the saddle points, or the local maxima of the Fermat potential.
They are named type I, II, and III ``images", respectively.
They correspond to the different number of caustics that the gravitational wave passes through: 0, 1, and 2 \cite{Ezquiaga:2020gdt}.
So the lensed waveform also depends on the type of the extreme point.
By performing the inverse Fourier transform, the lensed waveform in the time domain can be obtained.

Usually, one does not set any bound on the frequency range when taking the inverse Fourier transformation.
This works well for the oscillatory component of the gravitational wave, since this component, as its name indicates, is oscillating frequently enough,
so the low frequency Fourier modes contribute to the (unlensed) waveform negligibly.
This will not be true for the memory waveform.
The memory signal is composed of a huge portion of low frequency modes.
Thus, it is required to set some suitable frequency bounds when performing the inverse Fourier transform in the strong lensing regime.
In this way, we actually focus on the high frequency modes that are strongly lensed.
Of course, the lower frequency modes are also lensed, and could not be ignored if one had a detector, working at the low enough frequency range.
But in practice, an interferometer works in a relatively narrow frequency band.
As long as the lens is sufficiently massive, any signal within the frequency band can be treated as strongly lensed.

By choosing the appropriate frequency cutoff $\omega_c$, compatible with the curvature radius of the lens and the sensitivity band of the detector, the strongly lensed memory waveform in the time domain has several distinctive features.
First, the lensed memory waveform also oscillates at about $\omega_c$.
\textcolor{black}{Secondly, when plotted against time, the lensed memory waveform appears approximately symmetric about the arrival time of the image.}
For the type I or III image, the lensed memory signal is nearly odd under the reflection about the vertical line, but for the type II image, it is roughly even.
This is due to the fact that the unlensed memory waveform is largely the Heaviside step function.
Thirdly, at the maximum absolute amplitude, the slope of the type I image has the opposite sign from that of the type III.
This is crucial for distinguishing the type I image from the type III.

These symmetry features of the lensed memory waveform is actually universal, independent of the lens model, and the property of the binary system.
This is because the unlensed memory signal is dominated by a step function in the time domain.
If one considers the strong lensing of the step function signal, one obtains the already mentioned symmetry features, except that for the step function signal, the symmetry is exact.
This observation also suggests a possible way to speed up the search for the lensed gravitational wave signal, that is, using the step function to represent the actual (unlensed) memory signal.
The approximate memory waveform involves only two parameters, the maximal amplitude of the memory signal and the time delay.
Using this approximation to test whether a gravitational wave signal is lensed is expected to be fast.
It is also helpful for determining the type of the image.
Previous studies already analyzed the differences among the lensed oscillatory waveforms of the various image types and the unlensed one, and evaluated the challenges of identifying the image type \cite{Ezquiaga:2020gdt,Wang:2021kzt,Janquart:2021nus,Vijaykumar:2022dlp,Taylor:2024yjt}.
The method proposed in this work, using the morphology of the lensed memory waveform in the time domain, \textcolor{black}{is novel and interesting.}

Of course, one shall make sure that the step function approximates the unlensed memory waveform well.
So in this work, we simulated the actual memory waveforms produced by massive binary stars observable by Laser Interferometer Space Antenna (LISA)  \cite{Audley:2017drz,LISA:2024hlh}.
We calculated the mismatches between the actual memory waveforms and the step function, and found out that there are quite a lot of events whose memory waveforms can be well approximated by the step function.
Of course, the similar ideas can be applied to other interferometers.

This work is organized in the following way.
In Section~\ref{sec-gm}, the computation of the gravitational memory effect in the Bondi-Sachs formalism is reviewed.
In Section~\ref{sec-gle}, the strong gravitational lensing effect is discussed, and in particular, the transition from the wave optics to geometric optics is disclosed in Section~\ref{sec-wo-go}.
Then, the lensed memory waveform can be computed in Section~\ref{sec-glm}.
There, one first calculated the lensed memory waveform for the type I or III image in Section~\ref{sec-lmem-td-mm}, then derived the waveform for the type II image in Section~\ref{sec-lmem-td-s}.
An example was given to vividly illustrate the lensing effect on the memory waveform in Section~\ref{sec-eg}.
In the next section~\ref{sec-uni}, the universal characteristics of the lensed memory waveform were identified, and analyzed in detail.
Based on this, one may approximate the unlensed memory waveform as done in Section~\ref{sec-app-mem}, and thus, propose a method to determine the type of the image in Section~\ref{sec-ident}.
Finally, there is a conclusion~\ref{sec-con}.
Throughout, we used the natural units with $c=G=1$.

\section{Gravitational memory effects}
\label{sec-gm}

In the asymptotically flat spacetime, the gravitational memory effect can be conveniently described within the Bondi-Sachs formalism \cite{Bondi:1962px,Sachs:1962wk,Madler:2016xju}.
In this formalism, the Bondi-Sachs coordinate system $(u,r,\theta,\varphi)$ are used.
This coordinate system is constructed by considering a bunch of outgoing null geodesics, emanated at the successive times $u$.
Those null geodesics emitted at any time $u$ form a null cone.
On the null cone, the angular coordinates $(\theta,\varphi)$ are required to be constant along any null geodesic.
Therefore, one has $g^{uu}=g^{u\theta}=g^{u\varphi}=0$, and the metric can be parameterized in the following way \cite{Barnich:2010eb},
\begin{equation}
	\label{eq-met-bs-0}
	\begin{split}
		\ud s^2= & e^{2\p}\frac{V}{r}\ud u^2-2e^{2\p}\ud u\ud r         \\
		& +g_{AB}(\ud \zeta^A-U^A\ud u)(\ud \zeta^B-U^B\ud u),
	\end{split}
\end{equation}
where $\zeta^A=\theta,\phi$, and $p,V,g_{AB}$, and $U^A$ are metric functions.
One also requires that $\det(g_{AB})=r^4\sin^2\theta$.
This coordinate system is well suited for describing radiations.
The radial direction $r^a\equiv(\pd_r)^a$ is the null direction, and
\begin{equation}
	\label{eq-ngeo}
	r^b\nabla_br^a=2r^a\pd_r \p .
\end{equation}
So the radial coordinate lines are actually the null geodesics used to construct the coordinate system.
Of course, none of these geodesics is affinely parametrized.

By solving the vacuum Einstein's equation with the following boundary conditions \cite{Hou:2021oxe},
\begin{gather*}
	\label{eq-bc}
	\p=\mathcal O(r^{-1}),\quad V=-r+\mathcal O(r^0),\quad U^A=\mathcal O(r^{-2}),\\
	g_{AB}=r^2\gamma_{AB}+\mathcal O(r),
\end{gather*}
one can obtain the metric \cite{Flanagan:2015pxa},
\begin{equation*}
	\label{eq-met-bs}
	\begin{split}
		\ud s^2= & -\ud u^2-2\ud u\ud r+r^2\gamma_{AB}\ud \zeta^A\ud \zeta^B \\
		& +\frac{2m}{r}\ud u^2+\sD^Bc_{AB}\ud u\ud \zeta^A
		+rc_{AB}\ud \zeta^A\ud \zeta^B                                       \\
		& +\cdots.
	\end{split}
\end{equation*}
Here, $\gamma_{AB}$ is the round metric on a unit 2-sphere,
\begin{equation}
	\label{eq-def-gamma}
	\gamma_{AB}\ud\zeta^A\ud\zeta^B=\ud\theta^2+\sin^2\theta\ud\varphi^2,
\end{equation}
and $\sD_A$ is the associated covariant derivative.
The capital Latin indices are lowered and raised by $\gamma_{AB}$ and its inverse $\gamma^{AB}$, respectively.
The shear tensor $c_{AB}$ is traceless, i.e., $\gamma^{AB}c_{AB}=0$, and its time derivative, $N_{AB}=\pd_uc_{AB}$, is called the news tensor.
$m$ is called the Bondi mass aspect \cite{Bondi:1962px,Sachs:1962wk}.
The dots in Eq.~\eqref{eq-met-bs} represent higher order terms in $1/r$ that are irrelevant for the following discussion.

With this metric, one can easily calculate the geodesic deviation equation for two nearby test particles at the null infinity \cite{Wald:1984rg},
\begin{equation}
	\label{eq-gde}
	\ddot S^{\hat A}=\frac{1}{2r}\ddot c_{\hat A\hat B}S^{\hat B}+\cdots,
\end{equation}
\textcolor{black}{in the local tetrad basis $e_{\hat 0}=\pd_u,\,e_{\hat r}=\pd_u-\pd_r,\,e_{\hat 1}=r^{-1}\pd_\theta,\,e_{\hat 2}=(r\sin\theta)^{-1}\pd_\phi$ near $r\rightarrow+\infty$.}
Here, the deviation vector $S^a$ is taken to be in the angular direction, $S^a=S^A\delta^a_A$.
So as long as $c_{AB}$ depends on $u$ nontrivially, e.g., quasi-periodically, it represents the gravitational wave.
The dynamical degrees of freedom are encoded in $c_{AB}$.
Assume that before the arrival of the gravitational wave ($u>u_0$), the two test particles are at rest relative to each other.
Then integrating this equation twice results in
\begin{equation}
	\label{eq-def-mm}
	\Delta S^{\hat A}=\frac{\Delta c_{AB}}{2r}S_0^{\hat B},
\end{equation}
where $S_0^{\hat A}$ is the initial relative distance.
The difference $\Delta c_{AB}=c_{AB}(u_f)-c_{AB}(u_0)$ is usually called the memory tensor, with $u_f$ the time when the gravitational wave disappears.
This tensor measures the displacement memory effect.

Einstein's equation also determines the following evolution equation for $m$ \cite{Flanagan:2015pxa},
\begin{gather}
	\label{eq-m-evo}
	\dot m=\frac{1}{4}\sD_A\sD_BN^{AB}-\frac{1}{8}N_{AB}N^{AB},
\end{gather}
where $\sD^2\equiv\sD_A\sD^A$.
This equation has an important feature, that is, the right-hand side contains the linear terms in the dynamical degrees of freedom.
This feature enables the computation of the memory waveform.
Integrating Eq.~\eqref{eq-m-evo} and rearranging, one finds out that
\begin{equation}
	\label{eq-mem}
	\sD_A\sD_B\Delta c^{AB}=4\Delta m+\frac{1}{2}\int_{u_0}^{u_f}\ud uN_{AB}N^{AB}.
\end{equation}
Here, the first term on the right-hand side gives the linear memory, and the second contributes to the null memory \cite{Bieri:2013ada}.
In this work, we focus on the nonlinear memory.
Now, decompose $c_{AB}$ in the following way,
\begin{equation}
	\label{eq-decc}
	c_{AB}=\left(\sD_A\sD_B-\frac{\gamma_{AB}}{2}\sD^2\right)\Phi+\epsilon_{C(A}\sD_{B)}\sD^C\Psi,
\end{equation}
where $\Phi$ and $\Psi$ are scalar functions of $u$ and $x^A$ in general, representing the electric and magnetic parts of $c_{AB}$, respectively.
Substituting this into Eq.~\eqref{eq-mem}, one knows that
\begin{equation}
	\label{eq-dism-1}
	\sD^2(\sD^2+2)\Delta\Phi=\int_{u_0}^{u_f}\ud uN_{AB}N^{AB},
\end{equation}
where the linear memory contribution has already been neglected.
Note that the upper integration limit is $u_f$ in Eqs.~\eqref{eq-mem} and \eqref{eq-dism-1}, a fixed time.
It shall be replaced by an arbitrary retarded time $u$ in order to define the memory waveform \cite{Nichols:2017rqr,Mitman:2020pbt}, later.

Now, it is ready to plugin the waveform of the compact binary star system to compute the displacement and spin memory waveforms.
For this purpose, it is convenient to use the complex waveform \cite{Mitman:2020pbt}
\begin{equation}
	\label{eq-def-h}
	h\equiv\frac{1}{2r}\bar\gamma^A\bar\gamma^Bc_{AB}=\sum_{\ell=2}^\infty\sum_{m=-\ell}^\ell{}_{-2}Y_{\ell m}h_{\ell m},
\end{equation}
where $\gamma_A=-(1,i\sin\theta)$ is a complex vector field, $\bar \gamma_A$ is its complex conjugate, and ${}_{-2}Y_{\ell m}(\zeta^A)$ is the spin-weighted spherical harmonics of the spin weight $-2$.
$h_{\ell m}$ are the spherical harmonic modes.
In terms of $\gamma_A$ and $\bar\gamma_A$, one has $\gamma_{AB}=(\gamma_A\bar\gamma_B+\bar\gamma_A\gamma_B)/2$ and $\epsilon_{AB}=i(\gamma_A\bar\gamma_B-\gamma_B\bar\gamma_A)/2$.
One could also write $\Phi=\sum\Phi_{\ell m}Y_{\ell m}$  in terms of the scalar spherical harmonics.
Then, Eq.~\eqref{eq-dism-1} can be reexpressed as
\begin{equation}
	\label{eq-phi-lm}
	\begin{split}
		\Delta\Phi_{\ell m}(u)= & 2r^2\frac{(\ell-2)!}{(\ell+2)!}\widetilde\sum\mathcal C_{\ell m}(-2,\hat{\ell},\hat{m};2,\tilde{\ell},-\tilde{m}) \\
		& \times(-1)^{\tilde{m}}\int_{u_0}^u\dot h_{\hat{\ell}\hat{m}}\dot{h}^*_{\tilde{\ell}\tilde{m}}\ud u',
	\end{split}
\end{equation}
where the symbol $\widetilde\sum$ means to sum over $\hat{\ell},\hat{m}$ and $\tilde{\ell},\tilde{m}$, and
\begin{equation}
	\label{eq-def-cl}
	\begin{split}
		& \mathcal C_{\ell m}(\hat{s},\hat{\ell},\hat{m};\tilde{s},\tilde{\ell},\tilde{m})                                               \\
		\equiv & \oint\ud^2\zeta\sqrt\gamma({}_{\hat{s}}Y_{\hat{\ell}\hat{m}})({}_{\tilde{s}}Y_{\tilde{\ell}\tilde{m}})({}_{s}\bar Y_{\ell m}).
	\end{split}
\end{equation}
By its very definition, $\mathcal C_{\ell m}(\hat{s},\hat{\ell},\hat{m};\tilde{s},\tilde{\ell},\tilde{m})$ is nonvanishing as long as the following conditions are satisfied: $s=\hat{s}+\tilde{s}$, $m=\hat{m}+\tilde{m}$, and $\text{max}\{|\hat{\ell}-\tilde{\ell}|,|\hat{m}+\tilde{m}|,|\hat{s}+\tilde{s}|\}\le\ell\le \hat{\ell}+\tilde{\ell}$.
So the displacement memory waveform is determined by the following expression,
\begin{equation}
	\label{eq-mem-1}
	\begin{split}
		h_\text{D}(u,\zeta^A)= & \frac{1}{2r}\bar\gamma^A\bar\gamma^B\Delta c_{AB}|_{\Psi=0}                                              \\
		=                      & \frac{1}{2r}\sum_{\ell m}\sqrt{\frac{(\ell+2)!}{(\ell-2)!}}{}_{-2}Y_{lm}(\zeta^A)\Delta\Phi_{\ell m}(u),
	\end{split}
\end{equation}
where $\ell\ge2$.
\textcolor{black}{The plus and the cross polarization components are $h_{\text{D}+}=\Re h_{\text{D}}$ and $h_{\text{D}\times}=-\Im h_{\text{D}}$, respectively, where $\Re$ and $\Im$ denote the real and the imaginary parts.}

Now, one obtains the memory waveforms in the time domain.
It is also handy to calculate the Fourier transform, which will be used in the following discussion.
It is sufficient to compute the Fourier transform of the integral in Eq.~\eqref{eq-phi-lm}, which is
\begin{equation}
	\label{eq-f-dismm}
	\frac{2\pi}{if}\int f'(f'-f)\tilde h_{\hat{\ell}\hat{m}}(f')\tilde h^*_{\tilde{\ell}\tilde{m}}(f'-f)\ud f'.
\end{equation}
Here, the Fourier transformation is defined to be
\begin{equation}
	\label{eq-def-ft}
	\tilde h_{\ell m}(f)=\int_{-\infty}^{+\infty}h_{\ell m}(u)e^{-i2\pi fu}\ud u.
\end{equation}
To obtain Eq.~\eqref{eq-f-dismm}, one takes $u_0=-\infty$ in Eq.~\eqref{eq-mem-1}.
Therefore, once one substitutes the frequency space gravitational waveform, one gets the memory waveforms in the frequency domain directly.

\begin{figure}[h]
	\centering
	\includegraphics[width=0.45\textwidth]{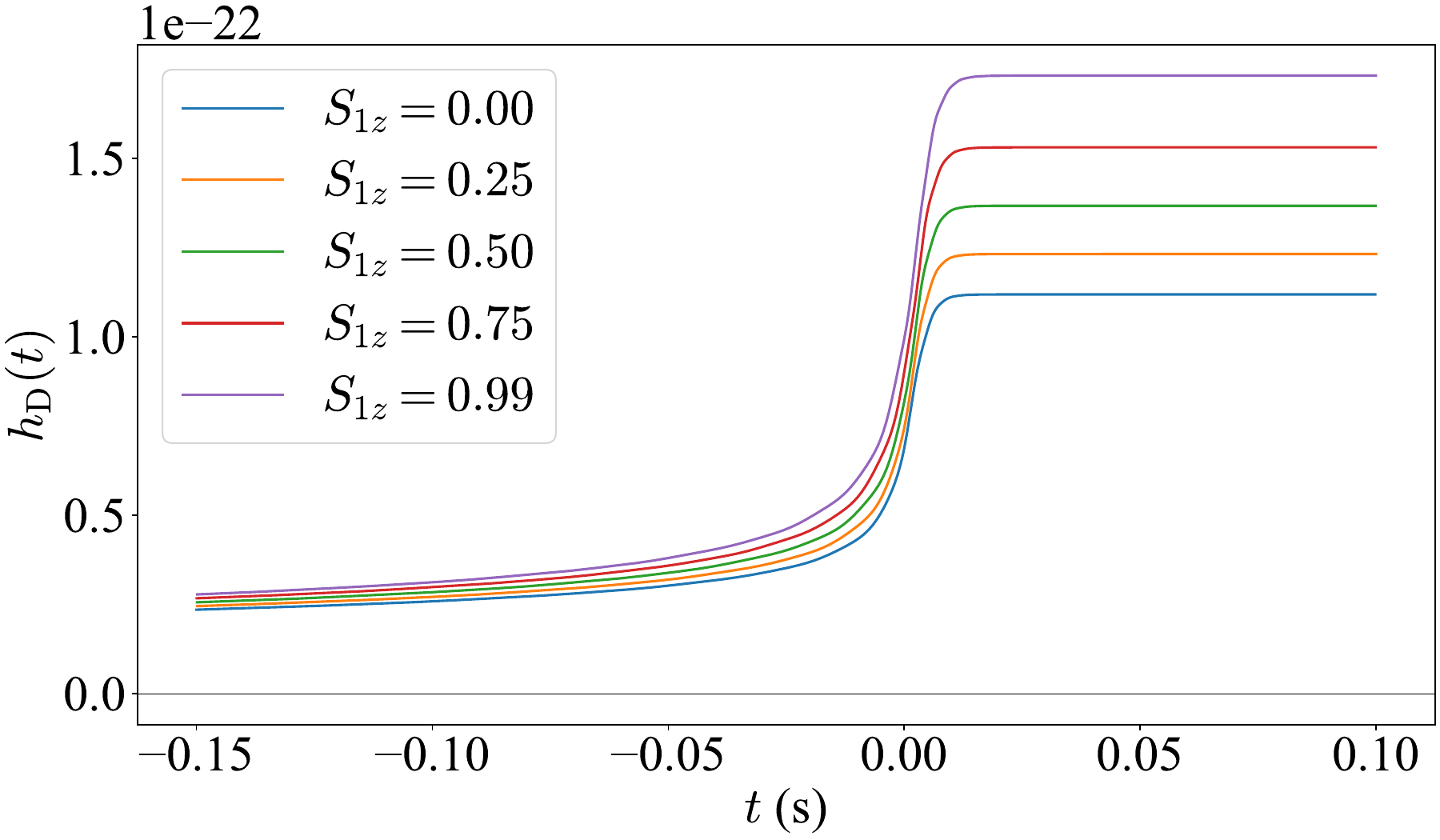}
	\includegraphics[width=0.45\textwidth]{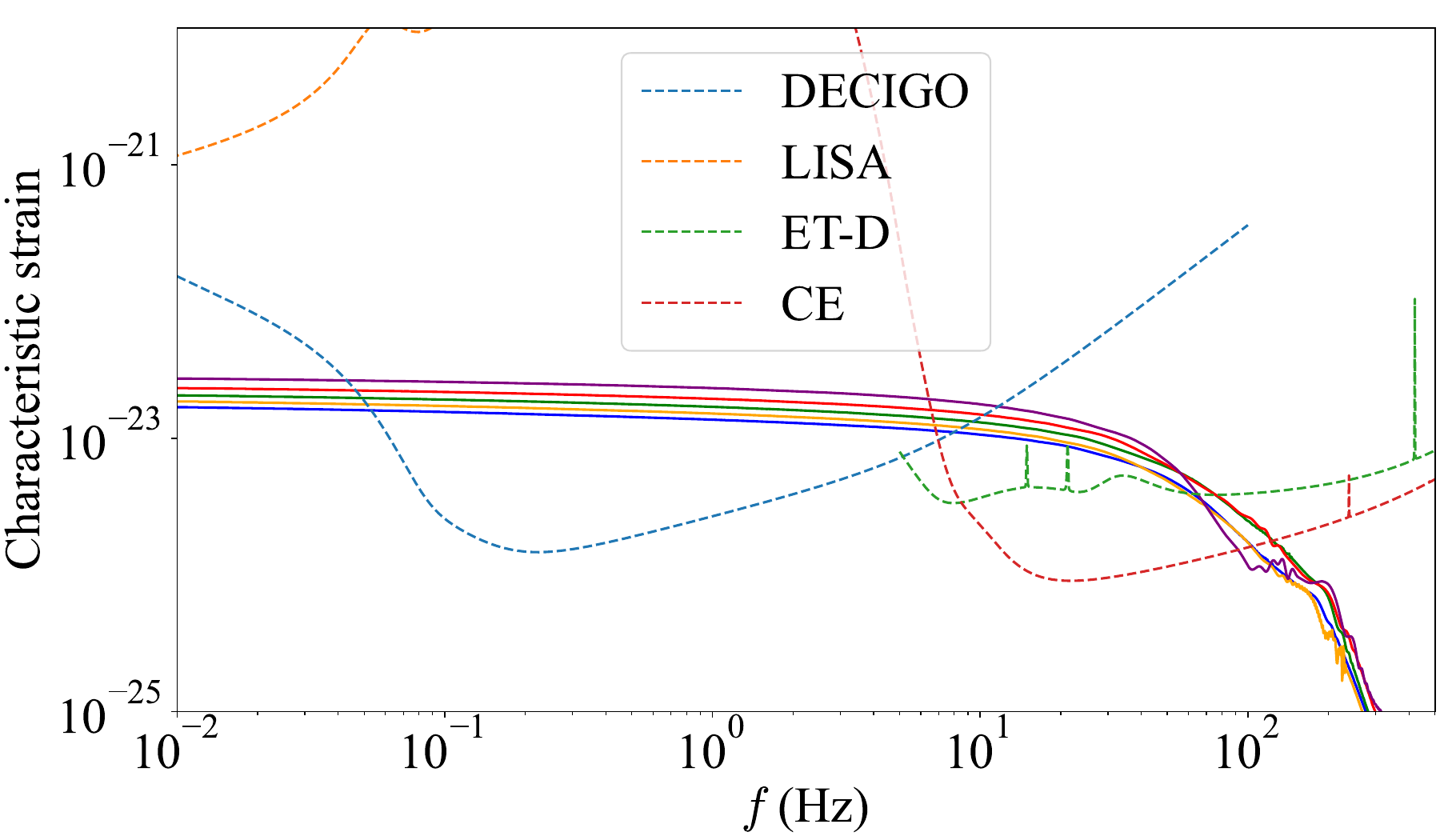}
	\caption{Memory waveforms.
		Upper panel: the time-domain memory waveform $h_{\text D+}$ for the plus polarization at different $S_{1z}$'s.
		Lower panel: the frequency-domain memory waveform $\tilde h_{\text D}$ corresponding to the time-domain waveforms.
		The sensitivity curves of DECIGO, LISA, CE and ET are also displayed.}
	\label{fig-mem}
\end{figure}

With PyCBC \cite{alex_nitz_2024_10473621}, one can simulate the memory waveforms both in the time and frequency domains.
For the purpose of demonstration, let us consider a GW150914-like binary black hole system \cite{gw150914}.
\textcolor{black}{We model the system as a spin algined binary
	The masses of the two black holes are $m_1=36M_\odot$ and $m_2=29M_\odot$, and the luminosity distance is taken to be 425 Mpc.
	We set the inclination angle $\iota=\pi/2$ to maximize the memory magnitude, and vary the spins.
	Here, $\iota$ is the angle between the line-of-sight and the angular momentum of the binary.}
Then, we scan over $S_{1z}$ and keep $S_{2z}=0$.
Figure~\ref{fig-mem} shows the memory waveforms.
The upper panel displays the time-domain displacement memory waveforms $h_{\text{D}+}$ at $S_{1z}=0,0.25,0.50,0.75$, and $0.99$ \cite{Boyle:2019kee}, and the lower panel shows the corresponding frequency-domain memory waveforms, that is, the characteristic strains $|f\tilde h_{\text D}|$, represented by the solid curves.
The dashed curves are the sensitivity curves, $\sqrt{f S_n}$, of DECIGO \cite{Seto:2001qf,decigo2019,Kawamura:2020pcg}, LISA \cite{Audley:2017drz}, Cosmic Explorer (CE) \cite{Evans:2016mbw} and Einstein Telescope (ET) \cite{Punturo:2010zza}.
The sensitivity curve of DECIGO is taken from Ref.~\cite{Hou:2024rgo}.
Those for CE and ET were computed using \verb+PyCBC+ \cite{alex_nitz_2024_10473621}.
LISA's power spectrum $S_n$ is given by \cite{Babak:2021mhe},
\begin{subequations}
	\label{eq-lisa-sn}
	\begin{gather}
		S_n=\frac{10}{3}\left[\frac{S_\text{I}(f)}{(2\pi f)^4}+S_\text{II}(f)\right]R(f),\\
		S_\text{I}=5.76\times10^{-48}\left[1+\left(\frac{f_1}{f}\right)^2\right]\text{Hz}^3,\\
		S_\text{II}=3.6\times10^{-41}\text{Hz}^{-1},\\
		R(f)=1+\left(\frac{f}{f_2}\right)^2,
	\end{gather}
\end{subequations}
where $f_1=0.4$ mHz, and $f_2=25$ mHz.
This power spectrum is very similar to the one designed for the 6-link configuration \cite{Babak:2021mhe}.
The oscillatory modes used for calculating these waveforms were generated assuming the IMRPhenomXPHM model \cite{Pratten:2020ceb}.
All available spherical modes for the oscillatory components were taken into account, and all permissible \footnote{According to Eqs.~\eqref{eq-phi-lm}, \eqref{eq-def-cl} and \eqref{eq-mem-1}, for a chosen pair of $h_{\hat\ell\hat m}$ and $h_{\tilde\ell\tilde m}$, $\mathcal C_{\ell m}(-2,\hat\ell,\hat m;2,\tilde\ell,\tilde m)$ might be vanishing for a lot of $(\ell,m)$ pairs.} spherical modes for the memory waveforms were also considered.
As shown by the upper panel, the final memory magnitude increases with $S_{1z}$.
According to the lower panel, one may conclude that LISA may not be able to detect the memory signals, while DECIGO, CE and ET are more likely to observe them.
One may also vary $\iota$ and keep other parameters fixed to study the change in the memory waveform as shown in Fig.~(2) in Ref.~\cite{Hou:2024rgo}.
There, the time-domain memory waveform indeed has the largest final magnitude at $\iota=\pi/2$.
\textcolor{black}{At $\iota$ near 0 or $\pi$, $h_{\text D+}$ oscillates while increasing, but $h_{\text D\times}$ is always oscillatory.
	$h_{\text D+}$ is an even function of $t$, while $h_{\text D\times}$ is odd, approximately.
	Both $ h_{\text D+}$ and $h_{\text D\times}$ are nonzero.
}

As one can see from the upper panel of Fig.~\ref{fig-mem}, the time-domain memory waveform at $\iota$'s near $\pi/2$ is quite featureless, and people may claim that the memory effect is basically a low frequency phenomenon \footnote{It might be possible that the approximate step-function like behavior of the time-domain memory signal is degenerate with the instrumental glitches or non-Gaussian noise.
	However, we will not discuss this possibility in the current work, as the focus is on the lensing effect.
	As discussed below, the step-function like behavior  disappears after the memory signal is strongly lensed.}.
This statement is actually consistent with the lower panel of Fig.~\ref{fig-mem}, where $|\tilde h_{\text D}|$ is larger at lower frequencies.
As in this work, we would like to consider the strong lensing effect of the memory signal, people may think this is not appropriate.
In fact, this is not a serious problem, as the strong lensing effect occurs when the wavelength of the signal is much smaller than the curvature radius of the lens.
For the memory signals that might be detected by DECIGO, CE and ET, there are plenty of large galaxies serving as the desired lenses.
Therefore, it is still possible to observe the strongly lensed memory signals, and in the next section, the gravitational lensing effect is to be reviewed.

\section{Gravitational lensing effect}
\label{sec-gle}

In the geometric optics limit, when the null radiation passes nearby a heavy object, the trajectory will be bent due to the gravitational pull exerted by the object, according to the equivalence principle \cite{Wald:1984rg}.
This is the gravitational lensing effect.
In the past, the lensing effect of the light was thoroughly studied.
Since the detection of the gravitational wave, the interest in its lensing effect has been revived \cite{Hou:2019dcm,Hou:2019wdg,Ezquiaga:2020gdt,Ezquiaga:2020dao,Xu:2021bfn}.
As pointed out in Ref.~\cite{Hou:2019wdg}, although there are some differences between the lensing effects of the light and the gravitational wave, they are largely similar.
In the following, we will briefly review the lensing effect of the light.

\begin{figure}[h]
	\centering
	\includegraphics[width=0.9\linewidth]{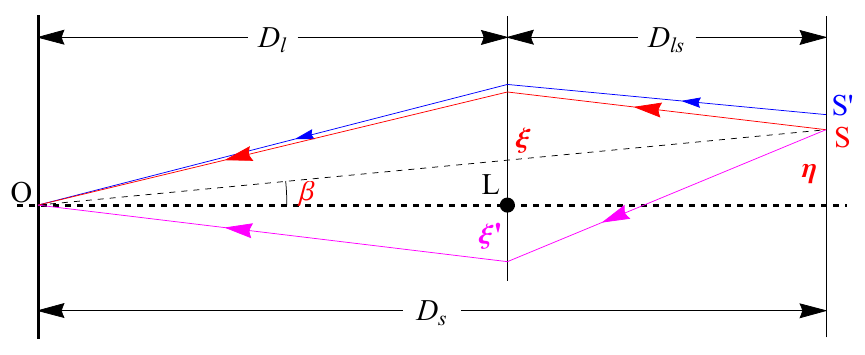}
	\caption{The typical geometry of a gravitational lensing system.
		O is the observer, L is the lens, and S, S' are two nearby sources.
		The vertical lines represent the observer, the lens, and the source planes, respectively.
		$D_l$, $D_s$, and $D_{ls}$ are the distances between these planes.
		The horizontal dashed line, connecting O and L, is the optical axis.
		$\vec\eta$ is the position of the source S relative to the intersection of the optical axis and the source plane.
		$\vec\xi$ is the position of the deflecting point of the red trajectory relative to the lens L on the lens plane, while $\vec\xi'$ is for the magenta trajectory.
		$\beta$ is the misalignment angle for S.
	}
	\label{fig-lge}
\end{figure}

Generally speaking, the trajectory of the photon is very far away from the central object, where the gravitational field can be approximated by the Newtonian potential $U$.
The metric is then given by
\begin{equation}
	\label{eq-met-le}
	\ud s^2\approx-\left(1+2U\right)\ud T^2+\left(1-2U\right)\delta_{ij}\ud X^i\ud X^j,
\end{equation}
in the quasi-global Minkowski coordinates $(T,X,Y,Z)$.
So the effective index of refraction of the gravitational field is $n\approx1-2U$, and according to the Fermat's principle \cite{gravlens1992}, the spatial direction $\vec e$ of the photon satisfies
\begin{equation}
	\label{eq-fmt-n}
	\frac{\ud \vec e}{\ud l}=2\vec e(\vec e\cdot\nabla U)-2\nabla U,
\end{equation}
where $l$ is the spatial length of the trajectory, and $\nabla$ represents the gradient, both defined with respect to the flat 3-metric $\delta_{ij}$.
If the initial direction of the photon is $\vec e_0$, and after passing by the lens, the photon changes direction, then, the final direction is
\begin{equation}
	\label{eq-ef}
	\vec e_f=\vec e_0+2\int [\vec e(\vec e\cdot\nabla U)-\nabla U]\ud l.
\end{equation}
One usually considers a geometrically-thin lens, whose mass is mainly distributed in a plane perpendicular to the photon trajectory.
If the surface mass density of the lens is $\Sigma(\vec \xi)$, the above equation can be rewritten as  \cite{gravlens1992},
\begin{equation}
	\label{eq-leq}
	\vec \eta=\frac{D_s}{D_l}\vec\xi-D_{ls}\vec\alpha(\vec\xi),
\end{equation}
where the deflection angle is given by
\begin{equation}
	\label{eq-defl}
	\vec\alpha(\vec\xi)=4\int\frac{\vec\xi-\vec\xi''}{|\vec\xi-\vec\xi''|^2}\Sigma(\vec\xi'')\ud^2\xi''.
\end{equation}
The quantities in these equations are displayed in Fig.~\ref{fig-lge}, where O represents the observer, L is the lens, and S, S' stand for two nearby sources.
The vertical lines represent the observer, the lens, and the source planes, respectively.
$D_l$, $D_s$, and $D_{ls}$ are the distances between these planes.
The horizontal dashed line, connecting O and L, is the optical axis.
Three photon trajectories are plotted.
The red and magenta ones are from S, while the blue one is from S'.
The blue trajectory will be useful for the later discussion.
The vector $\vec\eta$ measures the position of the source S relative to the intersection of the optical axis and the source plane.
$\vec\xi$ is the position of the deflecting point of the red photon trajectory relative to the lens L on the lens plane, while $\vec\xi'$ is for the magenta trajectory.
$\beta$ is called the misalignment angle for S.
$\vec\xi$ and $\vec\xi'$ determine the positions of the images.
There is a time delay between them, that is,
\begin{equation}
	\label{eq-td}
	\Delta t=\hat\phi(\vec\xi,\vec\eta)-\hat\phi(\vec\xi',\vec\eta),
\end{equation}
where the Fermat potential is defined as
\begin{equation}
	\label{eq-def-fp}
	\begin{split}
		\hat\phi(\vec\xi,\vec\eta)= & \frac{D_lD_s}{2D_{ls}}\left(\frac{\vec\xi}{D_l}-\frac{\vec\eta}{D_s}\right)^2 \\
		& -4\int\ud^2\xi''\Sigma(\vec\xi'')\ln\frac{|\vec\xi-\vec\xi''|}{\xi_0},
	\end{split}
\end{equation}
with $\xi_0$ a reference length scale.

For many lensing systems, the sources are at the cosmological distances from the lenses and the observer.
\textcolor{black}{However, the lens map still takes the same form as Eq.~\eqref{eq-leq}.}
But the distances $D_l$, $D_s$, and $D_{ls}$ are now replaced by the angular diameter distances.
More specifically, let $H_0$ be the Hubble constant, $\Omega_m$ and $\Omega_\Lambda$ be the matter and dark energy density parameters, respectively.
Assuming the spatial curvature of the universe is zero, the angular diameter distance between the observer and a celestial object at redshift $z$ is \cite{Weinberg:2008zzc},
\begin{equation}
	\label{eq-dang}
	D_A(z)=\frac{1}{H_0(1+z)}\int_0^z\frac{\ud z'}{\sqrt{\Omega_m(1+z')^3+\Omega_\Lambda}}.
\end{equation}
Then, $D_l=D_A(z_l)$, $D_s=D_A(z_s)$, and $D_{ls}=D_s-D_l(1+z_l)/(1+z_s)$, with $z_l$ and $z_s$ being the redshifts of the lens and source, respectively.
The time delay is also modified accordingly, that is,
\begin{equation}
	\label{eq-td-cos}
	\Delta t=\phi(\vec\xi,\vec\eta)-\phi(\vec\xi',\vec\eta),\quad \phi=(1+z_l)\hat\phi.
\end{equation}
In our work, we will set $H_0=70\text{ km s}^{-1} \text{Mpc}^{-1}$, $\Omega_m=0.3$ and $\Omega_\Lambda=0.7$ \cite{Weinberg:2008zzc}.

It is also convenient to reexpress these relations in terms of dimensionless variables.
Define
\begin{equation}
	\label{eq-def-xy}
	\vec\x=\frac{\vec\xi}{\xi_0},\quad \vec\y=\frac{D_l}{\xi_0 D_s}\vec\eta.
\end{equation}
The lens map is now
\begin{equation}
	\label{eq-leq-dim}
	\vec\y=\vec \x-\frac{D_lD_{ls}}{\xi_0D_s}\vec\alpha(\vec x),
\end{equation}
and the Fermat potential is
\begin{gather}
	\phi(\vec y,\vec x)=(1+z_l)\frac{\xi_0^2D_s}{D_lD_{ls}}\left[\frac{1}{2}(\vec x-\vec y)^2-\psi(\vec x)\right],	\label{eq-fp-dim}\\
	\psi(\vec x)=4\frac{D_lD_{ls}}{D_s}\int\ud^2x'\Sigma(\xi_0\vec x')\ln|\vec x-\vec x'|.
\end{gather}
In fact, Eq.~\eqref{eq-leq-dim} can be obtained by calculating $\nabla_{\vec x}\phi=0$.
In some literature, the overall factor in the front of the square brackets in Eq.~\eqref{eq-fp-dim} is omitted \cite{Liao:2019aqq}.

An extended source occupies a 2-dimensional region $\cS$ on the source plane, in which there are multiple point sources, such as S and S' shown in Fig.~\ref{fig-lge}.
The area of $\cS$ is
\begin{equation}
	\label{eq-area-s}
	S_s=\int_\cS\ud^2\eta.
\end{equation}
The image of the source on the lens plane is a 2-dimensional region $\cS'$, whose area is
\begin{equation}
	\label{eq-area-s-le}
	S_l=\int_{\cS'}\ud^2\xi.
\end{equation}
Assuming the areas are very small, one can check that
\begin{equation}
	\label{eq-area-ratio}
	S_s\approx\left|\det\left(\frac{\pd\vec\eta}{\pd\vec\xi}\right)\right|S_l=\left(\frac{D_s}{D_l}\right)^2\left|\det\left(\frac{\pd\vec y}{\pd\vec x}\right)\right|S_l,
\end{equation}
where $\pd\vec\eta/\pd\vec\xi$ and $\pd\vec y/\pd\vec x$ are  the Jacobian matrices.
Along the null geodesics, the photon number is conserved.
If the source emitted $\ud N$ photons with frequency $\nu_s$ during the time interval $\ud t_s$, then the same number of photons will pass the image $\cS'$ in the time interval $\ud t_l$ and the frequency becomes $\nu_l$.
At the observer, the received flux is
\begin{equation}
	\label{eq-flux}
	\cF_o=\nu_o\frac{4\pi}{S_l/D_l^2}\frac{\ud N}{\ud T_o},
\end{equation}
where $\nu_o$ is the photon frequency measured by the observer, $\ud t_o$ is the time period during which the observer receives the $\ud N$ photons, and $S_l/D_l^2$ is the solid angle subtended by the image $\cS'$ at the observer.
However, if there were no the lens, the flux would be
\begin{equation}
	\label{eq-flux-no}
	\cF'_o=\nu_o\frac{4\pi}{S_s/D_s^2}\frac{\ud N}{\ud T_o}.
\end{equation}
The ratio of the two fluxes gives the magnification factor,
\begin{equation}
	\label{eq-mag}
	\mu_I=\frac{\cF_o}{\cF'_o}=\frac{S_l}{S_s}\left(\frac{D_s}{D_l}\right)^2=\frac{1}{|\det(\pd\vec y/\pd\vec x)|}.
\end{equation}
Since in this work, the gravitational lensing of the gravitational wave is considered, the amplitude magnification factor is defined as
\begin{equation}
	\label{eq-amp-mag-gw}
	\mu=\sqrt{\mu_I}=\frac{1}{\sqrt{|\det(\pd\vec y/\pd\vec x)|}}.
\end{equation}
In many cases, $\mu>1$ for at least some of the images.

There are many lens models depending on the physical nature of the lens.
The simplest one is the point-mass model, that is, the lens is simply a spherical star, characterized by its mass $M_\text{L}$.
The length scale is usually chosen to be the so-called Einstein radius \cite{Takahashi:2003ix},
\begin{equation}
	\label{eq-er-pm}
	\xi_0=2\sqrt{M_\text{L}\frac{D_lD_{ls}}{D_s}}.
\end{equation}
The lens map is simple, which is given by
\begin{equation}
	\label{eq-lm-pm}
	y=x-\frac{1}{x},
\end{equation}
so there are always two images, at $x_{1,2}=(y\pm\sqrt{y^2+4})/2$.
The amplitude magnifications corresponding to these images are
\begin{equation}
	\label{eq-mus-pm}
	\mu_{1,2}=\frac{|x_{1,2}|}{\sqrt{x_1^2-x_2^2}}=\sqrt{\frac{1}{2}\pm\frac{y^2+2}{2y\sqrt{y^2+4}}},
\end{equation}
and the time delay is
\begin{equation}
	\label{eq-td-pm}
	\Delta T=4(1+z_l)M_\text{L}\left(\frac{x_1^2-x_2^2}{2}+\ln\left|\frac{x_1}{x_2}\right|\right).
\end{equation}
The Fermat potential is taken to be
\begin{equation}
	\label{eq-fp-pm}
	\phi(y,x)=4(1+z_l)M_\text{L}\left[\frac{(x-y)^2}{2}-\ln|x|\right],
\end{equation}
and so the image at $x_1$ is of type I and the one at $x_2$ is of type II.

A more realistic model is the singular isothermal sphere (SIS) model, in which the lens is an early-type galaxy.
The surface mass density of this model is given by \cite{gravlens1992}
\begin{equation}
	\label{eq-sis-md}
	\Sigma(\vec\xi)=\frac{\sigma^2}{2\xi},\quad \xi=|\vec\xi|,
\end{equation}
where $\sigma$ is the line-of-sight velocity dispersion.
For such a model, the length scale $\xi_0$ is usually taken to be
\begin{equation}
	\label{eq-def-xi0}
	\xi_0=4\pi\sigma^2\frac{D_lD_{ls}}{D_s}.
\end{equation}
So the lens map is
\begin{equation}
	\label{eq-leq-sis}
	y=x-\frac{x}{|x|}.
\end{equation}
Here, since the $\Sigma(\vec\xi)=\Sigma(\xi)$ is axisymmetric, only scalar quantities shown in the above equation, and one usually takes $y>0$.
When $y>1$, there can be only one image, at $x_*=y+1$, with the amplitude magnification factor given by
\begin{equation}
	\label{eq-mag-sis-1}
	\mu_*=\sqrt{\frac{1}{y}+1}.
\end{equation}
When $y\le1$, there are two images, at $x_\pm=y\pm1$.
The amplitude magnification factors are
\begin{equation}
	\label{eq-mag-sis-2}
	\mu_\pm=\sqrt{\frac{1}{y}\pm1}.
\end{equation}
The time delay between the two images is
\begin{equation}
	\label{eq-td-sis}
	\Delta T=32\pi^2\sigma^4(1+z_l)\frac{D_lD_{ls}}{D_s}y.
\end{equation}
Note that the Fermat potential can be taken to be \cite{Takahashi:2003ix}
\begin{equation}
	\label{eq-fp-sis}
	\phi(x,y)=16\pi^2\sigma^4(1+z_l)\frac{D_lD_{ls}}{D_s}\left[\frac{1}{2}(x-y)^2-|x|\right].
\end{equation}
When there is only one image, it corresponds to a local minimum of the Fermat potential; when there are two images, the one at $x_+$ is of type I, while the one at $x_-$ is of type II \cite{gravlens1992}.


In the above, we explicitly studied the photon's motion influenced by a massive celestial object to describe the lensing effect.
The motion of the gravitational wave also satisfies the similar properties.
Since the detection of gravitational waves relies on the precise waveform, one also needs to investigate how the waveform is lensed.
It turns out that this is not as trivial as one has expected.
In the following subsection, we will first review some basics of the wave optics, and then take the geometric optics limit to get the strongly lensed waveform.

\subsection{From wave optics to geometric optics}
\label{sec-wo-go}

In the wave optics, the propagation of the gravitational wave nearby the lens shall be derived by solving the perturbed Einstein's equation.
The metric is written as $g_{\mu\nu}=\hat g_{\mu\nu}+h_{\mu\nu}$, where $\hat g_{\mu\nu}$ is the background metric, i.e. Eq.~\eqref{eq-met-le}, and $h_{\mu\nu}$ is the small perturbation.
Following Ref.~\cite{Nakamura1999wo},  $h_{\mu\nu}$ can be expressed in terms of the polarization tensor $e_{\mu\nu}$ and a scalar field $\sH$ such that $h_{\mu\nu}=e_{\mu\nu}\sH$.
One can normalize $e_{\mu\nu}$ such that $e_{\mu\nu}e^{\mu\nu}=2$, so $\sH=e^{\mu\nu}h_{\mu\nu}/2$.
In the weak field limit, $e_{\mu\nu}$ is approximately a constant tensor \cite{Hou:2019wdg}.
Then, one has
\begin{equation}
	\label{eq-phi-eq}
	\pd_\mu(\sqrt{-\hat g}\hat g^{\mu\nu}\pd_\nu\sH)=0,
\end{equation}
with $\hat g$ the determinant of $\hat g_{\mu\nu}$.
By Eq.~\eqref{eq-met-le}, one can rewrite the above as
\begin{equation}
	\label{eq-phi-eq-2}
	(\nabla^2+\omega^2)\tilde\sH=4\omega^2U\tilde\sH,
\end{equation}
where one has already Fourier transformed $\sH$, i.e., $\sH(T,\vec X)=\int\ud\omega \tilde\sH(\omega,\vec X)e^{i\omega T}/2\pi$.
If there were no lens, the solution to Eq.~\eqref{eq-phi-eq} would be $\tilde\sH_0\propto e^{-i\omega R}/R$ with $R=|\vec X|$.
The presence of the lens modifies it, and one defines the amplification factor
\begin{equation}
	\label{eq-def-f}
	F(\omega,\vec X)\equiv\frac{\tilde \sH}{\tilde \sH_0}.
\end{equation}
It turns out that \cite{Nakamura1999wo,Takahashi:2003ix,Liao:2019aqq},
\begin{equation}
	\label{eq-fac}
	F(\omega,\vec X)=(1+z_l)\frac{\xi_0^2D_s}{D_lD_{ls}}\frac{i\omega}{2\pi}
	\int\ud^2x e^{-i\omega T_d(\vec\y,\vec\x)},
\end{equation}
where $T_d(\vec\y,\vec\x)=\phi(\vec\y,\vec\x)-\phi_m(\vec\y)$,
\begin{equation}
	\label{eq-def-phim-pm}
	\phi_m(\vec y)=4(1+z_l)M_\text{L}\left[\frac{(x_1-y)^2}{2}-\ln x_1\right],
\end{equation}
for the point-mass model, and
\begin{equation}
	\label{eq-def-phim}
	\phi_m(\vec y)=16\pi^2\sigma^4(1+z_l)\frac{D_lD_{ls}}{D_s}\left(y+\frac{1}{2}\right),
\end{equation}
for the SIS model.
One shall clearly distinguish $T_d$ from $\Delta T$ in the previous section.
In fact, $\Delta T=|T_d(\vec y,\vec x_1)-T_d(\vec y,\vec x_2)|$.

In the geometric optics limit, $\omega\rightarrow\infty$, so one has
\begin{equation}
	\label{eq-fac-inf}
	\lim_{\omega\to\infty}F(\omega,\vec X)=\sum_ak_a\mu_a e^{-i\omega T_d(\vec\y,\vec\x_a)},
\end{equation}
obtained using the saddle point approximation \cite{Nakamura1999wo}.
Here, the subscript $a$ labels different types of the images, corresponding to different extreme points $\vec\x_a$ of the Fermat potential $\phi(\vec\y,\vec\x_a)$.
$\mu_a$ is the amplitude magnification factor for the $a$-th image, and $k_a$ is a factor.
One can check that
\begin{equation}
	\label{eq-ka}
	k_a=\left\{
	\begin{array}{rl}
		1,                                       & \text{type I}   \\
		\displaystyle -i\frac{\omega}{|\omega|}, & \text{type II}  \\
		-1,                                      & \text{type III}
	\end{array}
	\right..
\end{equation}
That is, if $\vec\x_a$ is a local minimum of $\phi(\vec\y,\vec\x)$, then $k_a=1$; if it is a local maximum, then $k_a=-1$; \textcolor{black}{finally, if it is a saddle point, then $k_a=-i\omega/|\omega|$} \footnote{For the saddle point, one can compare this $k_a$ with that in Ref.~\cite{Takahashi:2003ix}, which is $e^{-i\pi/2}$.
	The factor $e^{-i\pi/2}$ violates the reality condition $F^*(\omega,\vec X)=F(-\omega,\vec X)$.}.
Such a $k_a$ guarantees that $F^*(\omega,\vec X)=F(-\omega,\vec X)$ even in the $\omega\rightarrow\infty$ limit.

Of course, in a realistic situation, $\omega$ cannot be infinite.
Usually, one requires \cite{Takahashi:2003ix}
\begin{equation}
	\label{eq-gol-c}
	4M_{\text Lz}\omega\gg1,
\end{equation}
in order to have the geometric optics approximation to hold.
Here, $M_{\text Lz}$ is just $(1+z_l)M_\text{L}$ for the point-mass model, while for the SIS model, it is
\begin{equation}
	\label{eq-def-mlz-sis}
	M_{\text Lz}\equiv4\pi^2\sigma^4(1+z_l)\frac{D_lD_{ls}}{D_s}.
\end{equation}
Therefore, for the ground-based interfereometers, $M_{\text Lz}\gtrsim 10^4M_\odot$ in order to have the strong lensing effect.
For LISA, Taiji and TianQin, one requires that $M_{\text Lz}\gtrsim 10^8M_\odot$ \cite{Hou:2019wdg}.
For DECIGO, the lower limit of $M_{\text Lz}$  is somewhere in between.

Now, one can consider the gravitational waveform if it is lensed.
Although there might be several images, let us consider them separately, and combine them together eventually.
It is the easiest to get the frequency-domain waveform,
\begin{equation}
	\label{eq-lgw-fd}
	\tilde h_{\mu\nu}(\omega,\vec X)=k_a\mu_a\tilde h_{\mu\nu}^{(0)}(\omega,\vec X)e^{-i\omega T_{d,a}},
\end{equation}
from the $a$-th image, where $T_{d,a}=T_d(\vec\y,\vec\x_a)$ for simplicity.
Therefore, the effect of the lensing is to magnify the amplitude by $\mu_a$, and shift the phase by $k_ae^{-i\omega T_{d,a}}$.
Evidently, the phase shift contains two contributions, one of which is due to the time delay $T_{d,a}$, and the other is $k_a$, related to the type of the image.
\textcolor{black}{$\mu_a$ is independent of the frequency, but the phase shift is frequency-dependent.}

The frequency-dependence of the phase shift makes it less trivial to compute the time-domain waveform, especially for the lensed signal from the saddle point.
Now, consider the case where the signal comes from the local minimum or the local maximum of the Fermat potential, and the time-domain waveform is given by
\begin{equation}
	\label{eq-lgw-td-mm}
	h_{\mu\nu}(T,\vec X)=k_m\mu_m h_{\mu\nu}^{(0)}(T-T_{d,m},\vec X),
\end{equation}
where the subscript $m$ stands for ``minimum'' or ``maximum''.
So in the time domain, the lensed signal is time shifted by $T_{d,m}$ relative to the would-be unlensed signal.
For the case of the local maximum, the lensed signal is flipped as $k_a=-1$ by  Eq.~\eqref{eq-ka}.
These signals arrive at the detector later than the would-be unlensed one, by $T_{d,m}$.

The lensed waveform from the saddle point is
\begin{equation}
	\label{eq-lgw-td-s}
	h_{\mu\nu}(T,\vec X)=\mu_s\int_{-\infty}^\infty\frac{\ud\tau}{\pi\tau}h^{(0)}_{\mu\nu}(T-T_{d,s}+\tau,\vec X),
\end{equation}
where the subscript $s$ stands for ``saddle''.
To get this result, one shall make use of the following result, i.e., for any function $\tilde\Phi(\omega)$,
\begin{equation}
	\label{eq-hf-if}
	\begin{split}
		\int_{0}^\infty\frac{\ud\omega}{2\pi}\tilde\sH(\omega)e^{i\omega T}
		= & \int_{-\infty}^\infty\frac{\ud\omega}{2\pi}\Theta(\omega)\tilde\sH(\omega)e^{i\omega T} \\
		= & \frac{\sH(T)}{2}+\int_{-\infty}^\infty\frac{\ud\tau}{i2\pi\tau}\sH(T+\tau),
	\end{split}
\end{equation}
where $\Theta(\omega)$ is the Heaviside step function, and
\begin{equation}
	\label{eq-ift-hsf}
	\Theta(\omega)=\int\ud\tau\left[\frac{\delta(\tau)}{2}+\frac{1}{i2\pi}\mathscr P\frac{1}{\tau}\right]e^{i\omega\tau},
\end{equation}
with $\mathscr P$ representing the principle part.
We have omitted the symbol $\mathscr P$ in Eq.~\eqref{eq-lgw-td-s} for simplicity.
Now, one can see that the lensed waveform Eq.~\eqref{eq-lgw-td-s} is more complicated.
This is because in the case of the saddle point, $k_s=i\omega/|\omega|$, referring to Eq.~\eqref{eq-ka}.
Then, the lensed waveform is not simply given by shifting the unlensed waveform and multiplying it by a numeric factor.
Instead, it is formally the interference of the time-shifted waveform, weighted by the reciprocal of the time shift, $1/\tau$, which is then multiplied by a numeric factor, $\mu_s/\pi$.

One may simplify Eq.~\eqref{eq-lgw-td-s}.
The oscillatory gravitational waveform $h_{\mu\nu}^{(0)}$ vanishes as long as the binary stars coalescence.
So there is a natural cutoff for the integral in Eq.~\eqref{eq-lgw-td-s}.
In an ideal situation, $h_{\mu\nu}^{(0)}$ becomes zero in the infinite past.
Moreover, due to the weighting factor $1/\tau$, the shifted waveform with $\tau$ close to zero contributes the most, so one may approximate this integral by
\begin{equation}
	\label{eq-lgw-td-s-app}
	h_{\mu\nu}(T,\vec X)\approx\mu_s\int_{-\tau_c}^{\tau_c}\frac{\ud\tau}{\pi\tau}h^{(0)}_{\mu\nu}(T-T_{d,s}+\tau,\vec X),
\end{equation}
for some small cutoff $\tau_c>0$.
In the limit of $\tau_c\rightarrow0$, one has
\begin{equation}
	\label{eq-lgw-td-s-app-0}
	h_{\mu\nu}(T,\vec X)\approx \frac{2\mu_s}{\pi}h^{(0)}_{\mu\nu}(T-T_{d,s},\vec X).
\end{equation}
This limit is appropriate for a highly oscillating signal.
One may use this to calculate the lensed waveform from the saddle point for simplicity.

If the time delay between some pair of images is shorter than the operating period of the interferometer, interference would happen and the beat pattern will be observed \cite{Hou:2019dcm}.
To get the expression for interference, one simply adds up Eqs.~\eqref{eq-lgw-td-mm} and \eqref{eq-lgw-td-s}.
For example, for a SIS lens, if there are two images at $\vec\x_+$ and $\vec\x_-$,
\begin{subequations}
	\begin{equation}
		\begin{split}
			\label{eq-th-2im-td}
			h_{\mu\nu}(T)= & \mu_+h^{(0)}_{\mu\nu}(T-T_{d,+})                                                    \\
			& +\mu_-\int_{-\infty}^\infty\frac{\ud\tau}{\pi\tau}h^{(0)}_{\mu\nu}(T-T_{d,-}+\tau),
		\end{split}
	\end{equation}
	where $T_{d,\pm}$ are evaluated at $\vec\x_\pm$.
	Its Fourier transform is
	\begin{equation}
		\tilde h_{\mu\nu}(\omega)=\left(\mu_+e^{-i\omega T_{d,+}}+i\mu_-\frac{\omega}{|\omega|}e^{-i\omega T_{d,-}}\right)\tilde h^{(0)}_{\mu\nu}.\label{eq-lh-2im}
	\end{equation}
\end{subequations}
One can derive similar expressions for the point-mass model.

Now, a comment is in order.
One shall realize that the low frequency modes of the \emph{oscillatory} components of the gravitational wave are much weaker than the high frequency ones.
So although they are also lensed and the geometric optics approximation does not hold for them, their contributions to the lensed waveform are negligible.
Therefore, Eqs.~\eqref{eq-lgw-td-mm} and \eqref{eq-lgw-td-s} are independent of a frequency cutoff $\omega_c$.
However, this will not be the case for the lensed memory waveform, as to be discussed in the following.

\section{Lensed memory waveforms}
\label{sec-glm}

To get the gravitationally lensed memory waveform, one applies the similar idea presented in Sec.~\ref{sec-wo-go}.
It is the simplest to write down the frequency-domain memory waveforms.
As in the previous section, we will also write down the lensed memory waveform for each image separately.
So for the $a$-th image, the lensed memory waveform is just
\begin{equation}
	\label{eq-lmem-fd-ka}
	\tilde h_{\text D}^{(\text L)}(\omega)=k_a\mu_a\tilde h_{\text D}(\omega)e^{-i\omega T_{d,a}},
\end{equation}
where $\tilde h_\text{D}$ is the unlensed memory waveform.
This takes exactly the same form as Eq.~\eqref{eq-lgw-fd}.
Of course, this equation holds for $\omega>\omega_c\gg1/4M_{\text Lz}$.
As an interferometer has a finite sensitivity band, one may require $\omega_c$ to be greater than the lower bound $\omega'_c$ of the band at the same time.
\textcolor{black}{To make the discussion simpler, we consider a massive enough lens such that $\omega'_c\gg1/4M_{\text Lz}$.
This ensures that the geometric optics approximation holds for all the frequency modes within the sensitivity band of the interferometer.}

It is a bit nontrivial to derive the time-domain memory waveform.
It has something to do with the very nature of the unlensed memory waveform, being a nonzero constant, roughly after the coalescence of the binary stars.
The time-domain memory waveform from the saddle point has to be carefully computed, otherwise one may get a diverging result.
To illustrate this divergence, let us directly make use of Eq.~\eqref{eq-lgw-td-s} to get,
\begin{equation}
	\label{eq-lmem-2im-td}
	h_\text{D}^{(\text L)}(T)=
	\mu_s\int_{-\infty}^\infty\frac{\ud\tau}{\pi\tau}h_\text{D}(T-T_{d,s}+\tau).
\end{equation}
This integral blows up logarithmically for any $T$.
This is because at a large enough $\tau\ge\tau_0$ for some $\tau_0$, the integrand is a nonvanishing constant, as shown in the upper panel in Fig.~\ref{fig-mem}.
The blowing up can also be explained by the fact that as the geometric optics approximation holds at a large enough angular frequency  $\omega\ge\omega_c$ with $\omega_c$ a suitable frequency cutoff, but in obtaining Eqs.~\eqref{eq-th-2im-td} and \eqref{eq-lmem-2im-td}, one  actually ignored this constraint.
This constraint does not cause any problem for the oscillatory part of the gravitational wave, as discussed in the last paragraph of the previous section, but it has to be properly handled for the memory effect.

Before moving on, we discuss briefly the necessity of investigating the time-domain lensed memory waveform.
Currently, the matched-filtering method \cite{Allen:2002jw,Allen:2005fk} uses the waveform in the frequency-domain, for the ground-based interferometer.
But when it comes to the space-borne detectors, one would have to face some issues if one continues to use the frequency-domain waveform.
This is because in order to take into account the response of detector arms, and the time-delay interferometry (TDI) technique \cite{Tinto:2014lxa}, one would have to find the correct transfer functions \cite{Marsat:2018oam,Marsat:2020rtl} to approximate.
Unfortunately, no such transfer function for the memory signal exists yet \cite{Garcia-Quiros:2025usi}.
Therefore, in the following, the lensed memory waveform will be derived in the time domain.

\subsection{Type I and type III images}
\label{sec-lmem-td-mm}

Firstly, consider the image of type I or type III.
To take the constraint mentioned above into account, one has to perform the following computation,
\begin{equation}
	\label{eq-lmem-td-mm}
	\begin{split}
		h_\text{D}^{(m)}(T)= & k_m\mu_m\int_{-\infty}^\infty\left[\Theta(\omega-\omega_c)+\Theta(-\omega-\omega_c)\right]     \\
		& \times   \tilde h_\text{D}(\omega)\exp[i\omega(T-T_{d,m})]\frac{\ud\omega}{2\pi}               \\
		=                    & k_m \mu_m \bigg[h_\text{D}(T-T_{d,m})                                                          \\
			& -\int_{-\infty}^\infty\ud\tau\frac{\sin\omega_c\tau}{\pi\tau}h_\text{D}(T-T_{d,m}+\tau)\bigg],
	\end{split}
\end{equation}
where Eq.~\eqref{eq-ift-hsf} has been applied.
The use of the Heaviside step function implements the constraint $|\omega|\ge\omega_c$.

So $h_\text{D}^{(m)}$ contains two parts.
The first part is proportional to $h_\text{D}$, which is simply the original memory signal, multiplied by the product $k_m\mu_m$ of the magnification factor and the phase factor, and then shifted in time by $T_{d,m}$.
This part looks like Eq.~\eqref{eq-lgw-td-mm}, but $h_\text{D}^{(m)}$ has one more part.

The second part, the last line, involves an integral, which is finite, even if $h_\text{D}$ becomes a nonzero  constant at a large enough $\tau$, because $1/\tau$ decays fast enough.
When $\omega_c\rightarrow0$, it disappears.
Since the weighting factor $\sin\omega_c\tau/\tau$ approaches $\omega_c$ as $\tau\rightarrow0$, the integrand is actually finite.
Although Eq.~\eqref{eq-lgw-td-s} is for the type II image of the oscillatory gravitational wave component, one may still compare it with the second part of Eq.~\eqref{eq-lmem-td-mm}.
One difference is the weighting factor.
The other difference is that one cannot set a small upper bound for the integral in Eq.~\eqref{eq-lmem-td-mm} as done for Eq.~\eqref{eq-lgw-td-s}.
This can be seen by reexpressing Eq.~\eqref{eq-lmem-td-mm}.
Let us assume the unlensed memory waveform reaches its maximum magnitude after $T_p$, then, one has
\begin{equation}
	\label{eq-lmem-td-mm-1}
	\begin{split}
		h_\text{D}^{(m)}(T)= & k_m \mu_m \bigg\{h_\text{D}(T-T_{d,m})                                                    \\
		& -\int_{-\infty}^{\tau_c}\ud\tau\frac{\sin\omega_c\tau}{\pi\tau}h_\text{D}(T-T_{d,m}+\tau) \\
		& -\left[\frac{1}{2}-\frac{1}{\pi}\text{Si}(\omega_c\tau_c)\right]h_{\text{max}}\bigg\},
	\end{split}
\end{equation}
where $\tau_c=T_p+T_{d,m}-T$, $\text{Si}(x)$ is the sine-integral function, and $h_{\text{max}}$ represents the final magnitude of the unlensed memory waveform.
By the property of $\text{Si}(x)$ \footnote{Please refer to \href{https://dlmf.nist.gov/}{NIST Digital Library of Mathematical Functions}.}, for a finite $\tau_c$, the above square brackets is of order one.
Therefore, one cannot place a small cutoff $\tau_c>0$.
In addition, although the middle line is complicated, the last line of Eq.~\eqref{eq-lmem-td-mm-1} is a oscillating function of $\tau_c$, i.e., $T$.
It oscillates around zero when $T-T_{d,m}>T_p$, and the oscillation becomes smaller and smaller with $T$.

In summary, Eq.~\eqref{eq-lmem-td-mm} gives the desired lensed memory waveform for type I and type III images.
It contains a monotonic increasing component, like the original memory waveform, and a second piece which is oscillating.
So the total waveform is oscillatory.

\textcolor{black}{As discussed, since we are interested in the strong lensing effect, we introduced the frequency cutoff $\omega_c$.
	When the signal is received by the interferometer, there would be an effective frequency cutoff $\omega'_c$ due to the finite sensitivity band of the interferometer.
	However, the effects of these two kinds of cutoffs are different.
	The cutoff $\omega_c$ completely erases the lower frequency modes of the memory signal, while $\omega'_c$ just marks a vague boundary, below which the interferometer is not sensitive enough. }

\subsection{Type II images}
\label{sec-lmem-td-s}

Secondly, consider the lensed memory signal from the type II image,
\begin{equation}
	\label{eq-lmem-td-s}
	\begin{split}
		h_\text{D}^{(s)}(T)= & \mu_s\int_{-\infty}^\infty[i\Theta(\omega-\omega_c)-i\Theta(-\omega-\omega_c)]\times         \\
		& \tilde h_\text{D}(\omega)\exp[i\omega(T-T_{d,s})]\frac{\ud\omega}{2\pi}                      \\
		=                    & \mu_s\int_{-\infty}^\infty\ud\tau\frac{\cos\omega_c\tau}{\pi\tau}h_\text{D}(T-T_{d,s}+\tau),
	\end{split}
\end{equation}
where in the first line, there is a minus sign before the second Heaviside step function, because of the special property of $k_a$ shown in Eq.~\eqref{eq-ka}.
Unlike Eq.~\eqref{eq-lmem-td-mm} for type I and type III images, this equation has just one part.
As one can check, this integral is finite unless $\omega_c=0$.
When $\omega_c=0$, it becomes Eq.~\eqref{eq-lgw-td-s}.
Compared with Eq.~\eqref{eq-lgw-td-s}, the weighting factor $\cos\omega_c\tau/\tau$ is different.
This factor peaks near $\tau=0$, and is unbounded, but the integral is still finite.
This can be checked by doing the following integral,
\begin{equation}
	\label{eq-chk-fi}
	\begin{split}
		& \int_{-\epsilon}^\epsilon\ud\tau\frac{\cos\omega_c\tau}{\pi\tau}h_\text{D}(T-T_{d,s}+\tau)                                 \\
		& \approx \epsilon\frac{\cos\omega_c\epsilon}{\pi\epsilon}[h_{\text D}(T-T_{d,s}+\epsilon)-h_{\text D}(T-T_{d,s}-\epsilon)],
	\end{split}
\end{equation}
where $0<\epsilon\ll1$.
Obviously, this is finite, so is Eq.~\eqref{eq-lmem-td-s}.

One cannot truncate the integral at some small cutoff $\tau_c>0$, either.
Let us rewrite Eq.~\eqref{eq-lmem-td-s} in the following form.
\begin{equation}
	\label{eq-lmem-td-s-1}
	\begin{split}
		h_\text{D}^{(s)}= & \mu_s\bigg[\int_{-\infty}^{\tau_c}\ud\tau\frac{\cos\omega_c\tau}{\pi\tau}h_\text{D}(T-T_{d,s}+\tau) \\
		& -\frac{1}{\pi}\text{Ci}(|\omega_c\tau_c|)h_{\text{max}}\bigg],
	\end{split}
\end{equation}
where $\text{Ci}(x)$ is the cosine integral function.
The second line is of the same order as $h_{\text{max}}$, i.e., the same order of the first line, so one cannot drop it.
As a final remark, the last line in Eq.~\eqref{eq-lmem-td-s-1} is also an oscillating function with a decreasing amplitude as long as $T-T_{d,s}>T_p$.
Therefore, like Eq.~\eqref{eq-lmem-td-mm}, the lensed memory waveform for type II images is also oscillatory.\\

According to the discussion in Sec.~\ref{sec-wo-go}, $\omega_c$ can be taken to be $\gg1/4M_{\text Lz}$ defined in Eq.~\eqref{eq-gol-c}.

\subsection{An example}
\label{sec-eg}

Now, one can plot the lensed memory waveforms to view.
For the purpose of demonstration, let us consider a lensing system whose lens is a point mass with $M_\text{L}=10^5M_\odot$ at $z_l=0.5$.
From Eq.~\eqref{eq-gol-c}, one knows that
\begin{equation}
	\label{eq-omc-c}
	\omega_c/2\pi\gg0.054 \text{ Hz}.
\end{equation}
We will set $\omega_c/2\pi=1$ Hz.
Let the binary system be at $z_s=1$, with masses $m_1=36M_\odot$ and $m_2=29M_\odot$.
Assume the spins are zero, and the inclination angle $\iota=\pi/2$.
In order to visualize the impact of the lensing effect on the memory waveform more clearly, we will consider a rather small time delay, on the order of seconds, so that we can view both of the lensed waveforms and their interference.
This requires the misalignment angle $\beta$ to be small, for example, $\beta=2.5\times10^{-4}$ arc sec.
Then, one can find out that $T_{d,1}=0$ and $T_{d,2}\approx2.8$ sec, so the time delay is about $\Delta T\approx2.8$ sec, which is short enough.
The amplitude magnification factors are $\mu_1\approx1.13$ and $\mu_2\approx0.89$.

\begin{figure}[h]
	\centering
	\includegraphics[width=0.45\textwidth]{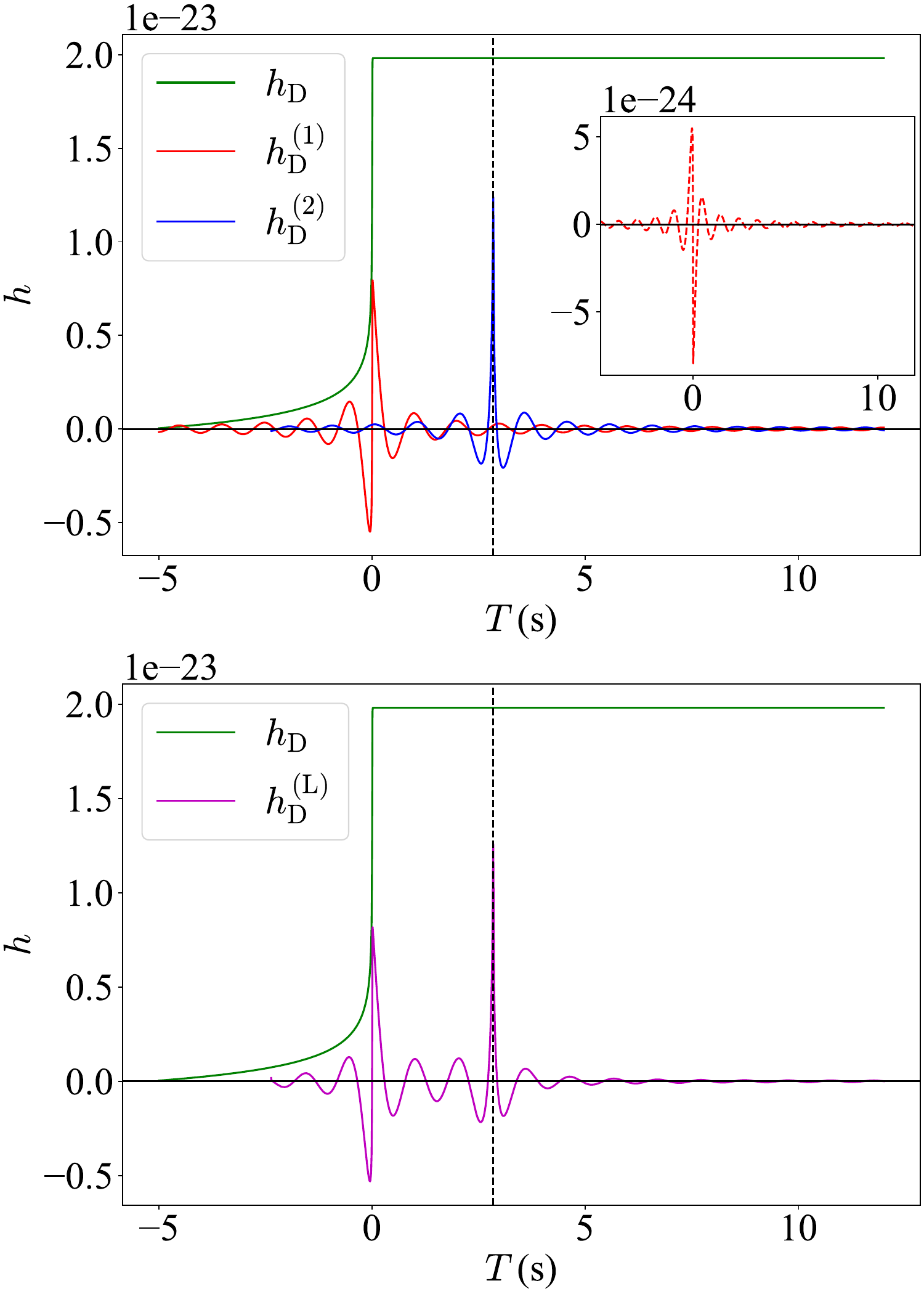}
	\caption{The lensed memory waveforms in the time domain.
		Upper panel: The lensed memory waveforms $h_\text{D}^{(1)}$ and $h_{\text{D}}^{(2)}$ from the two images at $x_1$ and $x_2$, respectively.
		If the first image were of type III, one would get the lensed waveform in the inset.
		Lower panel: The unlensed memory waveform $h_\text{D}$ and the interfered memory waveform $h_\text{D}^{(\text L)}$.
		In both panels, the vertical line is at $T=\Delta T\approx2.8$ s.}
	\label{fig-lmem}
\end{figure}
The lensed memory waveforms are plotted in Fig.~\ref{fig-lmem}.
In the upper panel, the lensed memory waveforms $h_\text{D}^{(1)}$ and $h_{\text{D}}^{(2)}$ from the two different images at $x_1$ and $x_2$ are drawn separately.
In the lower panel, the interfered memory waveform $h_\text{D}^{(\text L)}=h_\text{D}^{(1)}+h_\text{D}^{(2)}$ is displayed.
The unlensed memory waveform $h_\text{D}$ (the green curve) is also plotted in both panels to compare.
Although $\mu_1\ge1$, $h_\text{D}^{(1)}$ is quite small, because only higher frequency modes are strongly lensed, and the (unlensed) memory signal $h_\text{D}$ is dominated by the lower frequency components.
In addition, both $h_\text{D}^{(1)}$ and $h_\text{D}^{(2)}$ become oscillatory, consistent with Eqs.~\eqref{eq-lmem-td-mm-1} and \eqref{eq-lmem-td-s}.
They oscillate approximately at the frequency of $1.0$ Hz.
Both signals peak at certain times.
$h_\text{D}^{(1)}$ peaks at the very same time when the unlensed gravitational wave chirps, as $T_{d,1}=0$, while $h_\text{D}^{(2)}$ peaks at a later time, i.e., $T_{d,2}\approx2.8$ sec.
One can also find out that the red curve $h_\text{D}^{(1)}$ is nearly an odd function of time.
If one shifts the blue curve $h_\text{D}^{(2)}$ to the left by $T_{d,2}$, it becomes an even function, approximately.

The interference between $h_\text{D}^{(1)}$ and $h_\text{D}^{(2)}$ gives the magenta curve $h_\text{D}^{(\text L)}$ in the lower panel.
Unlike the oscillatory component of the gravitational wave, no beat pattern is observed \cite{Hou:2019dcm}.
If the time delay were much larger, the ground-based interferometer would detect $h_\text{D}^{(1)}$ and $h_\text{D}^{(2)}$ separately, and no interference would be observed.

For the point-mass or SIS model specifically considered in this work, the lensed image is either of type I or of type II.
For other lensing models, there might be images of type III.
The lensed memory waveforms from these images are similar to those from the type I except that they are flipped, as shown by the dashed curve in Fig.~\ref{fig-lmem}.

\textcolor{black}{
	One can vary $\omega_c$ to check how the lensed memory waveform changes.
	As $\omega_c$ increases, the lensed memory waveform becomes more and more oscillatory, and its amplitude becomes smaller and smaller, as shown in Fig.~\ref{fig-omc}.
	\begin{figure}[b]
		\centering
		\includegraphics[width=0.45\textwidth]{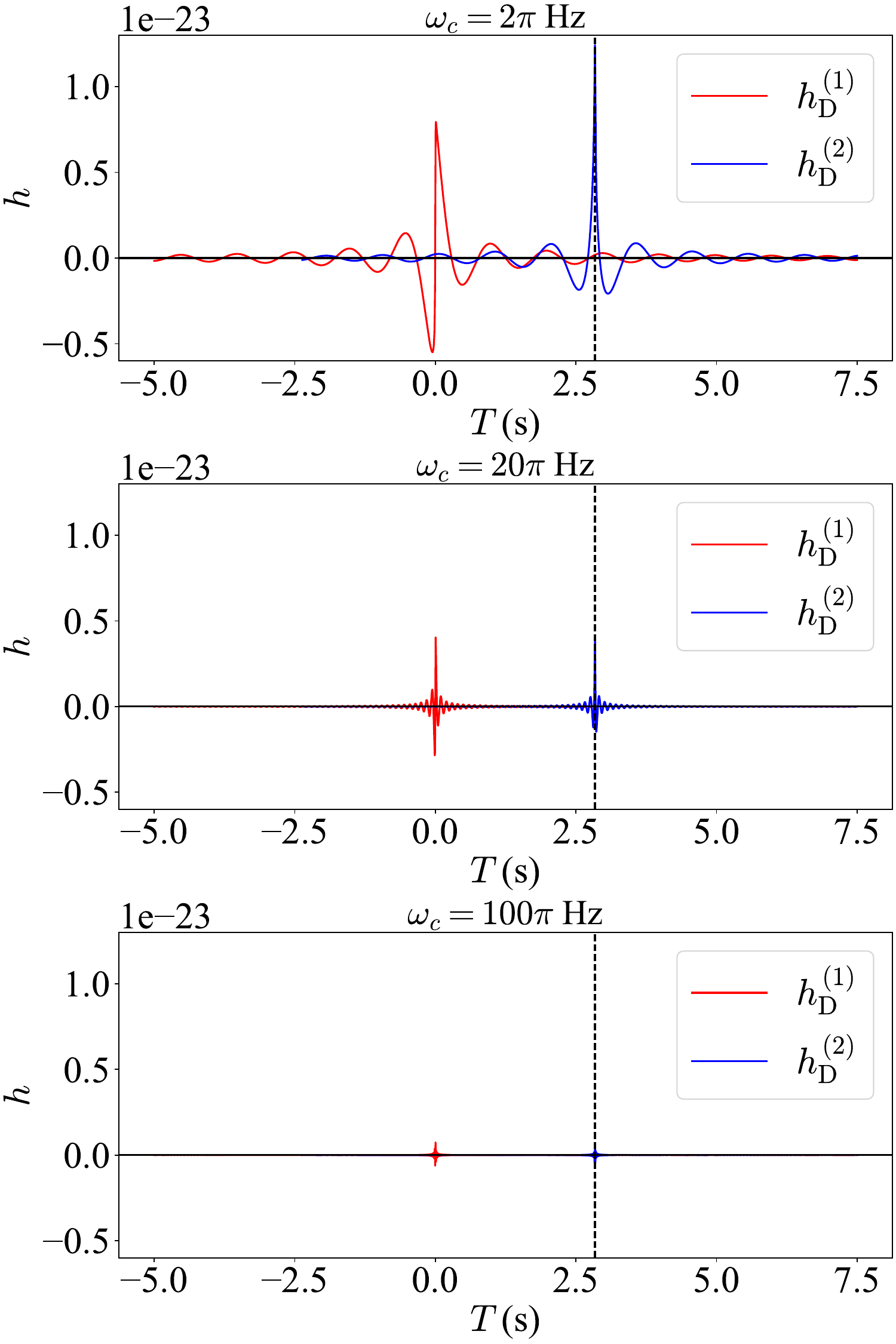}
		\caption{The lensed memory waveforms $h_\text{D}^{(1)}$ and $h_\text{D}^{(2)}$ for $\omega_c=2\pi$ Hz, $20\pi$ Hz, and $50\pi$ Hz.
			For the ease of comparison, we copied the red and blue curves in Fig.~\ref{fig-lmem}.
		}
		\label{fig-omc}
	\end{figure}
	This result is reasonable, since with the increase in $\omega_c$, less and less higher frequency modes of the memory signal are (strongly) lensed.
	The SNR's can also be calculated.
	If one measures the memory signals with ET-D or CE, then the results are shown in Table.~\ref{tab-snrs-1}.
	In this table, the SNR's of the lensed memory signals $h_\text{D}^{(1)}$ and $h_\text{D}^{(2)}$ of the two images, and their interfered signal $h_\text{D}^{(\text L)}$ at $\omega_c/2\pi=1$ Hz, $10$ Hz, and $50$ Hz are listed.
	We also calculated the SNRs of the unlensed memory signal $h_\text{D}$.
	The numbers outside of the parentheses are for ET-D, and those in the parentheses are for CE.
	\begin{table}
		\begin{tabular}{c|c|c|c}
			\hline\hline
			                         & $\omega_c/2\pi=1$ Hz & $\omega_c/2\pi=10$ Hz & $\omega_c/2\pi=50$ Hz \\
			\hline
			$h_\text{D}^{(1)}$       & 17.6(52.2)           & 13.1(51.3)            & 3.9(11.3)             \\
			$h_\text{D}^{(2)}$       & 13.9(41.2)           & 10.3(40.5)            & 2.9(8.9)              \\
			$h_\text{D}^{(\text L)}$ & 22.4(66.5)           & 16.5(65.3)            & 4.7(14.4)             \\
			$h_\text{D}$             & 15.6(46.1)           & 11.5(45.3)            & 3.3(10.0)             \\
			\hline
		\end{tabular}
		\caption{The SNR's of lensed memory signals $h_\text{D}^{(1)}$, $h_\text{D}^{(2)}$, and $h_\text{D}^{(\text L)}$ for $\omega_c/2\pi=1$ Hz, $10$ Hz, and $50$ Hz.
			The numbers outside of the parentheses are for ET-D, and those in the parentheses are for CE.
			The SNR's of the unlensed memory signal $h_\text{D}$ are also listed for comparison.}
		\label{tab-snrs-1}
	\end{table}
	It is clear from this table that the SNR's decrease with the increase in $\omega_c$.
	The SNR's of the second image are generally smaller than those of the first.
	That the magnification factor $\mu_2\approx0.89$ is less than $\mu_1\approx1.13$ may contribute to this result.
	This may also largely explain the numbers in the last row of the table are between those of $h_\text{D}^{(1)}$ and $h_\text{D}^{(2)}$.
	Since $h_\text{D}^{(\text L)}=h_\text{D}^{(1)}+h_\text{D}^{(2)}$, its SNR is greater than that of $h_\text{D}$.
	Due to the interference between $h_\text{D}^{(1)}$ and $h_\text{D}^{(2)}$ in the time-domain, the SNR of $h_\text{D}^{(\text L)}$ is not simply the sum of those of $h_\text{D}^{(1)}$ and $h_\text{D}^{(2)}$.
}

\textcolor{black}{
As discussed previously, in order for the geometric optics approximation to hold well, Eq.~\eqref{eq-gol-c} shall be satisfied.
$\omega_c$'s considered in Fig.~\ref{fig-omc} and Table.~\ref{tab-snrs-1} all fulfill this requirement.
The choice of $\omega_c$ is somewhat arbitrary, but as long as $\omega_c$ greater than $1/4M_{\text Lz}$ by one order of magnitude or so, one practically does not have to worry about the validity of the geometric optics approximation \cite{Liao:2019aqq}.
As computed in Eq.~\eqref{eq-omc-c}, $\omega_c/2\pi=1$ Hz is a reasonable choice.
}

\section{Universality of the lensed memory waveforms}
\label{sec-uni}

In the previous section, the lensed memory waveforms were displayed, provided that the lens is a point mass.
As discussed, the lensed memory waveforms from different types of the images all peak at certain times, around $T_{d,a}$.
If one shifts waveforms such that they all peak at $T=0$, one finds out that the lensed memory waveform corresponding to type I or type III image looks like an odd function, while the lensed waveform the type II image is roughly an even function.
In addition, the slope at $T=0$ is positive for the type I image, and it is negative for the type III image.
These characteristics can be expressed in the following way,
\begin{gather}
	h_\text{D}^{(m)}(2T_{d,m}-T)\approx -h_\text{D}^{(m)}(T),\\
	h_\text{D}^{(s)}(2T_{d,s}-T)\approx h_\text{D}^{(s)}(T),\\
	h_\text{D}^{'(\text{I})}(T_{d,\text{I}})>0,\quad h_\text{D}^{'(\text{III})}(T_{d,\text{III}})	<0,
\end{gather}
where the prime symbol means to take the derivative, and one does not shift the waveforms.
Actually, these features are universal.
They are independent of the lensed model and the binary star system, too.

This can be shown in the following way.
The \textcolor{black}{\emph{unlensed}} memory waveform $h_\text{D}(T)$ in the time-domain can be split into two parts,
\begin{equation}
	\label{eq-hd-split}
	h_\text{D}(T)=h_H(T)+h'_\text{D}(T),\quad h_H(T)=h_\text{max}\Theta(T),
\end{equation}
where $h_\text{max}$ is the maximal memory magnitude.
$h'_{\text{D}}(T)$ is a function of $T$ that dies out at $T=\pm\infty$.
It is smaller than $h_\text{max}$ by at least one order of magnitude.
Now, the Fourier transform of $h_H(T)$ is
\begin{equation}
	\label{eq-ft-hh}
	\tilde h_H(\omega)=\left[\frac{\delta(\omega)}{2}+\frac{1}{i2\pi\omega}\right]h_\text{max}.
\end{equation}
If this signal is strongly lensed with a frequency cutoff $\omega_c$, one can easily show that the lensed $h_H$ is
\begin{equation}
	\label{eq-lhh-td-mm}
	h_{H,\text{L}}^{(m)}(T)=k_m\mu_m h_\text{max}\left[\frac{|T|}{4\pi T}-\frac{\text{Si}(\omega_cT)}{2\pi^2}\right],
\end{equation}
if it corresponds to a type I or type III image.
If it corresponds to a type II image, one can show that
\begin{equation}
	\label{eq-lhh-td-s}
	h_{H,\text{L}}^{(s)}(T)=k_s\mu_sh_\text{max}\left[\frac{-\text{Ci}(\omega_c|T|)}{2\pi^2}\right].
\end{equation}
In fact, these results can be obtained by replacing $h_\text{D}$ by $h_H$ in Eqs.~\eqref{eq-lmem-td-mm} and \eqref{eq-lmem-td-s}.

\begin{figure}[h]
	\centering
	\includegraphics[width=0.45\textwidth]{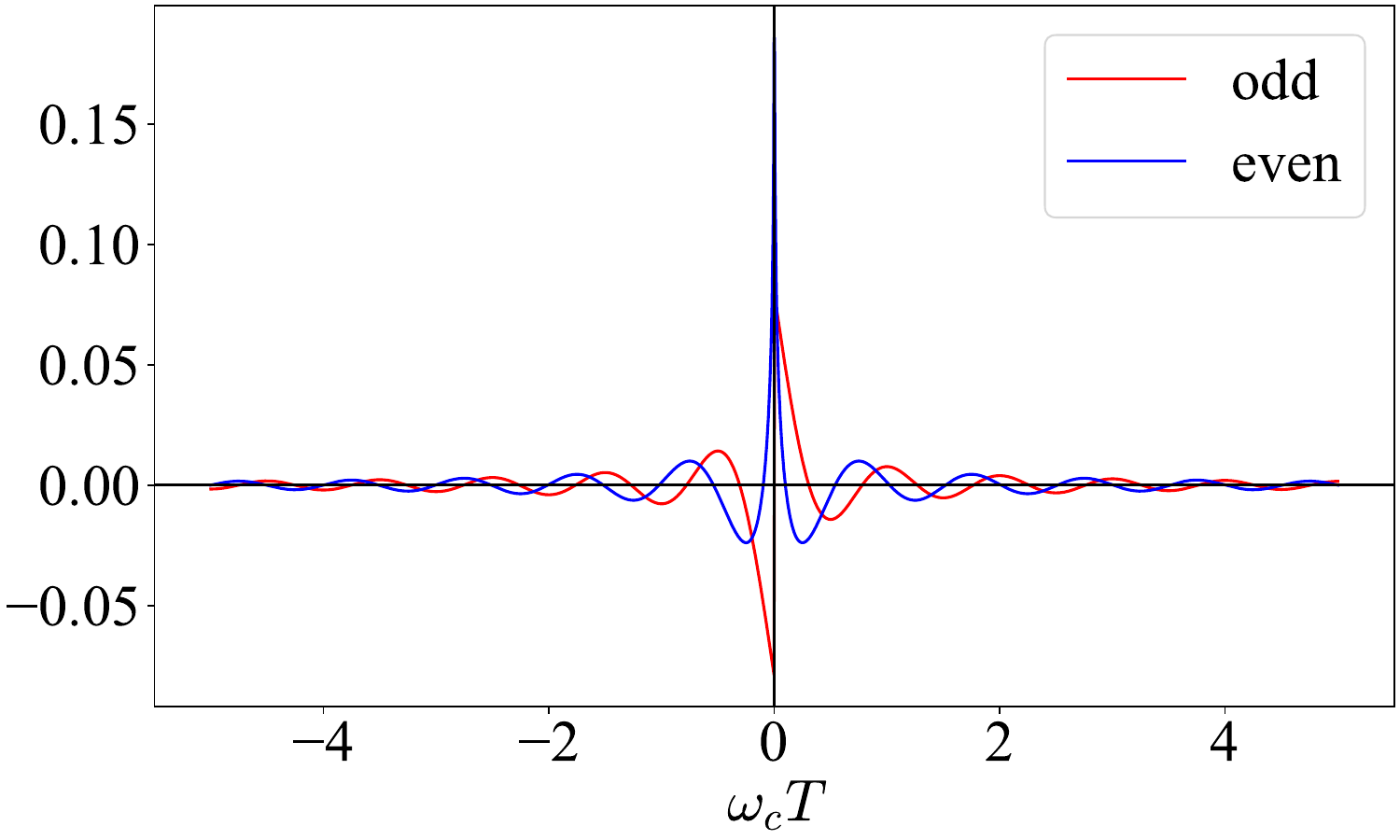}
	\caption{The expressions in the squared brackets of Eqs.~\eqref{eq-lhh-td-mm} and \eqref{eq-lhh-td-s}.
		The red and blue curves correspond to the squared brackets in Eqs.~\eqref{eq-lhh-td-mm} and \eqref{eq-lhh-td-s}, respectively.}
	\label{fig-inhev}
\end{figure}

One can easily plot the expressions in the squared brackets as shown in Fig.~\ref{fig-inhev}, where the red and blue curves correspond to the squared brackets in Eqs.~\eqref{eq-lhh-td-mm} and \eqref{eq-lhh-td-s}, respectively.
One immediately recognizes the similarities shared by the curves in the same color in Figs.~\ref{fig-lmem} and \ref{fig-inhev}.
Clearly, in Eqs.~\eqref{eq-lhh-td-mm} and \eqref{eq-lhh-td-s}, the squared brackets are universal, irrelevant to the details of the lens model and the binary system.
They can be used to explain the morphology of the lensed memory waveform observed in the previous section based on a specific example.
Since the expression in the squared brackets of Eq.~\eqref{eq-lhh-td-mm} is odd, the lensed memory waveform from the type I or type III image is also odd.
Similarly, Eq.~\eqref{eq-lhh-td-s} is clearly an even function, so the lensed memory waveform from the type II image is even.
Finally, by Fig.~\ref{fig-inhev}, the slope at $T=0$ for the type I image is positive as $k_\text{I}=1$, while it is negative for the type III image since $k_\text{III}=-1$.

There is also an intriguing characteristic of the blue curve, i.e., it has a sharp peak at $T=0$.
The direction in which the peak points to is upward in this plot.
A similar sharp peak can also be found in Fig.~\ref{fig-lmem} where the actual lensed memory waveforms were plotted.
This peak can be used to define ``a positive direction'' of the plot.
The slope at $T=0$ for the type I or III image is actually relative to this direction.
That is, if in a plot, the peak of the type II image points in the positive direction of the plot, then the slope at $T=0$ for the type I image is positive, and the slope for the type III image is negative.
However, if in a plot, the peak of the type II image points to the negative direction of the plot, then the slope for the type I image is negative, and for the type III image positive.
That one relates the sign of the slope of the type I or III image to the direction of the peak of the type II is important, because in the real detection, the observed signal shall be multiplied by the appropriate antenna pattern functions.
No matter what the types are, all images from the same binary star system shall also be multiplied by the same antenna pattern function $F_+$ \footnote{Here, by ``same'', we mean the same functional form for the antenna pattern function, but not the same value, if the space-borne interferometer is used to observe the lensed memory signal.}.
These functions have the property that when the polarization angle $\psi$ changes to $\psi\pm\pi/2$, they change their signs \cite{Poisson2014,Zhang:2020khm}.
So if one does not use the peak of the type II image to define a positive direction of the time-domain waveform plot, one cannot distinguish the type I image from the type III.
\textcolor{black}{A remark is in order.
Since $F_+$ actually changes in time for an space-borne interferometer, the observed lensed memory waveform $F_+(T)h_\text{D}^{(a)}(T)$ does not have exactly the same shape shown in Figs.~\ref{fig-lmem} and \ref{fig-omc}.
It seems that one cannot determine the direction of the peak of the type II image, or the slope of the type I or III image, by analyzing the observed lensed memory waveform.
However, this is not necessarily true.
One shall realize that $F_+(T)$ changes with time very slowly \cite{Zhang:2020khm}, on the order of $10^{-5}$ Hz, way below the sensitivity bands of many space-borne interferometers.
For the lensed memory signal that is detectible by these interferometers, $F_+(T)$ will not appreciably affect the morphology of the lensed memory waveform when the signal $h_\text{D}^{(a)}$ changes the fastest (e.g, when $T=0$ for the type I image, and $T=T_{d,2}\approx2.8$ sec for the type II image in Fig.~\ref{fig-lmem}).
}

\textcolor{black}{This observation suggests a possible novel way to determine the type of the lensed image, at least distinguishing the type II image from the type I and III.}
That is, one can use the morphology of the time-domain lensed memory waveform to tell.
Basically, one can construct the lensed memory waveforms of the three types, and try to match the observed signal.
Of course, in the real situation, this process is rather complicated.
One would have to generate a sufficiently large number of unlensed memory waveforms, and then lens them using Eqs.~\eqref{eq-lmem-td-mm} and \eqref{eq-lmem-td-s}.
\textcolor{black}{This gives us the desired waveform template.}
The lensed memory waveform would have a lot of parameters.
In addition to those parameterizing the unlensed memory waveforms, there are more: the amplitude amplification factor, the phase factor, the time delay, and the frequency cutoff.
\textcolor{black}{Then, one can use this waveform template to perform the parameter estimation, which is expected to take a very long time.}

A simpler approach can be taken.
Instead of generating the lensed memory waveform with the full expression \eqref{eq-hd-split}, one may simply use its approximation, i.e., the first term $h_H$ of Eq.~\eqref{eq-hd-split}.
It contains two parameters, $h_\text{max}$ and the moment when its value jumps, which is zero in this special case.
\textcolor{black}{The strong lensing effect would introduce 3 more parameters, in addition to $\omega_c$,} mentioned earlier.
Thus, it would be much easier to use the lensed memory waveform to tell the type of the lensed image.
For this to work, one would have to quantify the difference between $h_\text{D}$ and $h_H$.

\subsection{Approximating the memory waveform}
\label{sec-app-mem}

In order to make sure it is feasible to approximate the memory waveform, one may calculate the mismatch between the actual memory waveform $h_\text{D}(T)$ and the approximation $h_H(T)$.
To compute the mismatch, one shall first obtain the fitting factor (FF) between two signals $h_1$ and $h_2$, given by
\begin{equation}
	\label{eq-def-ff}
	\text{FF}=\text{max}_{\Delta\theta^a}\frac{\langle h_1|h_2\rangle}{\|h_1\|\|h_2\|}.
\end{equation}
Here, $\theta^a$ represents the collection of parameters that the waveforms depend on, such as the masses $m_1$ and $m_2$, the spins $\vec S_1$ and $\vec S_2$, and the distance $d_L$.
$\Delta\theta^a$ means to take the arbitrary variations in these parameters, and the maximal value is FF.
The angle brackets stand for the following inner product,
\begin{equation}
	\label{eq-def-inpro}
	\langle h_1|h_2\rangle=4\Re\int_0^\infty\frac{\tilde h_1^*(f)\tilde h_2(f)}{S_n(f)}\ud f,
\end{equation}
where $S_n(f)$ is one-sided noise power spectrum of the interferometer.
Finally, the norm $\|h\|$ is defined as $\sqrt{\langle h|h\rangle}$.
Once one computes FF, one gets the mismatch \cite{Apostolatos:1995pj},
\begin{equation}
	\label{eq-def-mm}
	\mathfrak M=1-\text{FF}.
\end{equation}
In this work, we considered LISA \cite{Audley:2017drz}, and its power spectrum is Eq.~\eqref{eq-lisa-sn}.

\begin{figure}[h]
	\centering
	\includegraphics[width=0.45\textwidth]{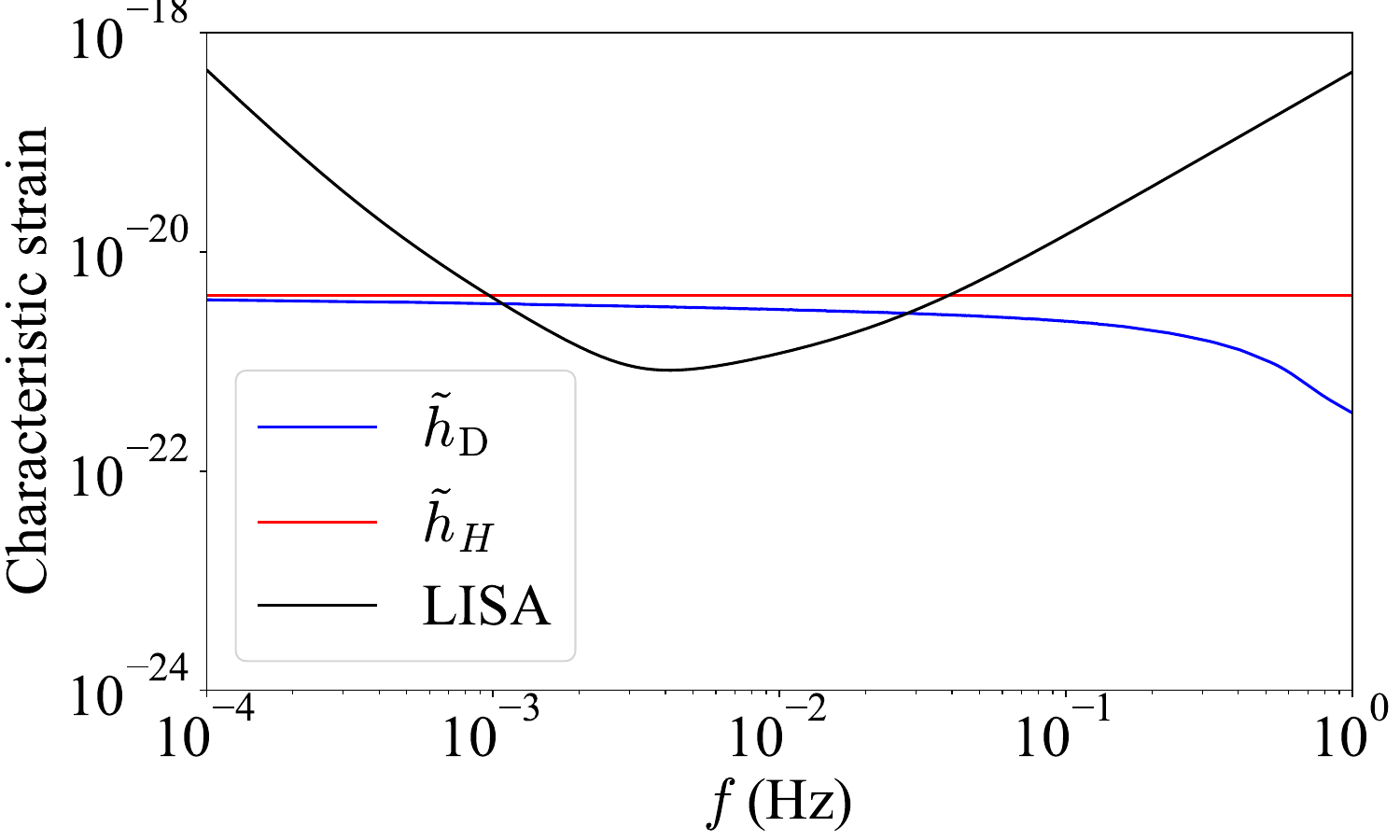}
	\caption{The actual memory waveform $\tilde h_\text{D}$ and the approximate one $\tilde h_H$.
		The black curve is LISA's noise curve.}
	\label{fig-mm}
\end{figure}

Now, to calculate $\mathfrak M$ between $h_\text{D}(T)$ and $h_H(T)$, one shall first simulate $\tilde h_\text{D}(f)$ with the method presented in Sec.~\ref{sec-gm}.
One can also use Eq.~\eqref{eq-f-dismm} to obtain $\lim_{f\rightarrow0}f\tilde h_\text{D}(f)$, which can be shown to be $h_\text{max}/i4\pi^2$.
Then, the approximate memory waveform in the frequency domain is
\begin{equation}
	\label{eq-cl-apm}
	\tilde h_H=\frac{\lim_{f'\rightarrow0}f'\tilde h_\text{D}(f')}{f},
\end{equation}
where the $\delta$-function part is ignored, as any interferometer could only detect signals with nonzero frequency.
Since we construct $\tilde h_H$ based on $\tilde h_\text{D}$, we will not vary any parameter of the memory signal when computing FF.
In Fig.~\ref{fig-mm}, we plotted $|f\tilde h_\text{D}(f)|$ generated by a binary system with $m_1=3600M_\odot$ and $m_2=3000M_\odot$, sharing all other parameters with the one in the previous section.
The approximate waveform $|f\tilde h_H(f)|$ is a horizontal line.
We also plotted LISA's sensitivity curve $\sqrt{fS_n(f)}$.
One can show that the mismatch between these two signals is about $10^{-3}$.
In doing the calculation, we set the frequency range from $10^{-4}$ Hz to 0 in Eq.~\eqref{eq-def-ff}.
Of course, one shall really compare the lensed $\tilde h_\text{D}$ and the lensed $\tilde h_H$.
However, since they share the same factor due to the lensing effect, i.e., $k_a\mu_a e^{-i\omega T_{d}}$, the mismatch remains the same as the one calculated for the unlensed waveforms.
\textcolor{black}{This is because,
	\begin{equation}
		\begin{split}
			\langle h_\text{D}^{(a)}|h_\text{H}^{(a)}\rangle	= & \langle k_a\mu_a e^{-i\omega T_{d}}h_\text{D}|k_a\mu_a e^{-i\omega T_{d}}h_\text{H}\rangle                                          \\
			=                                                  & 4\Re\int_0^\infty\frac{(k_a\mu_a e^{-i\omega T_{d}}\tilde h_\text{D})^*(k_a\mu_a e^{-i\omega T_{d}}\tilde h_\text{H})}{S_n(f)}\ud f \\
			=                                                  & \mu_a^2\langle h_\text{D}|h_\text{H}\rangle,
		\end{split}
	\end{equation}
	referring to Eq.~\eqref{eq-def-inpro}, and $|k_a|^2=1$.
	Similarly, one has $\|h_\text{D}^{(a)}\|^2=\mu_a^2\|h_\text{D}\|^2$ and $\|h_\text{H}^{(a)}\|^2=\mu_a^2\|h_\text{H}\|^2$, so the mismatch $\mathfrak M$ is unchanged.
}

\begin{figure}[h]
	\centering
	\includegraphics[width=0.45\textwidth]{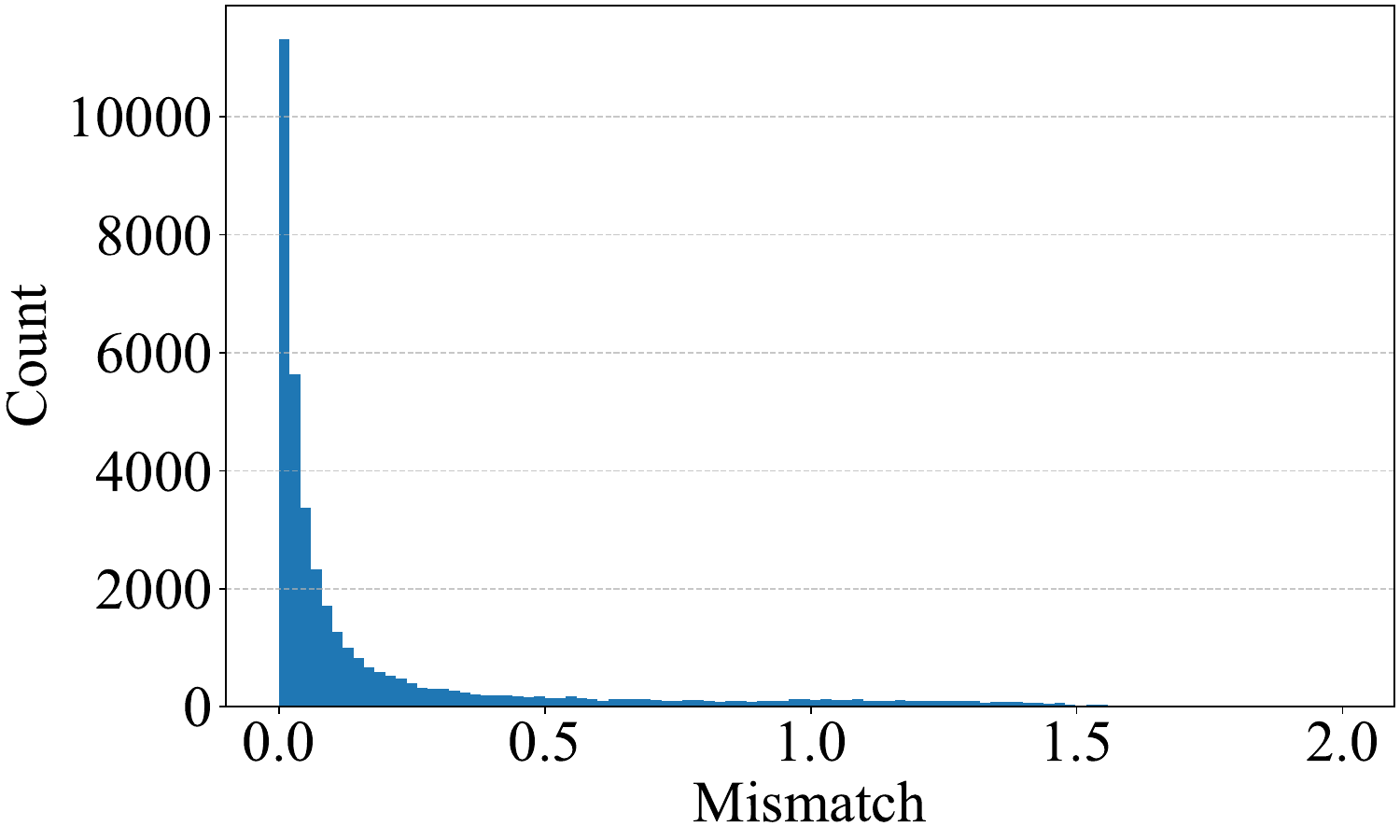}
	\caption{The histogram of the mismatches between the unlensed $\tilde h_\text{D}$ and the lensed $\tilde h_H$ for gravitational wave events observable by LISA.}
	\label{fig-mmhist}
\end{figure}

We would like to consider more mismatches for some possible sources of LISA.
We sampled the massive binary star systems based on the population model provided in Ref.~\cite{Barausse:2023yrx}, which was constructed after the recent Pulsar Timing Array observations \cite{Tarafdar:2022toa,EPTA:2023fyk,NANOGrav:2023gor,Reardon:2023gzh,Xu:2023wog}.
More specifically, we used ``HS-nod-SN-high-accr (B+20)'' model.
In the sampled binaries, the masses range from $10^3M_\odot$ to $10^6M_\odot$.
The redshift can be as high as $19\sim20$.
We sampled the source redshift, component masses in the source frame, and the six spin parameters from the population model, converted the redshift into a luminosity distance using our fiducial cosmology, and drew the extrinsic parameters -- right ascension, declination, inclination $(\iota)$, polarization angle $(\psi)$, and coalescence phase $(\phi_c)$ -- from isotropic priors (uniform on the sphere for sky position, uniform in $\cos\iota$, $\psi\sim\mathcal U[0,\pi)$, $\phi_c\sim\mathcal U[0,2\pi)$).
The coalescence time was sampled uniformly within a contiguous four-year observing window.
Then, the oscillatory components of the gravitational wave were also obtained with IMRPhenomXPHM using PyCBC, and the frequency-domain memory waveforms were computed using the equations displayed in Sec~\ref{sec-gm}.
The approximate memory waveforms could be calculated with Eq.~\eqref{eq-cl-apm}.
And eventually, the mismatches were determined.
In the following histogram~\ref{fig-mmhist}, we plotted the distribution of the mismatches.
As one can see, most of the events have a small mismatch.
In fact, about 49\% of the mismatches are less than 0.05, and 17\% are less than 0.01.
Therefore, there are quite a few gravitational wave events whose memory waveforms can be well approximated by $\tilde h_H$.
It would be easier to search for these memory signals.

\begin{figure}[h]
	\centering
	\includegraphics[width=0.45\textwidth]{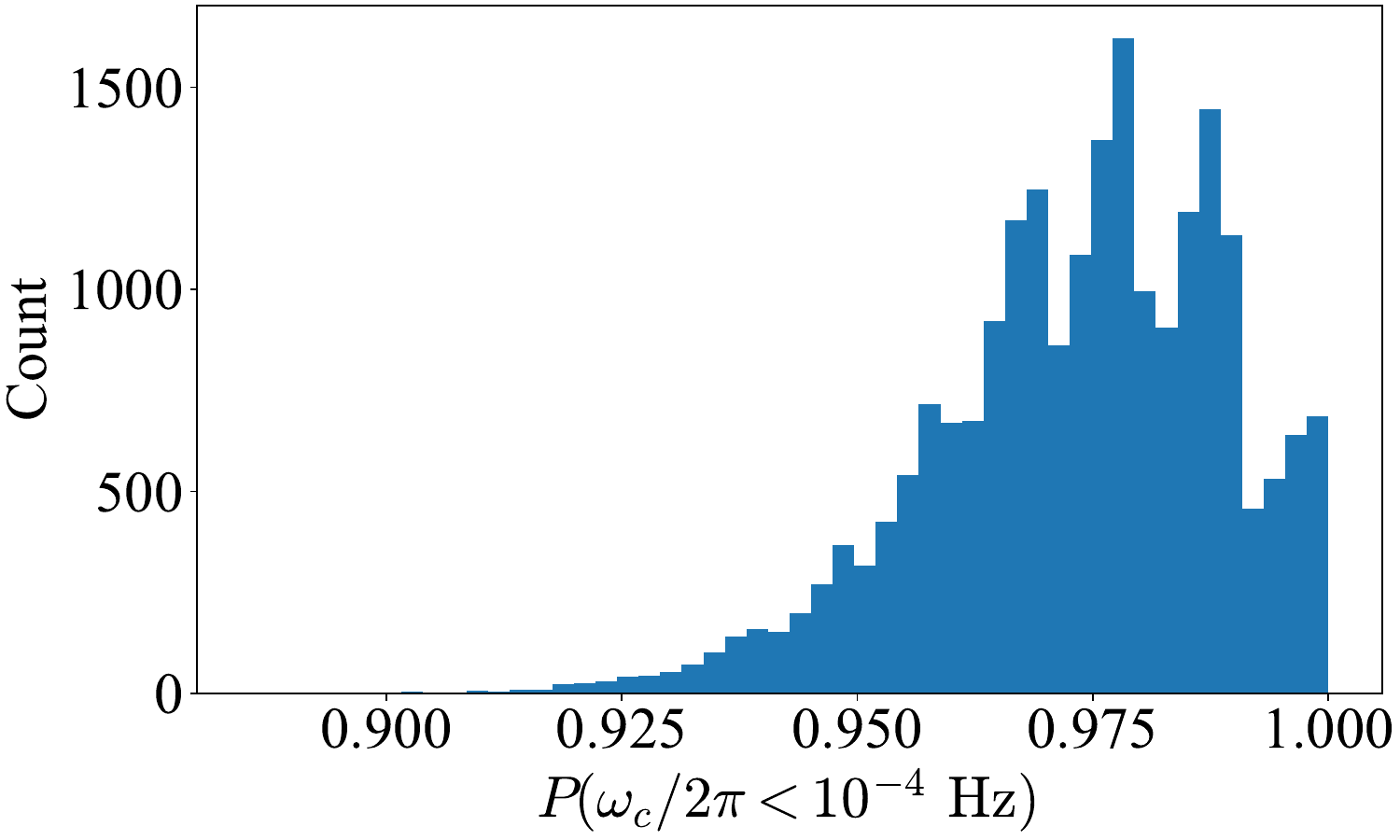}
	\caption{The histogram of the probability of having a smaller $\omega_c/2\pi<10^{-4}$ Hz.}
	\label{fig-probs}
\end{figure}

Of course, we estimated the mismatch between the exact memory waveform $h_\text{D}$ and the approximated one $h_H$.
It is not exactly the mismatch between the lensed waveforms, because although FF does not depend on the common factor multiplying $h_\text{D}$ and $h_H$, it is related to the frequency cutoff $\omega_c$.
Indeed, in Eqs.~\eqref{eq-def-ff} and \eqref{eq-def-inpro}, the lower integration limit shall be $\text{max}\{\omega_c/2\pi,10^{-4}\text{ Hz}\}$.
Therefore, in Appendix~\ref{app-cf}, we estimated the distribution of $\omega_c$ for each $z_s$ of the binary system sampled above.
We then calculated the probability of having a smaller $\omega_c/2\pi<10^{-4}$ Hz.
The histogram in Fig.~\ref{fig-probs} shows the distribution of the probability.
As one can see nearly all gravitational wave events have a huge probability $(0.95,1)$ such that $\omega_c/2\pi<10^{-4}$ Hz.
So for these events, one can simply set the lower integration limit in Eq.~\eqref{eq-def-inpro} to $10^{-4}$ Hz.
This means that the computed mismatches between $h_\text{D}$ and $h_H$ can be used to well approximate those for the lensed waveforms.

\subsection{Identifying the image types}
\label{sec-ident}

From the above discussion, one may propose the following strategy for identifying the image type.
At the very first step, one shall make use of Eq.~\eqref{eq-lhh-td-s} or its Fourier transform to match the detected signal.
The best scenario is that they match well, so the gravitational wave must be lensed, and the image is of type II.
Then, one shall use the accurate waveforms to further estimate the parameters of the source.
When using Eq.~\eqref{eq-lhh-td-s}, one shall combine all constant factors into a single factor.
It really does not matter whether one can measure each individual factor accurately in the very step in which one merely wants to determine if the image is of type II.
In addition, Eq.~\eqref{eq-lhh-td-s} shall also be modified slightly by replacing $T$ by $T-T_c$, where $T_c$ is the coalescence time.

Of course, there is always the possibility of the worse situation.
That is, if Eq.~\eqref{eq-lhh-td-s} does not match the detected signal well, the gravitational wave is either unlensed, or of type I or III.
In this case, one may have to use the more conventional methods \cite{Hannuksela:2019kle,McIsaac:2019use,Li:2019osa,Ezquiaga:2023xfe} for determining whether the event is lensed or not, for example, searching for multiple events in a catalog that share similar intrinsic parameters and sky locations \cite{Hannuksela:2019kle,LIGOScientific:2021izm,LIGOScientific:2023bwz,Janquart:2023mvf}.
And if it is lensed, the image type has to be determined later using the method proposed in Ref.~\cite{Janquart:2021nus}, for instance.
It shall be emphasized that if one knows the event is lensed somehow, one cannot use Eq.~\eqref{eq-lhh-td-mm} or the exact waveform to tell if the image is of type I or III.
This is because the antenna pattern functions, even for the space-borne interferometers, change sign if the polarization angle $\psi\rightarrow\psi\pm\pi/2$ \cite{Poisson2014,Zhang:2020khm}.
Therefore, one cannot use Eq.~\eqref{eq-lhh-td-mm} to easily distinguish the type I image from the type III.

This sounds pessimistic.
However, one shall realize that the time delays between multiple images are generally a few weeks or a few months.
The space-borne interferometers usually run for a couple of months or several years.
So if a gravitational wave is strongly lensed, it is possible that the interferometer detects all images.
One may get a lensed memory signal like the magenta curve in the lower panel of Fig.~\ref{fig-lmem}.
To search for such kind of signal, one would have to construct an appropriate template which contains multiple terms given by Eqs.~\eqref{eq-lhh-td-mm} and \eqref{eq-lhh-td-s} with suitable overall factors and time shifts.
If this kind of template matches the detected signal well, one can proceed to identify the types of the detected images.
For that, one shall make use of the direction of the peak of the type II image as a reference, if there exists at least one type II image.
So, one shall first identify the type II image, after drawing the detected memory signal in a plot.
Then, one finds the direction in which the peak points.
Finally, one can determine the types of the remaining images, according to the previous discussion.
That is, if the peak of the type II image points upward, the type I image shall have a positive slope at its own peak, while the type III image would have a negative slope.
On the contrary, if the peak of the type II image points downward, the type I image has a negative slope, and the type III positive.
This could be a very feasible method, because there could be at least 90\% of the multiple-image events whose second brightest images are of the type II \cite{Wang:2021kzt}.

Let us discuss the advantages and disadvantages of this method.
The main advantages of this method include the faster speed and the less computational cost.
As discussed previously, one does not need the exact memory waveform, which contains 15 parameters for a generic binary system.
It is sufficient to use the approximation $h_H$ with only 2 parameters.
It would be much easier and faster to simulate the approximated memory waveform.
The conventional methods rely on the accurate oscillatory gravitational waveforms, which also involve 15 parameters.
There are additional parameters to describe the lensing effect.
So these methods are definitely more time-consuming.
The big disadvantage of the current method is that the memory signal is weaker than the oscillatory part of the gravitational wave, the method proposed here may not be more efficient compared to the conventional ones.
Indeed, the signal-to-noise ratio (SNR) of the memory signal is smaller than that of the oscillatory part.
But in some cases, the memory SNR can be as large as a few tens or even one hundred for LISA \cite{Gasparotto:2023fcg}.
So the memory SNR may not be terribly small.
However, this does not imply that one shall use this method exclusively.
Instead, one may combine this method with the conventional ones to cross-check the results.
After all, this method is fast and cheap.

\section{Conclusion}
\label{sec-con}

In this work, we discussed the impact of the strong lensing effect on the memory waveform.
After passing by the lens, the originally monotonic memory waveform becomes oscillatory.
The morphology of the lensed memory waveform in the time domain is closely related to the type of the image.
For the type I and III images, corresponding to the local minima and maxima of the Fermat potential, the lensed memory waveform is approximately an odd function with respect to a symmetry axis.
At the axis, the type I image has a positive slope, while the type II image has a negative slope.
For the type II image, corresponding to the saddle point of the Fermat potential, the lensed memory waveform is nearly an even function.
These characteristics of the morphology largely owe to the fact that the unlensed memory waveform is dominated by a Heaviside step function component.
Therefore, the lensed memory waveform has the universal property, independent of the lens model and the parameters of the binary system.

These universal features may be used to help distinguish the type of the lensed gravitational wave, which has advantages.
This may greatly speed up the processes of identifying the lensed gravitational wave event, and determining the type of the image.
One may also combine this method with the conventional ones for the sake of cross check.
Once one knows the image type, one can use the correct lensed oscillatory gravitational waveform to estimate the parameters of the source.
Although the detection rate of the sufficiently loud lensed memory signal is expected to be low, as long as the space-borne interferometer observes such an event, one can quickly recognize it and make use of it for various physics purposes, such as detecting dark matter \cite{Cutler:2009qv,Camera:2013xfa,Congedo:2018wfn,Jung:2017flg,Tambalo:2022wlm} and measuring the cosmological constant \cite{Sereno:2010dr,Sereno:2011ty,Liao:2017ioi}.
It would be also interesting to apply this idea to other interferometers, which would be carried out in the future.

\begin{acknowledgments}

	This work was supported by the National Key Research and Development Program of China, grant No.~2024YFC2207400, the National Natural Science Foundation of China under grant Nos.~11633001 and 11920101003, and the Strategic Priority Research Program of the Chinese Academy of Sciences, grant No.~XDB23000000.
	R. Z. was also supported by Key Scientific Research Project of Universities in Henan Province under Grant No.~25B140006, and Young Backbone Teacher Program of Henan University of Technology under Grant No.~21421218.
	S. H. was also supported by the National Natural Science Foundation of China under Grant No.~12205222.

\end{acknowledgments}

\appendix

\section{Estimate the frequency cutoffs}
\label{app-cf}

In Section~\ref{sec-app-mem}, we computed the mismatches between the exact memory waveform $h_\text{D}$ and the approximate $h_H$.
This does not actually give us the mismatch between the lensed waveforms.
Although FF does not depend on any overall factor multiplying both $h_\text{D}$ and $h_H$, it depends on the limits of the integration, referring to Eqs.~\eqref{eq-def-ff} and \eqref{eq-def-inpro}.
More specifically, since we consider the strong lensing effect, which occurs when $\omega\gg\omega_c=1/4M_{\text Lz}$, the lower limit in Eq.~\eqref{eq-def-inpro} shall be $\text{max}\{\omega_c/2\pi,10^{-4}\text{ Hz}\}$.
So one has to determine $\omega_c$, which according to Eq.~\eqref{eq-def-mlz-sis}, is related to $z_l,z_s$, and $\sigma$.
For the binary systems considered in Section~\ref{sec-app-mem}, $z_s$'s were already sampled.
One has to sample $z_l$ and $\sigma$.

\textcolor{black}{Here, if we take $n$ to be the lens number density, then the differential number density of the lens population is assumed to be \cite{Choi:2006qg}}
\begin{equation}
	\label{eq-msf}
	\frac{\ud n}{\ud\sigma}=\frac{n_*}{\sigma_*}\frac{\gamma}{\Gamma(\alpha/\gamma)}\left( \frac{\sigma}{\sigma_*} \right)^{\alpha-1}\exp\left[ -\left( \frac{\sigma}{\sigma_*} \right)^\gamma \right],
\end{equation}
where $\Gamma(x)$ is the gamma function, and
\begin{gather*}
	n_*=8.0\times10^{-3}h^3\text{ Mpc}^{-3},\quad\sigma_*=161\pm5\text{ km/s},\\
	\alpha=2.32\pm0.10,\quad\gamma=2.67\pm0.07.
\end{gather*}
This basically tells us the probability density of $\sigma$, which can be used to sample $\sigma$.

In order to sample $z_l$, one needs to compute the differential optical depth $\ud\tau_o/\ud z_l$.
For this, one knows that \citep{Sereno:2010dr}
\begin{equation}
	\label{eq-dtau}
	\frac{\pd^2\tau}{\pd z_{l}\pd\sigma}=\frac{\ud n}{\ud\sigma}s^*_\text{cr}\frac{\ud t}{\ud z_{l}}.
\end{equation}
Here, $t$ is the cosmological comoving time, and $s_\text{cr}^*$ is the cross-section of the lensing \cite{Biesiada:2014kwa},
\begin{equation}
	\label{eq-def-scr}
	s_\text{cr}^*=16\pi^3\sigma^4\left(\frac{D_lD_{ls}}{D_s}\right)^2y_\text{max}^2.
\end{equation}
Following \cite{Piorkowska-Kurpas:2020mst}, we set $y_\text{max}=1/2$, so that the lensed memory signal has a large enough SNR.
Then, one can integrate Eq.~\eqref{eq-dtau} to get
\begin{equation}
	\label{eq-dtau-zl}
	\frac{\ud\tau}{\ud z_{l}}=\frac{16\pi^3n_*\sigma_*^4D_l^2D_{ls}^2}{(1+z_{l})H(z_{l})D_s^2}\frac{\Gamma((4+\alpha)/\gamma)}{\Gamma(\alpha/\gamma)}y_\text{max}^2.
\end{equation}
One can use this equation to sample $z_l$ for a particular $z_s$.

Now, one can substitute $z_l,z_s$, and $\sigma$ into Eq.~\eqref{eq-def-mlz-sis} and finally, compute $\omega_c$.
For each $z_s$, we computed $10^4$ $\omega_c$'s and estimated the probability of obtaining a smaller $\omega_c/2\pi<10^{-4}$ Hz.
The distribution of all probabilities for the sampled binary systems is shown in Fig.~\ref{fig-probs} in the main text.

\bibliography{membeat_v6.bbl}

\begin{thebibliography}{107}%
\makeatletter
\providecommand \@ifxundefined [1]{%
 \@ifx{#1\undefined}
}%
\providecommand \@ifnum [1]{%
 \ifnum #1\expandafter \@firstoftwo
 \else \expandafter \@secondoftwo
 \fi
}%
\providecommand \@ifx [1]{%
 \ifx #1\expandafter \@firstoftwo
 \else \expandafter \@secondoftwo
 \fi
}%
\providecommand \natexlab [1]{#1}%
\providecommand \enquote  [1]{``#1''}%
\providecommand \bibnamefont  [1]{#1}%
\providecommand \bibfnamefont [1]{#1}%
\providecommand \citenamefont [1]{#1}%
\providecommand \href@noop [0]{\@secondoftwo}%
\providecommand \href [0]{\begingroup \@sanitize@url \@href}%
\providecommand \@href[1]{\@@startlink{#1}\@@href}%
\providecommand \@@href[1]{\endgroup#1\@@endlink}%
\providecommand \@sanitize@url [0]{\catcode `\\12\catcode `\$12\catcode
  `\&12\catcode `\#12\catcode `\^12\catcode `\_12\catcode `\%12\relax}%
\providecommand \@@startlink[1]{}%
\providecommand \@@endlink[0]{}%
\providecommand \url  [0]{\begingroup\@sanitize@url \@url }%
\providecommand \@url [1]{\endgroup\@href {#1}{\urlprefix }}%
\providecommand \urlprefix  [0]{URL }%
\providecommand \Eprint [0]{\href }%
\providecommand \doibase [0]{https://doi.org/}%
\providecommand \selectlanguage [0]{\@gobble}%
\providecommand \bibinfo  [0]{\@secondoftwo}%
\providecommand \bibfield  [0]{\@secondoftwo}%
\providecommand \translation [1]{[#1]}%
\providecommand \BibitemOpen [0]{}%
\providecommand \bibitemStop [0]{}%
\providecommand \bibitemNoStop [0]{.\EOS\space}%
\providecommand \EOS [0]{\spacefactor3000\relax}%
\providecommand \BibitemShut  [1]{\csname bibitem#1\endcsname}%
\let\auto@bib@innerbib\@empty
\bibitem [{\citenamefont {Abbott}\ \emph {et~al.}(2016)\citenamefont {Abbott}
  \emph {et~al.}}]{gw150914}%
  \BibitemOpen
  \bibfield  {author} {\bibinfo {author} {\bibfnamefont {B.~P.}\ \bibnamefont
  {Abbott}} \emph {et~al.} (\bibinfo {collaboration} {Virgo, LIGO
  Scientific}),\ }\bibfield  {title} {\bibinfo {title} {{Observation of
  Gravitational Waves from a Binary Black Hole Merger}},\ }\href
  {https://doi.org/10.1103/PhysRevLett.116.061102} {\bibfield  {journal}
  {\bibinfo  {journal} {Phys. Rev. Lett.}\ }\textbf {\bibinfo {volume} {116}},\
  \bibinfo {pages} {061102} (\bibinfo {year} {2016})},\ \Eprint
  {https://arxiv.org/abs/1602.03837} {arXiv:1602.03837 [gr-qc]} \BibitemShut
  {NoStop}%
\bibitem [{\citenamefont {Einstein}(1916)}]{Einstein:1916cc}%
  \BibitemOpen
  \bibfield  {author} {\bibinfo {author} {\bibfnamefont {A.}~\bibnamefont
  {Einstein}},\ }\bibfield  {title} {\bibinfo {title} {{Approximative
  Integration of the Field Equations of Gravitation}},\ }\href@noop {}
  {\bibfield  {journal} {\bibinfo  {journal} {Sitzungsber. Preuss. Akad. Wiss.
  Berlin (Math. Phys.)}\ }\textbf {\bibinfo {volume} {1916}},\ \bibinfo {pages}
  {688} (\bibinfo {year} {1916})}\BibitemShut {NoStop}%
\bibitem [{\citenamefont {Einstein}(1918)}]{Einstein:1918btx}%
  \BibitemOpen
  \bibfield  {author} {\bibinfo {author} {\bibfnamefont {A.}~\bibnamefont
  {Einstein}},\ }\bibfield  {title} {\bibinfo {title} {{\"Uber
  Gravitationswellen}},\ }\href@noop {} {\bibfield  {journal} {\bibinfo
  {journal} {Sitzungsber. Preuss. Akad. Wiss. Berlin (Math. Phys.)}\ }\textbf
  {\bibinfo {volume} {1918}},\ \bibinfo {pages} {154} (\bibinfo {year}
  {1918})}\BibitemShut {NoStop}%
\bibitem [{\citenamefont {Wald}(1984)}]{Wald:1984rg}%
  \BibitemOpen
  \bibfield  {author} {\bibinfo {author} {\bibfnamefont {R.~M.}\ \bibnamefont
  {Wald}},\ }\href {https://doi.org/10.7208/chicago/9780226870373.001.0001}
  {\emph {\bibinfo {title} {{General Relativity}}}}\ (\bibinfo  {publisher}
  {University of Chicago Press},\ \bibinfo {address} {Chicago, IL},\ \bibinfo
  {year} {1984})\BibitemShut {NoStop}%
\bibitem [{\citenamefont {{Schneider}}\ \emph {et~al.}(1992)\citenamefont
  {{Schneider}}, \citenamefont {{Ehlers}},\ and\ \citenamefont
  {{Falco}}}]{gravlens1992}%
  \BibitemOpen
  \bibfield  {author} {\bibinfo {author} {\bibfnamefont {P.}~\bibnamefont
  {{Schneider}}}, \bibinfo {author} {\bibfnamefont {J.}~\bibnamefont
  {{Ehlers}}},\ and\ \bibinfo {author} {\bibfnamefont {E.~E.}\ \bibnamefont
  {{Falco}}},\ }\href {https://doi.org/10.1007/978-3-662-03758-4} {\emph
  {\bibinfo {title} {Gravitational Lenses}}}\ (\bibinfo  {publisher} {Springer,
  Berlin, Heidelberg},\ \bibinfo {year} {1992})\ p.\ \bibinfo {pages}
  {560}\BibitemShut {NoStop}%
\bibitem [{\citenamefont {Zel'dovich}\ and\ \citenamefont
  {Polnarev}(1974)}]{Zeldovich:1974gvh}%
  \BibitemOpen
  \bibfield  {author} {\bibinfo {author} {\bibfnamefont {Y.~B.}\ \bibnamefont
  {Zel'dovich}}\ and\ \bibinfo {author} {\bibfnamefont {A.~G.}\ \bibnamefont
  {Polnarev}},\ }\bibfield  {title} {\bibinfo {title} {{Radiation of
  gravitational waves by a cluster of superdense stars}},\ }\href@noop {}
  {\bibfield  {journal} {\bibinfo  {journal} {Sov. Astron.}\ }\textbf {\bibinfo
  {volume} {18}},\ \bibinfo {pages} {17} (\bibinfo {year} {1974})}\BibitemShut
  {NoStop}%
\bibitem [{\citenamefont {Braginsky}\ and\ \citenamefont
  {Grishchuk}(1985)}]{Braginsky:1986ia}%
  \BibitemOpen
  \bibfield  {author} {\bibinfo {author} {\bibfnamefont {V.}~\bibnamefont
  {Braginsky}}\ and\ \bibinfo {author} {\bibfnamefont {L.}~\bibnamefont
  {Grishchuk}},\ }\bibfield  {title} {\bibinfo {title} {{Kinematic Resonance
  and Memory Effect in Free Mass Gravitational Antennas}},\ }\href@noop {}
  {\bibfield  {journal} {\bibinfo  {journal} {Sov. Phys. JETP}\ }\textbf
  {\bibinfo {volume} {62}},\ \bibinfo {pages} {427} (\bibinfo {year}
  {1985})}\BibitemShut {NoStop}%
\bibitem [{\citenamefont {Christodoulou}(1991)}]{Christodoulou1991}%
  \BibitemOpen
  \bibfield  {author} {\bibinfo {author} {\bibfnamefont {D.}~\bibnamefont
  {Christodoulou}},\ }\bibfield  {title} {\bibinfo {title} {Nonlinear nature of
  gravitation and gravitational-wave experiments},\ }\href
  {https://doi.org/10.1103/PhysRevLett.67.1486} {\bibfield  {journal} {\bibinfo
   {journal} {Phys. Rev. Lett.}\ }\textbf {\bibinfo {volume} {67}},\ \bibinfo
  {pages} {1486} (\bibinfo {year} {1991})}\BibitemShut {NoStop}%
\bibitem [{\citenamefont {Thorne}(1992)}]{Thorne:1992sdb}%
  \BibitemOpen
  \bibfield  {author} {\bibinfo {author} {\bibfnamefont {K.~S.}\ \bibnamefont
  {Thorne}},\ }\bibfield  {title} {\bibinfo {title} {{Gravitational-wave bursts
  with memory: The Christodoulou effect}},\ }\href
  {https://doi.org/10.1103/PhysRevD.45.520} {\bibfield  {journal} {\bibinfo
  {journal} {Phys. Rev. D}\ }\textbf {\bibinfo {volume} {45}},\ \bibinfo
  {pages} {520} (\bibinfo {year} {1992})}\BibitemShut {NoStop}%
\bibitem [{\citenamefont {Hou}\ \emph {et~al.}(2019)\citenamefont {Hou},
  \citenamefont {Fan},\ and\ \citenamefont {Zhu}}]{Hou:2019wdg}%
  \BibitemOpen
  \bibfield  {author} {\bibinfo {author} {\bibfnamefont {S.}~\bibnamefont
  {Hou}}, \bibinfo {author} {\bibfnamefont {X.-L.}\ \bibnamefont {Fan}},\ and\
  \bibinfo {author} {\bibfnamefont {Z.-H.}\ \bibnamefont {Zhu}},\ }\bibfield
  {title} {\bibinfo {title} {{Gravitational Lensing of Gravitational Waves:
  Rotation of Polarization Plane}},\ }\href
  {https://doi.org/10.1103/PhysRevD.100.064028} {\bibfield  {journal} {\bibinfo
   {journal} {Phys. Rev. D}\ }\textbf {\bibinfo {volume} {100}},\ \bibinfo
  {pages} {064028} (\bibinfo {year} {2019})},\ \Eprint
  {https://arxiv.org/abs/1907.07486} {arXiv:1907.07486 [gr-qc]} \BibitemShut
  {NoStop}%
\bibitem [{\citenamefont {Hou}\ \emph {et~al.}(2020)\citenamefont {Hou},
  \citenamefont {Fan}, \citenamefont {Liao},\ and\ \citenamefont
  {Zhu}}]{Hou:2019dcm}%
  \BibitemOpen
  \bibfield  {author} {\bibinfo {author} {\bibfnamefont {S.}~\bibnamefont
  {Hou}}, \bibinfo {author} {\bibfnamefont {X.-L.}\ \bibnamefont {Fan}},
  \bibinfo {author} {\bibfnamefont {K.}~\bibnamefont {Liao}},\ and\ \bibinfo
  {author} {\bibfnamefont {Z.-H.}\ \bibnamefont {Zhu}},\ }\bibfield  {title}
  {\bibinfo {title} {{Gravitational Wave Interference via Gravitational
  Lensing: Measurements of Luminosity Distance, Lens Mass, and Cosmological
  Parameters}},\ }\href {https://doi.org/10.1103/PhysRevD.101.064011}
  {\bibfield  {journal} {\bibinfo  {journal} {Phys. Rev. D}\ }\textbf {\bibinfo
  {volume} {101}},\ \bibinfo {pages} {064011} (\bibinfo {year} {2020})},\
  \Eprint {https://arxiv.org/abs/1911.02798} {arXiv:1911.02798 [gr-qc]}
  \BibitemShut {NoStop}%
\bibitem [{\citenamefont {Favata}(2009{\natexlab{a}})}]{Favata:2008ti}%
  \BibitemOpen
  \bibfield  {author} {\bibinfo {author} {\bibfnamefont {M.}~\bibnamefont
  {Favata}},\ }\bibfield  {title} {\bibinfo {title} {{Gravitational-wave memory
  revisited: memory from the merger and recoil of binary black holes}},\ }\href
  {https://doi.org/10.1088/1742-6596/154/1/012043} {\bibfield  {journal}
  {\bibinfo  {journal} {J. Phys. Conf. Ser.}\ }\textbf {\bibinfo {volume}
  {154}},\ \bibinfo {pages} {012043} (\bibinfo {year} {2009}{\natexlab{a}})},\
  \Eprint {https://arxiv.org/abs/0811.3451} {arXiv:0811.3451 [astro-ph]}
  \BibitemShut {NoStop}%
\bibitem [{\citenamefont {Favata}(2009{\natexlab{b}})}]{Favata:2008yd}%
  \BibitemOpen
  \bibfield  {author} {\bibinfo {author} {\bibfnamefont {M.}~\bibnamefont
  {Favata}},\ }\bibfield  {title} {\bibinfo {title} {{Post-Newtonian
  corrections to the gravitational-wave memory for quasi-circular, inspiralling
  compact binaries}},\ }\href {https://doi.org/10.1103/PhysRevD.80.024002}
  {\bibfield  {journal} {\bibinfo  {journal} {Phys. Rev. D}\ }\textbf {\bibinfo
  {volume} {80}},\ \bibinfo {pages} {024002} (\bibinfo {year}
  {2009}{\natexlab{b}})},\ \Eprint {https://arxiv.org/abs/0812.0069}
  {arXiv:0812.0069 [gr-qc]} \BibitemShut {NoStop}%
\bibitem [{\citenamefont {Favata}(2009{\natexlab{c}})}]{Favata:2009ii}%
  \BibitemOpen
  \bibfield  {author} {\bibinfo {author} {\bibfnamefont {M.}~\bibnamefont
  {Favata}},\ }\bibfield  {title} {\bibinfo {title} {{Nonlinear
  gravitational-wave memory from binary black hole mergers}},\ }\href
  {https://doi.org/10.1088/0004-637X/696/2/L159} {\bibfield  {journal}
  {\bibinfo  {journal} {Astrophys. J. Lett.}\ }\textbf {\bibinfo {volume}
  {696}},\ \bibinfo {pages} {L159} (\bibinfo {year} {2009}{\natexlab{c}})},\
  \Eprint {https://arxiv.org/abs/0902.3660} {arXiv:0902.3660 [astro-ph.SR]}
  \BibitemShut {NoStop}%
\bibitem [{\citenamefont {Favata}(2011)}]{Favata:2011qi}%
  \BibitemOpen
  \bibfield  {author} {\bibinfo {author} {\bibfnamefont {M.}~\bibnamefont
  {Favata}},\ }\bibfield  {title} {\bibinfo {title} {{The Gravitational-wave
  memory from eccentric binaries}},\ }\href
  {https://doi.org/10.1103/PhysRevD.84.124013} {\bibfield  {journal} {\bibinfo
  {journal} {Phys. Rev. D}\ }\textbf {\bibinfo {volume} {84}},\ \bibinfo
  {pages} {124013} (\bibinfo {year} {2011})},\ \Eprint
  {https://arxiv.org/abs/1108.3121} {arXiv:1108.3121 [gr-qc]} \BibitemShut
  {NoStop}%
\bibitem [{\citenamefont {Bondi}\ \emph {et~al.}(1962)\citenamefont {Bondi},
  \citenamefont {van~der Burg},\ and\ \citenamefont {Metzner}}]{Bondi:1962px}%
  \BibitemOpen
  \bibfield  {author} {\bibinfo {author} {\bibfnamefont {H.}~\bibnamefont
  {Bondi}}, \bibinfo {author} {\bibfnamefont {M.~G.~J.}\ \bibnamefont {van~der
  Burg}},\ and\ \bibinfo {author} {\bibfnamefont {A.~W.~K.}\ \bibnamefont
  {Metzner}},\ }\bibfield  {title} {\bibinfo {title} {{Gravitational waves in
  general relativity. 7. Waves from axisymmetric isolated systems}},\ }\href
  {https://doi.org/10.1098/rspa.1962.0161} {\bibfield  {journal} {\bibinfo
  {journal} {Proc. Roy. Soc. Lond. A}\ }\textbf {\bibinfo {volume} {269}},\
  \bibinfo {pages} {21} (\bibinfo {year} {1962})}\BibitemShut {NoStop}%
\bibitem [{\citenamefont {Sachs}(1962{\natexlab{a}})}]{Sachs:1962wk}%
  \BibitemOpen
  \bibfield  {author} {\bibinfo {author} {\bibfnamefont {R.~K.}\ \bibnamefont
  {Sachs}},\ }\bibfield  {title} {\bibinfo {title} {{Gravitational waves in
  general relativity. 8. Waves in asymptotically flat space-times}},\ }\href
  {https://doi.org/10.1098/rspa.1962.0206} {\bibfield  {journal} {\bibinfo
  {journal} {Proc. Roy. Soc. Lond. A}\ }\textbf {\bibinfo {volume} {270}},\
  \bibinfo {pages} {103} (\bibinfo {year} {1962}{\natexlab{a}})}\BibitemShut
  {NoStop}%
\bibitem [{\citenamefont {Mädler}\ and\ \citenamefont
  {Winicour}(2016)}]{Madler:2016xju}%
  \BibitemOpen
  \bibfield  {author} {\bibinfo {author} {\bibfnamefont {T.}~\bibnamefont
  {Mädler}}\ and\ \bibinfo {author} {\bibfnamefont {J.}~\bibnamefont
  {Winicour}},\ }\bibfield  {title} {\bibinfo {title} {{Bondi-Sachs
  Formalism}},\ }\href {https://doi.org/10.4249/scholarpedia.33528} {\bibfield
  {journal} {\bibinfo  {journal} {Scholarpedia}\ }\textbf {\bibinfo {volume}
  {11}},\ \bibinfo {pages} {33528} (\bibinfo {year} {2016})},\ \Eprint
  {https://arxiv.org/abs/1609.01731} {arXiv:1609.01731 [gr-qc]} \BibitemShut
  {NoStop}%
\bibitem [{\citenamefont {Sachs}(1962{\natexlab{b}})}]{Sachs1962asgr}%
  \BibitemOpen
  \bibfield  {author} {\bibinfo {author} {\bibfnamefont {R.}~\bibnamefont
  {Sachs}},\ }\bibfield  {title} {\bibinfo {title} {Asymptotic symmetries in
  gravitational theory},\ }\href
  {http://link.aps.org/doi/10.1103/PhysRev.128.2851} {\bibfield  {journal}
  {\bibinfo  {journal} {Phys. Rev.}\ }\textbf {\bibinfo {volume} {128}},\
  \bibinfo {pages} {2851} (\bibinfo {year} {1962}{\natexlab{b}})}\BibitemShut
  {NoStop}%
\bibitem [{\citenamefont {{Strominger}}(2014)}]{Strominger2014bms}%
  \BibitemOpen
  \bibfield  {author} {\bibinfo {author} {\bibfnamefont {A.}~\bibnamefont
  {{Strominger}}},\ }\bibfield  {title} {\bibinfo {title} {{On BMS invariance
  of gravitational scattering}},\ }\href
  {https://doi.org/10.1007/JHEP07(2014)152} {\bibfield  {journal} {\bibinfo
  {journal} {JHEP}\ }\textbf {\bibinfo {volume} {07}},\ \bibinfo {eid} {152}},\
  \Eprint {https://arxiv.org/abs/1312.2229} {arXiv:1312.2229 [hep-th]}
  \BibitemShut {NoStop}%
\bibitem [{\citenamefont {Strominger}\ and\ \citenamefont
  {Zhiboedov}(2016)}]{Strominger:2014pwa}%
  \BibitemOpen
  \bibfield  {author} {\bibinfo {author} {\bibfnamefont {A.}~\bibnamefont
  {Strominger}}\ and\ \bibinfo {author} {\bibfnamefont {A.}~\bibnamefont
  {Zhiboedov}},\ }\bibfield  {title} {\bibinfo {title} {{Gravitational Memory,
  BMS Supertranslations and Soft Theorems}},\ }\href
  {https://doi.org/10.1007/JHEP01(2016)086} {\bibfield  {journal} {\bibinfo
  {journal} {JHEP}\ }\textbf {\bibinfo {volume} {01}},\ \bibinfo {pages}
  {086}},\ \Eprint {https://arxiv.org/abs/1411.5745} {arXiv:1411.5745 [hep-th]}
  \BibitemShut {NoStop}%
\bibitem [{\citenamefont {Flanagan}\ and\ \citenamefont
  {Nichols}(2017)}]{Flanagan:2015pxa}%
  \BibitemOpen
  \bibfield  {author} {\bibinfo {author} {\bibfnamefont {E.~E.}\ \bibnamefont
  {Flanagan}}\ and\ \bibinfo {author} {\bibfnamefont {D.~A.}\ \bibnamefont
  {Nichols}},\ }\bibfield  {title} {\bibinfo {title} {{Conserved charges of the
  extended Bondi-Metzner-Sachs algebra}},\ }\href
  {https://doi.org/10.1103/PhysRevD.95.044002} {\bibfield  {journal} {\bibinfo
  {journal} {Phys. Rev. D}\ }\textbf {\bibinfo {volume} {95}},\ \bibinfo
  {pages} {044002} (\bibinfo {year} {2017})},\ \Eprint
  {https://arxiv.org/abs/1510.03386} {arXiv:1510.03386 [hep-th]} \BibitemShut
  {NoStop}%
\bibitem [{\citenamefont {Favata}(2010)}]{Favata:2010zu}%
  \BibitemOpen
  \bibfield  {author} {\bibinfo {author} {\bibfnamefont {M.}~\bibnamefont
  {Favata}},\ }\bibfield  {title} {\bibinfo {title} {{The gravitational-wave
  memory effect}},\ }\href {https://doi.org/10.1088/0264-9381/27/8/084036}
  {\bibfield  {journal} {\bibinfo  {journal} {Class. Quant. Grav.}\ }\textbf
  {\bibinfo {volume} {27}},\ \bibinfo {pages} {084036} (\bibinfo {year}
  {2010})},\ \Eprint {https://arxiv.org/abs/1003.3486} {arXiv:1003.3486
  [gr-qc]} \BibitemShut {NoStop}%
\bibitem [{\citenamefont {Lasky}\ \emph {et~al.}(2016)\citenamefont {Lasky},
  \citenamefont {Thrane}, \citenamefont {Levin}, \citenamefont {Blackman},\
  and\ \citenamefont {Chen}}]{Lasky:2016knh}%
  \BibitemOpen
  \bibfield  {author} {\bibinfo {author} {\bibfnamefont {P.~D.}\ \bibnamefont
  {Lasky}}, \bibinfo {author} {\bibfnamefont {E.}~\bibnamefont {Thrane}},
  \bibinfo {author} {\bibfnamefont {Y.}~\bibnamefont {Levin}}, \bibinfo
  {author} {\bibfnamefont {J.}~\bibnamefont {Blackman}},\ and\ \bibinfo
  {author} {\bibfnamefont {Y.}~\bibnamefont {Chen}},\ }\bibfield  {title}
  {\bibinfo {title} {{Detecting gravitational-wave memory with LIGO:
  implications of GW150914}},\ }\href
  {https://doi.org/10.1103/PhysRevLett.117.061102} {\bibfield  {journal}
  {\bibinfo  {journal} {Phys. Rev. Lett.}\ }\textbf {\bibinfo {volume} {117}},\
  \bibinfo {pages} {061102} (\bibinfo {year} {2016})},\ \Eprint
  {https://arxiv.org/abs/1605.01415} {arXiv:1605.01415 [astro-ph.HE]}
  \BibitemShut {NoStop}%
\bibitem [{\citenamefont {Hou}\ \emph {et~al.}(2025)\citenamefont {Hou},
  \citenamefont {Zhao}, \citenamefont {Cao},\ and\ \citenamefont
  {Zhu}}]{Hou:2024rgo}%
  \BibitemOpen
  \bibfield  {author} {\bibinfo {author} {\bibfnamefont {S.}~\bibnamefont
  {Hou}}, \bibinfo {author} {\bibfnamefont {Z.-C.}\ \bibnamefont {Zhao}},
  \bibinfo {author} {\bibfnamefont {Z.}~\bibnamefont {Cao}},\ and\ \bibinfo
  {author} {\bibfnamefont {Z.-H.}\ \bibnamefont {Zhu}},\ }\bibfield  {title}
  {\bibinfo {title} {{Space-Borne Interferometers to Detect Thousands of Memory
  Signals Emitted by Stellar-Mass Binary Black Holes}},\ }\href
  {https://doi.org/10.1088/0256-307X/42/10/101101} {\bibfield  {journal}
  {\bibinfo  {journal} {Chin. Phys. Lett.}\ }\textbf {\bibinfo {volume} {42}},\
  \bibinfo {pages} {101101} (\bibinfo {year} {2025})},\ \Eprint
  {https://arxiv.org/abs/2411.18053} {arXiv:2411.18053 [gr-qc]} \BibitemShut
  {NoStop}%
\bibitem [{\citenamefont {Misner}\ \emph {et~al.}(1973)\citenamefont {Misner},
  \citenamefont {Thorne},\ and\ \citenamefont {Wheeler}}]{mtw}%
  \BibitemOpen
  \bibfield  {author} {\bibinfo {author} {\bibfnamefont {C.~W.}\ \bibnamefont
  {Misner}}, \bibinfo {author} {\bibfnamefont {K.~S.}\ \bibnamefont {Thorne}},\
  and\ \bibinfo {author} {\bibfnamefont {J.~A.}\ \bibnamefont {Wheeler}},\
  }\href@noop {} {\emph {\bibinfo {title} {{Gravitation}}}}\ (\bibinfo
  {publisher} {W. H. Freeman},\ \bibinfo {address} {San Francisco},\ \bibinfo
  {year} {1973})\BibitemShut {NoStop}%
\bibitem [{\citenamefont {Poisson}\ and\ \citenamefont
  {Will}(2014)}]{Poisson2014}%
  \BibitemOpen
  \bibfield  {author} {\bibinfo {author} {\bibfnamefont {E.}~\bibnamefont
  {Poisson}}\ and\ \bibinfo {author} {\bibfnamefont {C.~M.}\ \bibnamefont
  {Will}},\ }\href {https://doi.org/10.1017/CBO9781139507486} {\emph {\bibinfo
  {title} {Gravity: Newtonian, Post-Newtonian, Relativistic}}}\ (\bibinfo
  {publisher} {Cambridge University Press},\ \bibinfo {year}
  {2014})\BibitemShut {NoStop}%
\bibitem [{\citenamefont {Abbott}\ \emph {et~al.}(2019)\citenamefont {Abbott}
  \emph {et~al.}}]{LIGOScientific:2018mvr}%
  \BibitemOpen
  \bibfield  {author} {\bibinfo {author} {\bibfnamefont {B.~P.}\ \bibnamefont
  {Abbott}} \emph {et~al.} (\bibinfo {collaboration} {Virgo and LIGO Scientific
  Collaborations}),\ }\bibfield  {title} {\bibinfo {title} {{GWTC-1: A
  Gravitational-Wave Transient Catalog of Compact Binary Mergers Observed by
  LIGO and Virgo during the First and Second Observing Runs}},\ }\href
  {https://doi.org/10.1103/PhysRevX.9.031040} {\bibfield  {journal} {\bibinfo
  {journal} {Phys. Rev. X}\ }\textbf {\bibinfo {volume} {9}},\ \bibinfo {pages}
  {031040} (\bibinfo {year} {2019})},\ \Eprint
  {https://arxiv.org/abs/1811.12907} {arXiv:1811.12907 [astro-ph.HE]}
  \BibitemShut {NoStop}%
\bibitem [{\citenamefont {Abbott}\ \emph
  {et~al.}(2021{\natexlab{a}})\citenamefont {Abbott} \emph
  {et~al.}}]{Abbott:2020niy}%
  \BibitemOpen
  \bibfield  {author} {\bibinfo {author} {\bibfnamefont {R.}~\bibnamefont
  {Abbott}} \emph {et~al.} (\bibinfo {collaboration} {LIGO Scientific,
  Virgo}),\ }\bibfield  {title} {\bibinfo {title} {{GWTC-2: Compact Binary
  Coalescences Observed by LIGO and Virgo During the First Half of the Third
  Observing Run}},\ }\href {https://doi.org/10.1103/PhysRevX.11.021053}
  {\bibfield  {journal} {\bibinfo  {journal} {Phys. Rev. X}\ }\textbf {\bibinfo
  {volume} {11}},\ \bibinfo {pages} {021053} (\bibinfo {year}
  {2021}{\natexlab{a}})},\ \Eprint {https://arxiv.org/abs/2010.14527}
  {arXiv:2010.14527 [gr-qc]} \BibitemShut {NoStop}%
\bibitem [{\citenamefont {Abbott}\ \emph
  {et~al.}(2024{\natexlab{a}})\citenamefont {Abbott} \emph
  {et~al.}}]{LIGOScientific:2021usb}%
  \BibitemOpen
  \bibfield  {author} {\bibinfo {author} {\bibfnamefont {R.}~\bibnamefont
  {Abbott}} \emph {et~al.} (\bibinfo {collaboration} {LIGO Scientific,
  VIRGO}),\ }\bibfield  {title} {\bibinfo {title} {{GWTC-2.1: Deep extended
  catalog of compact binary coalescences observed by LIGO and Virgo during the
  first half of the third observing run}},\ }\href
  {https://doi.org/10.1103/PhysRevD.109.022001} {\bibfield  {journal} {\bibinfo
   {journal} {Phys. Rev. D}\ }\textbf {\bibinfo {volume} {109}},\ \bibinfo
  {pages} {022001} (\bibinfo {year} {2024}{\natexlab{a}})},\ \Eprint
  {https://arxiv.org/abs/2108.01045} {arXiv:2108.01045 [gr-qc]} \BibitemShut
  {NoStop}%
\bibitem [{\citenamefont {Abbott}\ \emph {et~al.}(2023)\citenamefont {Abbott}
  \emph {et~al.}}]{KAGRA:2021vkt}%
  \BibitemOpen
  \bibfield  {author} {\bibinfo {author} {\bibfnamefont {R.}~\bibnamefont
  {Abbott}} \emph {et~al.} (\bibinfo {collaboration} {KAGRA, VIRGO, LIGO
  Scientific}),\ }\bibfield  {title} {\bibinfo {title} {{GWTC-3: Compact Binary
  Coalescences Observed by LIGO and Virgo during the Second Part of the Third
  Observing Run}},\ }\href {https://doi.org/10.1103/PhysRevX.13.041039}
  {\bibfield  {journal} {\bibinfo  {journal} {Phys. Rev. X}\ }\textbf {\bibinfo
  {volume} {13}},\ \bibinfo {pages} {041039} (\bibinfo {year} {2023})},\
  \Eprint {https://arxiv.org/abs/2111.03606} {arXiv:2111.03606 [gr-qc]}
  \BibitemShut {NoStop}%
\bibitem [{\citenamefont {Abbott}\ \emph
  {et~al.}(2021{\natexlab{b}})\citenamefont {Abbott} \emph
  {et~al.}}]{LIGOScientific:2021izm}%
  \BibitemOpen
  \bibfield  {author} {\bibinfo {author} {\bibfnamefont {R.}~\bibnamefont
  {Abbott}} \emph {et~al.} (\bibinfo {collaboration} {LIGO Scientific,
  VIRGO}),\ }\bibfield  {title} {\bibinfo {title} {{Search for Lensing
  Signatures in the Gravitational-Wave Observations from the First Half of
  LIGO{\textendash}Virgo{\textquoteright}s Third Observing Run}},\ }\href
  {https://doi.org/10.3847/1538-4357/ac23db} {\bibfield  {journal} {\bibinfo
  {journal} {Astrophys. J.}\ }\textbf {\bibinfo {volume} {923}},\ \bibinfo
  {pages} {14} (\bibinfo {year} {2021}{\natexlab{b}})},\ \Eprint
  {https://arxiv.org/abs/2105.06384} {arXiv:2105.06384 [gr-qc]} \BibitemShut
  {NoStop}%
\bibitem [{\citenamefont {Abbott}\ \emph
  {et~al.}(2024{\natexlab{b}})\citenamefont {Abbott} \emph
  {et~al.}}]{LIGOScientific:2023bwz}%
  \BibitemOpen
  \bibfield  {author} {\bibinfo {author} {\bibfnamefont {R.}~\bibnamefont
  {Abbott}} \emph {et~al.} (\bibinfo {collaboration} {LIGO Scientific, KAGRA,
  VIRGO}),\ }\bibfield  {title} {\bibinfo {title} {{Search for
  Gravitational-lensing Signatures in the Full Third Observing Run of the
  LIGO{\textendash}Virgo Network}},\ }\href
  {https://doi.org/10.3847/1538-4357/ad3e83} {\bibfield  {journal} {\bibinfo
  {journal} {Astrophys. J.}\ }\textbf {\bibinfo {volume} {970}},\ \bibinfo
  {pages} {191} (\bibinfo {year} {2024}{\natexlab{b}})},\ \Eprint
  {https://arxiv.org/abs/2304.08393} {arXiv:2304.08393 [gr-qc]} \BibitemShut
  {NoStop}%
\bibitem [{\citenamefont {Sereno}\ \emph {et~al.}(2010)\citenamefont {Sereno},
  \citenamefont {Sesana}, \citenamefont {Bleuler}, \citenamefont {Jetzer},
  \citenamefont {Volonteri},\ and\ \citenamefont {Begelman}}]{Sereno:2010dr}%
  \BibitemOpen
  \bibfield  {author} {\bibinfo {author} {\bibfnamefont {M.}~\bibnamefont
  {Sereno}}, \bibinfo {author} {\bibfnamefont {A.}~\bibnamefont {Sesana}},
  \bibinfo {author} {\bibfnamefont {A.}~\bibnamefont {Bleuler}}, \bibinfo
  {author} {\bibfnamefont {P.}~\bibnamefont {Jetzer}}, \bibinfo {author}
  {\bibfnamefont {M.}~\bibnamefont {Volonteri}},\ and\ \bibinfo {author}
  {\bibfnamefont {M.~C.}\ \bibnamefont {Begelman}},\ }\bibfield  {title}
  {\bibinfo {title} {{Strong lensing of gravitational waves as seen by LISA}},\
  }\href {https://doi.org/10.1103/PhysRevLett.105.251101} {\bibfield  {journal}
  {\bibinfo  {journal} {Phys. Rev. Lett.}\ }\textbf {\bibinfo {volume} {105}},\
  \bibinfo {pages} {251101} (\bibinfo {year} {2010})},\ \Eprint
  {https://arxiv.org/abs/1011.5238} {arXiv:1011.5238 [astro-ph.CO]}
  \BibitemShut {NoStop}%
\bibitem [{\citenamefont {Pi\'orkowska}\ \emph {et~al.}(2013)\citenamefont
  {Pi\'orkowska}, \citenamefont {Biesiada},\ and\ \citenamefont
  {Zhu}}]{Piorkowska:2013eww}%
  \BibitemOpen
  \bibfield  {author} {\bibinfo {author} {\bibfnamefont {A.}~\bibnamefont
  {Pi\'orkowska}}, \bibinfo {author} {\bibfnamefont {M.}~\bibnamefont
  {Biesiada}},\ and\ \bibinfo {author} {\bibfnamefont {Z.-H.}\ \bibnamefont
  {Zhu}},\ }\bibfield  {title} {\bibinfo {title} {{Strong gravitational lensing
  of gravitational waves in Einstein Telescope}},\ }\href
  {https://doi.org/10.1088/1475-7516/2013/10/022} {\bibfield  {journal}
  {\bibinfo  {journal} {JCAP}\ }\textbf {\bibinfo {volume} {1310}},\ \bibinfo
  {pages} {022}},\ \Eprint {https://arxiv.org/abs/1309.5731} {arXiv:1309.5731
  [astro-ph.CO]} \BibitemShut {NoStop}%
\bibitem [{\citenamefont {Ding}\ \emph {et~al.}(2015)\citenamefont {Ding},
  \citenamefont {Biesiada},\ and\ \citenamefont {Zhu}}]{Ding:2015uha}%
  \BibitemOpen
  \bibfield  {author} {\bibinfo {author} {\bibfnamefont {X.}~\bibnamefont
  {Ding}}, \bibinfo {author} {\bibfnamefont {M.}~\bibnamefont {Biesiada}},\
  and\ \bibinfo {author} {\bibfnamefont {Z.-H.}\ \bibnamefont {Zhu}},\
  }\bibfield  {title} {\bibinfo {title} {{Strongly lensed gravitational waves
  from intrinsically faint double compact binaries—prediction for the
  Einstein Telescope}},\ }\href {https://doi.org/10.1088/1475-7516/2015/12/006}
  {\bibfield  {journal} {\bibinfo  {journal} {JCAP}\ }\textbf {\bibinfo
  {volume} {1512}}\bibfield  {number} {\bibinfo  {number} { (12)},\ \bibinfo
  {pages} {006}},\ }\Eprint {https://arxiv.org/abs/1508.05000}
  {arXiv:1508.05000 [astro-ph.HE]} \BibitemShut {NoStop}%
\bibitem [{\citenamefont {Pi\'orkowska-Kurpas}\ \emph
  {et~al.}(2021)\citenamefont {Pi\'orkowska-Kurpas}, \citenamefont {Hou},
  \citenamefont {Biesiada}, \citenamefont {Ding}, \citenamefont {Cao},
  \citenamefont {Fan}, \citenamefont {Kawamura},\ and\ \citenamefont
  {Zhu}}]{Piorkowska-Kurpas:2020mst}%
  \BibitemOpen
  \bibfield  {author} {\bibinfo {author} {\bibfnamefont {A.}~\bibnamefont
  {Pi\'orkowska-Kurpas}}, \bibinfo {author} {\bibfnamefont {S.}~\bibnamefont
  {Hou}}, \bibinfo {author} {\bibfnamefont {M.}~\bibnamefont {Biesiada}},
  \bibinfo {author} {\bibfnamefont {X.}~\bibnamefont {Ding}}, \bibinfo {author}
  {\bibfnamefont {S.}~\bibnamefont {Cao}}, \bibinfo {author} {\bibfnamefont
  {X.}~\bibnamefont {Fan}}, \bibinfo {author} {\bibfnamefont {S.}~\bibnamefont
  {Kawamura}},\ and\ \bibinfo {author} {\bibfnamefont {Z.-H.}\ \bibnamefont
  {Zhu}},\ }\bibfield  {title} {\bibinfo {title} {{Inspiraling double compact
  object detection and lensing rate -- forecast for DECIGO and B-DECIGO}},\
  }\href {https://doi.org/10.3847/1538-4357/abd482} {\bibfield  {journal}
  {\bibinfo  {journal} {Astrophys. J.}\ }\textbf {\bibinfo {volume} {908}},\
  \bibinfo {pages} {196} (\bibinfo {year} {2021})},\ \Eprint
  {https://arxiv.org/abs/2005.08727} {arXiv:2005.08727 [astro-ph.HE]}
  \BibitemShut {NoStop}%
\bibitem [{\citenamefont {Hou}\ \emph {et~al.}(2021)\citenamefont {Hou},
  \citenamefont {Li}, \citenamefont {Yu}, \citenamefont {Biesiada},
  \citenamefont {Fan}, \citenamefont {Kawamura},\ and\ \citenamefont
  {Zhu}}]{Hou:2020mpr}%
  \BibitemOpen
  \bibfield  {author} {\bibinfo {author} {\bibfnamefont {S.}~\bibnamefont
  {Hou}}, \bibinfo {author} {\bibfnamefont {P.}~\bibnamefont {Li}}, \bibinfo
  {author} {\bibfnamefont {H.}~\bibnamefont {Yu}}, \bibinfo {author}
  {\bibfnamefont {M.}~\bibnamefont {Biesiada}}, \bibinfo {author}
  {\bibfnamefont {X.-L.}\ \bibnamefont {Fan}}, \bibinfo {author} {\bibfnamefont
  {S.}~\bibnamefont {Kawamura}},\ and\ \bibinfo {author} {\bibfnamefont
  {Z.-H.}\ \bibnamefont {Zhu}},\ }\bibfield  {title} {\bibinfo {title}
  {{Lensing rates of gravitational wave signals displaying beat patterns
  detectable by DECIGO and B-DECIGO}},\ }\href
  {https://doi.org/10.1103/PhysRevD.103.044005} {\bibfield  {journal} {\bibinfo
   {journal} {Phys. Rev. D}\ }\textbf {\bibinfo {volume} {103}},\ \bibinfo
  {pages} {044005} (\bibinfo {year} {2021})},\ \Eprint
  {https://arxiv.org/abs/2009.08116} {arXiv:2009.08116 [gr-qc]} \BibitemShut
  {NoStop}%
\bibitem [{\citenamefont {Gao}\ \emph {et~al.}(2022)\citenamefont {Gao},
  \citenamefont {Chen}, \citenamefont {Hu}, \citenamefont {Zhang},\ and\
  \citenamefont {Huang}}]{Gao:2021sxw}%
  \BibitemOpen
  \bibfield  {author} {\bibinfo {author} {\bibfnamefont {Z.}~\bibnamefont
  {Gao}}, \bibinfo {author} {\bibfnamefont {X.}~\bibnamefont {Chen}}, \bibinfo
  {author} {\bibfnamefont {Y.-M.}\ \bibnamefont {Hu}}, \bibinfo {author}
  {\bibfnamefont {J.-D.}\ \bibnamefont {Zhang}},\ and\ \bibinfo {author}
  {\bibfnamefont {S.-J.}\ \bibnamefont {Huang}},\ }\bibfield  {title} {\bibinfo
  {title} {{A higher probability of detecting lensed supermassive black hole
  binaries by LISA}},\ }\href {https://doi.org/10.1093/mnras/stac365}
  {\bibfield  {journal} {\bibinfo  {journal} {Mon. Not. Roy. Astron. Soc.}\
  }\textbf {\bibinfo {volume} {512}},\ \bibinfo {pages} {1} (\bibinfo {year}
  {2022})},\ \Eprint {https://arxiv.org/abs/2102.10295} {arXiv:2102.10295
  [astro-ph.CO]} \BibitemShut {NoStop}%
\bibitem [{\citenamefont {Yang}\ \emph {et~al.}(2021)\citenamefont {Yang},
  \citenamefont {Wu}, \citenamefont {Liao}, \citenamefont {Ding}, \citenamefont
  {You}, \citenamefont {Cao}, \citenamefont {Biesiada},\ and\ \citenamefont
  {Zhu}}]{Yang:2021viz}%
  \BibitemOpen
  \bibfield  {author} {\bibinfo {author} {\bibfnamefont {L.}~\bibnamefont
  {Yang}}, \bibinfo {author} {\bibfnamefont {S.}~\bibnamefont {Wu}}, \bibinfo
  {author} {\bibfnamefont {K.}~\bibnamefont {Liao}}, \bibinfo {author}
  {\bibfnamefont {X.}~\bibnamefont {Ding}}, \bibinfo {author} {\bibfnamefont
  {Z.}~\bibnamefont {You}}, \bibinfo {author} {\bibfnamefont {Z.}~\bibnamefont
  {Cao}}, \bibinfo {author} {\bibfnamefont {M.}~\bibnamefont {Biesiada}},\ and\
  \bibinfo {author} {\bibfnamefont {Z.-H.}\ \bibnamefont {Zhu}},\ }\bibfield
  {title} {\bibinfo {title} {{Event rate predictions of strongly lensed
  gravitational waves with detector networks and more realistic templates}},\
  }\href {https://doi.org/10.1093/mnras/stab3298} {\bibfield  {journal}
  {\bibinfo  {journal} {Mon. Not. Roy. Astron. Soc.}\ }\textbf {\bibinfo
  {volume} {509}},\ \bibinfo {pages} {3772} (\bibinfo {year} {2021})},\ \Eprint
  {https://arxiv.org/abs/2105.07011} {arXiv:2105.07011 [astro-ph.GA]}
  \BibitemShut {NoStop}%
\bibitem [{\citenamefont {Lin}\ \emph {et~al.}(2023)\citenamefont {Lin},
  \citenamefont {Zhang}, \citenamefont {Dai}, \citenamefont {Huang},\ and\
  \citenamefont {Mei}}]{Lin:2023ccz}%
  \BibitemOpen
  \bibfield  {author} {\bibinfo {author} {\bibfnamefont {X.-y.}\ \bibnamefont
  {Lin}}, \bibinfo {author} {\bibfnamefont {J.-d.}\ \bibnamefont {Zhang}},
  \bibinfo {author} {\bibfnamefont {L.}~\bibnamefont {Dai}}, \bibinfo {author}
  {\bibfnamefont {S.-J.}\ \bibnamefont {Huang}},\ and\ \bibinfo {author}
  {\bibfnamefont {J.}~\bibnamefont {Mei}},\ }\bibfield  {title} {\bibinfo
  {title} {{Detecting strong gravitational lensing of gravitational waves with
  TianQin}},\ }\href {https://doi.org/10.1103/PhysRevD.108.064020} {\bibfield
  {journal} {\bibinfo  {journal} {Phys. Rev. D}\ }\textbf {\bibinfo {volume}
  {108}},\ \bibinfo {pages} {064020} (\bibinfo {year} {2023})},\ \Eprint
  {https://arxiv.org/abs/2304.04800} {arXiv:2304.04800 [gr-qc]} \BibitemShut
  {NoStop}%
\bibitem [{\citenamefont {H\"ubner}\ \emph {et~al.}(2020)\citenamefont
  {H\"ubner}, \citenamefont {Talbot}, \citenamefont {Lasky},\ and\
  \citenamefont {Thrane}}]{Hubner:2019sly}%
  \BibitemOpen
  \bibfield  {author} {\bibinfo {author} {\bibfnamefont {M.}~\bibnamefont
  {H\"ubner}}, \bibinfo {author} {\bibfnamefont {C.}~\bibnamefont {Talbot}},
  \bibinfo {author} {\bibfnamefont {P.~D.}\ \bibnamefont {Lasky}},\ and\
  \bibinfo {author} {\bibfnamefont {E.}~\bibnamefont {Thrane}},\ }\bibfield
  {title} {\bibinfo {title} {{Measuring gravitational-wave memory in the first
  LIGO/Virgo gravitational-wave transient catalog}},\ }\href
  {https://doi.org/10.1103/PhysRevD.101.023011} {\bibfield  {journal} {\bibinfo
   {journal} {Phys. Rev. D}\ }\textbf {\bibinfo {volume} {101}},\ \bibinfo
  {pages} {023011} (\bibinfo {year} {2020})},\ \Eprint
  {https://arxiv.org/abs/1911.12496} {arXiv:1911.12496 [astro-ph.HE]}
  \BibitemShut {NoStop}%
\bibitem [{\citenamefont {Grant}\ and\ \citenamefont
  {Nichols}(2023)}]{Grant:2022bla}%
  \BibitemOpen
  \bibfield  {author} {\bibinfo {author} {\bibfnamefont {A.~M.}\ \bibnamefont
  {Grant}}\ and\ \bibinfo {author} {\bibfnamefont {D.~A.}\ \bibnamefont
  {Nichols}},\ }\bibfield  {title} {\bibinfo {title} {{Outlook for detecting
  the gravitational-wave displacement and spin memory effects with current and
  future gravitational-wave detectors}},\ }\href
  {https://doi.org/10.1103/PhysRevD.107.064056} {\bibfield  {journal} {\bibinfo
   {journal} {Phys. Rev. D}\ }\textbf {\bibinfo {volume} {107}},\ \bibinfo
  {pages} {064056} (\bibinfo {year} {2023})},\ \bibinfo {note} {[Erratum:
  Phys.Rev.D 108, 029901 (2023)]},\ \Eprint {https://arxiv.org/abs/2210.16266}
  {arXiv:2210.16266 [gr-qc]} \BibitemShut {NoStop}%
\bibitem [{\citenamefont {Zhao}\ and\ \citenamefont
  {Cao}(2022)}]{Zhao:2021zlr}%
  \BibitemOpen
  \bibfield  {author} {\bibinfo {author} {\bibfnamefont {Z.-C.}\ \bibnamefont
  {Zhao}}\ and\ \bibinfo {author} {\bibfnamefont {Z.}~\bibnamefont {Cao}},\
  }\bibfield  {title} {\bibinfo {title} {{Stochastic gravitational wave
  background due to gravitational wave memory}},\ }\href
  {https://doi.org/10.1007/s11433-022-1965-y} {\bibfield  {journal} {\bibinfo
  {journal} {Sci. China Phys. Mech. Astron.}\ }\textbf {\bibinfo {volume}
  {65}},\ \bibinfo {pages} {119511} (\bibinfo {year} {2022})},\ \Eprint
  {https://arxiv.org/abs/2111.13883} {arXiv:2111.13883 [gr-qc]} \BibitemShut
  {NoStop}%
\bibitem [{\citenamefont {Sun}\ \emph {et~al.}(2023)\citenamefont {Sun},
  \citenamefont {Shi}, \citenamefont {Zhang},\ and\ \citenamefont
  {Mei}}]{Sun:2022pvh}%
  \BibitemOpen
  \bibfield  {author} {\bibinfo {author} {\bibfnamefont {S.}~\bibnamefont
  {Sun}}, \bibinfo {author} {\bibfnamefont {C.}~\bibnamefont {Shi}}, \bibinfo
  {author} {\bibfnamefont {J.-d.}\ \bibnamefont {Zhang}},\ and\ \bibinfo
  {author} {\bibfnamefont {J.}~\bibnamefont {Mei}},\ }\bibfield  {title}
  {\bibinfo {title} {{Detecting the gravitational wave memory effect with
  TianQin}},\ }\href {https://doi.org/10.1103/PhysRevD.107.044023} {\bibfield
  {journal} {\bibinfo  {journal} {Phys. Rev. D}\ }\textbf {\bibinfo {volume}
  {107}},\ \bibinfo {pages} {044023} (\bibinfo {year} {2023})},\ \Eprint
  {https://arxiv.org/abs/2207.13009} {arXiv:2207.13009 [gr-qc]} \BibitemShut
  {NoStop}%
\bibitem [{\citenamefont {Gasparotto}\ \emph {et~al.}(2023)\citenamefont
  {Gasparotto}, \citenamefont {Vicente}, \citenamefont {Blas}, \citenamefont
  {Jenkins},\ and\ \citenamefont {Barausse}}]{Gasparotto:2023fcg}%
  \BibitemOpen
  \bibfield  {author} {\bibinfo {author} {\bibfnamefont {S.}~\bibnamefont
  {Gasparotto}}, \bibinfo {author} {\bibfnamefont {R.}~\bibnamefont {Vicente}},
  \bibinfo {author} {\bibfnamefont {D.}~\bibnamefont {Blas}}, \bibinfo {author}
  {\bibfnamefont {A.~C.}\ \bibnamefont {Jenkins}},\ and\ \bibinfo {author}
  {\bibfnamefont {E.}~\bibnamefont {Barausse}},\ }\bibfield  {title} {\bibinfo
  {title} {{Can gravitational-wave memory help constrain binary black-hole
  parameters? A LISA case study}},\ }\href
  {https://doi.org/10.1103/PhysRevD.107.124033} {\bibfield  {journal} {\bibinfo
   {journal} {Phys. Rev. D}\ }\textbf {\bibinfo {volume} {107}},\ \bibinfo
  {pages} {124033} (\bibinfo {year} {2023})},\ \Eprint
  {https://arxiv.org/abs/2301.13228} {arXiv:2301.13228 [gr-qc]} \BibitemShut
  {NoStop}%
\bibitem [{\citenamefont {Inchausp{\'e}}\ \emph {et~al.}(2025)\citenamefont
  {Inchausp{\'e}}, \citenamefont {Gasparotto}, \citenamefont {Blas},
  \citenamefont {Heisenberg}, \citenamefont {Zosso},\ and\ \citenamefont
  {Tiwari}}]{Inchauspe:2024ibs}%
  \BibitemOpen
  \bibfield  {author} {\bibinfo {author} {\bibfnamefont {H.}~\bibnamefont
  {Inchausp{\'e}}}, \bibinfo {author} {\bibfnamefont {S.}~\bibnamefont
  {Gasparotto}}, \bibinfo {author} {\bibfnamefont {D.}~\bibnamefont {Blas}},
  \bibinfo {author} {\bibfnamefont {L.}~\bibnamefont {Heisenberg}}, \bibinfo
  {author} {\bibfnamefont {J.}~\bibnamefont {Zosso}},\ and\ \bibinfo {author}
  {\bibfnamefont {S.}~\bibnamefont {Tiwari}},\ }\bibfield  {title} {\bibinfo
  {title} {{Measuring gravitational wave memory with LISA}},\ }\href
  {https://doi.org/10.1103/PhysRevD.111.044044} {\bibfield  {journal} {\bibinfo
   {journal} {Phys. Rev. D}\ }\textbf {\bibinfo {volume} {111}},\ \bibinfo
  {pages} {044044} (\bibinfo {year} {2025})},\ \Eprint
  {https://arxiv.org/abs/2406.09228} {arXiv:2406.09228 [gr-qc]} \BibitemShut
  {NoStop}%
\bibitem [{\citenamefont {{Nakamura}}\ and\ \citenamefont
  {{Deguchi}}(1999)}]{Nakamura1999wo}%
  \BibitemOpen
  \bibfield  {author} {\bibinfo {author} {\bibfnamefont {T.~T.}\ \bibnamefont
  {{Nakamura}}}\ and\ \bibinfo {author} {\bibfnamefont {S.}~\bibnamefont
  {{Deguchi}}},\ }\bibfield  {title} {\bibinfo {title} {{Wave Optics in
  Gravitational Lensing}},\ }\href {https://doi.org/10.1143/PTPS.133.137}
  {\bibfield  {journal} {\bibinfo  {journal} {Progress of Theoretical Physics
  Supplement}\ }\textbf {\bibinfo {volume} {133}},\ \bibinfo {pages} {137}
  (\bibinfo {year} {1999})}\BibitemShut {NoStop}%
\bibitem [{\citenamefont {Takahashi}\ and\ \citenamefont
  {Nakamura}(2003)}]{Takahashi:2003ix}%
  \BibitemOpen
  \bibfield  {author} {\bibinfo {author} {\bibfnamefont {R.}~\bibnamefont
  {Takahashi}}\ and\ \bibinfo {author} {\bibfnamefont {T.}~\bibnamefont
  {Nakamura}},\ }\bibfield  {title} {\bibinfo {title} {{Wave effects in
  gravitational lensing of gravitational waves from chirping binaries}},\
  }\href {https://doi.org/10.1086/377430} {\bibfield  {journal} {\bibinfo
  {journal} {Astrophys. J.}\ }\textbf {\bibinfo {volume} {595}},\ \bibinfo
  {pages} {1039} (\bibinfo {year} {2003})},\ \Eprint
  {https://arxiv.org/abs/astro-ph/0305055} {arXiv:astro-ph/0305055 [astro-ph]}
  \BibitemShut {NoStop}%
\bibitem [{\citenamefont {Ezquiaga}\ \emph {et~al.}(2021)\citenamefont
  {Ezquiaga}, \citenamefont {Holz}, \citenamefont {Hu}, \citenamefont {Lagos},\
  and\ \citenamefont {Wald}}]{Ezquiaga:2020gdt}%
  \BibitemOpen
  \bibfield  {author} {\bibinfo {author} {\bibfnamefont {J.~M.}\ \bibnamefont
  {Ezquiaga}}, \bibinfo {author} {\bibfnamefont {D.~E.}\ \bibnamefont {Holz}},
  \bibinfo {author} {\bibfnamefont {W.}~\bibnamefont {Hu}}, \bibinfo {author}
  {\bibfnamefont {M.}~\bibnamefont {Lagos}},\ and\ \bibinfo {author}
  {\bibfnamefont {R.~M.}\ \bibnamefont {Wald}},\ }\bibfield  {title} {\bibinfo
  {title} {{Phase effects from strong gravitational lensing of gravitational
  waves}},\ }\href {https://doi.org/10.1103/PhysRevD.103.064047} {\bibfield
  {journal} {\bibinfo  {journal} {Phys. Rev. D}\ }\textbf {\bibinfo {volume}
  {103}},\ \bibinfo {pages} {064047} (\bibinfo {year} {2021})},\ \Eprint
  {https://arxiv.org/abs/2008.12814} {arXiv:2008.12814 [gr-qc]} \BibitemShut
  {NoStop}%
\bibitem [{\citenamefont {Wang}\ \emph {et~al.}(2021)\citenamefont {Wang},
  \citenamefont {Lo}, \citenamefont {Li},\ and\ \citenamefont
  {Chen}}]{Wang:2021kzt}%
  \BibitemOpen
  \bibfield  {author} {\bibinfo {author} {\bibfnamefont {Y.}~\bibnamefont
  {Wang}}, \bibinfo {author} {\bibfnamefont {R.~K.~L.}\ \bibnamefont {Lo}},
  \bibinfo {author} {\bibfnamefont {A.~K.~Y.}\ \bibnamefont {Li}},\ and\
  \bibinfo {author} {\bibfnamefont {Y.}~\bibnamefont {Chen}},\ }\bibfield
  {title} {\bibinfo {title} {{Identifying Type II Strongly Lensed
  Gravitational-Wave Images in Third-Generation Gravitational-Wave
  Detectors}},\ }\href {https://doi.org/10.1103/PhysRevD.103.104055} {\bibfield
   {journal} {\bibinfo  {journal} {Phys. Rev. D}\ }\textbf {\bibinfo {volume}
  {103}},\ \bibinfo {pages} {104055} (\bibinfo {year} {2021})},\ \Eprint
  {https://arxiv.org/abs/2101.08264} {arXiv:2101.08264 [gr-qc]} \BibitemShut
  {NoStop}%
\bibitem [{\citenamefont {Janquart}\ \emph {et~al.}(2021)\citenamefont
  {Janquart}, \citenamefont {Seo}, \citenamefont {Hannuksela}, \citenamefont
  {Li},\ and\ \citenamefont {Broeck}}]{Janquart:2021nus}%
  \BibitemOpen
  \bibfield  {author} {\bibinfo {author} {\bibfnamefont {J.}~\bibnamefont
  {Janquart}}, \bibinfo {author} {\bibfnamefont {E.}~\bibnamefont {Seo}},
  \bibinfo {author} {\bibfnamefont {O.~A.}\ \bibnamefont {Hannuksela}},
  \bibinfo {author} {\bibfnamefont {T.~G.~F.}\ \bibnamefont {Li}},\ and\
  \bibinfo {author} {\bibfnamefont {C.~V.~D.}\ \bibnamefont {Broeck}},\
  }\bibfield  {title} {\bibinfo {title} {{On the Identification of Individual
  Gravitational-wave Image Types of a Lensed System Using Higher-order
  Modes}},\ }\href {https://doi.org/10.3847/2041-8213/ac3bcf} {\bibfield
  {journal} {\bibinfo  {journal} {Astrophys. J. Lett.}\ }\textbf {\bibinfo
  {volume} {923}},\ \bibinfo {pages} {L1} (\bibinfo {year} {2021})},\ \Eprint
  {https://arxiv.org/abs/2110.06873} {arXiv:2110.06873 [gr-qc]} \BibitemShut
  {NoStop}%
\bibitem [{\citenamefont {Vijaykumar}\ \emph {et~al.}(2023)\citenamefont
  {Vijaykumar}, \citenamefont {Mehta},\ and\ \citenamefont
  {Ganguly}}]{Vijaykumar:2022dlp}%
  \BibitemOpen
  \bibfield  {author} {\bibinfo {author} {\bibfnamefont {A.}~\bibnamefont
  {Vijaykumar}}, \bibinfo {author} {\bibfnamefont {A.~K.}\ \bibnamefont
  {Mehta}},\ and\ \bibinfo {author} {\bibfnamefont {A.}~\bibnamefont
  {Ganguly}},\ }\bibfield  {title} {\bibinfo {title} {{Detection and parameter
  estimation challenges of type-II lensed binary black hole signals}},\ }\href
  {https://doi.org/10.1103/PhysRevD.108.043036} {\bibfield  {journal} {\bibinfo
   {journal} {Phys. Rev. D}\ }\textbf {\bibinfo {volume} {108}},\ \bibinfo
  {pages} {043036} (\bibinfo {year} {2023})},\ \Eprint
  {https://arxiv.org/abs/2202.06334} {arXiv:2202.06334 [gr-qc]} \BibitemShut
  {NoStop}%
\bibitem [{\citenamefont {Taylor}\ \emph {et~al.}(2025)\citenamefont {Taylor},
  \citenamefont {Davis},\ and\ \citenamefont {Lo}}]{Taylor:2024yjt}%
  \BibitemOpen
  \bibfield  {author} {\bibinfo {author} {\bibfnamefont {K.}~\bibnamefont
  {Taylor}}, \bibinfo {author} {\bibfnamefont {D.}~\bibnamefont {Davis}},\ and\
  \bibinfo {author} {\bibfnamefont {R.~K.~L.}\ \bibnamefont {Lo}},\ }\bibfield
  {title} {\bibinfo {title} {{Phase consistency test to identify type II
  strongly lensed gravitational-wave signals using a single event}},\ }\href
  {https://doi.org/10.1103/m16d-hwkz} {\bibfield  {journal} {\bibinfo
  {journal} {Phys. Rev. D}\ }\textbf {\bibinfo {volume} {112}},\ \bibinfo
  {pages} {024035} (\bibinfo {year} {2025})},\ \Eprint
  {https://arxiv.org/abs/2412.15148} {arXiv:2412.15148 [gr-qc]} \BibitemShut
  {NoStop}%
\bibitem [{\citenamefont {Amaro-Seoane}\ \emph {et~al.}(2017)\citenamefont
  {Amaro-Seoane} \emph {et~al.}}]{Audley:2017drz}%
  \BibitemOpen
  \bibfield  {author} {\bibinfo {author} {\bibfnamefont {P.}~\bibnamefont
  {Amaro-Seoane}} \emph {et~al.} (\bibinfo {collaboration} {LISA}),\ }\bibfield
   {title} {\bibinfo {title} {{Laser Interferometer Space Antenna}},\
  }\href@noop {} {\bibfield  {journal} {\bibinfo  {journal} {ArXiv}\ }
  (\bibinfo {year} {2017})},\ \Eprint {https://arxiv.org/abs/1702.00786}
  {arXiv:1702.00786 [astro-ph.IM]} \BibitemShut {NoStop}%
\bibitem [{\citenamefont {Colpi}\ \emph {et~al.}(2024)\citenamefont {Colpi}
  \emph {et~al.}}]{LISA:2024hlh}%
  \BibitemOpen
  \bibfield  {author} {\bibinfo {author} {\bibfnamefont {M.}~\bibnamefont
  {Colpi}} \emph {et~al.} (\bibinfo {collaboration} {LISA}),\ }\bibfield
  {title} {\bibinfo {title} {{LISA Definition Study Report}},\ }\href@noop {}
  {\bibfield  {journal} {\bibinfo  {journal} {arXiv}\ } (\bibinfo {year}
  {2024})},\ \Eprint {https://arxiv.org/abs/2402.07571} {arXiv:2402.07571
  [astro-ph.CO]} \BibitemShut {NoStop}%
\bibitem [{\citenamefont {Barnich}\ and\ \citenamefont
  {Troessaert}(2010)}]{Barnich:2010eb}%
  \BibitemOpen
  \bibfield  {author} {\bibinfo {author} {\bibfnamefont {G.}~\bibnamefont
  {Barnich}}\ and\ \bibinfo {author} {\bibfnamefont {C.}~\bibnamefont
  {Troessaert}},\ }\bibfield  {title} {\bibinfo {title} {{Aspects of the
  BMS/CFT correspondence}},\ }\href {https://doi.org/10.1007/JHEP05(2010)062}
  {\bibfield  {journal} {\bibinfo  {journal} {JHEP}\ }\textbf {\bibinfo
  {volume} {05}},\ \bibinfo {pages} {062}},\ \Eprint
  {https://arxiv.org/abs/1001.1541} {arXiv:1001.1541 [hep-th]} \BibitemShut
  {NoStop}%
\bibitem [{\citenamefont {Hou}\ \emph {et~al.}(2022)\citenamefont {Hou},
  \citenamefont {Zhu},\ and\ \citenamefont {Zhu}}]{Hou:2021oxe}%
  \BibitemOpen
  \bibfield  {author} {\bibinfo {author} {\bibfnamefont {S.}~\bibnamefont
  {Hou}}, \bibinfo {author} {\bibfnamefont {T.}~\bibnamefont {Zhu}},\ and\
  \bibinfo {author} {\bibfnamefont {Z.-H.}\ \bibnamefont {Zhu}},\ }\bibfield
  {title} {\bibinfo {title} {{Asymptotic analysis of Chern-Simons modified
  gravity and its memory effects}},\ }\href
  {https://doi.org/10.1103/PhysRevD.105.024025} {\bibfield  {journal} {\bibinfo
   {journal} {Phys. Rev. D}\ }\textbf {\bibinfo {volume} {105}},\ \bibinfo
  {pages} {024025} (\bibinfo {year} {2022})},\ \Eprint
  {https://arxiv.org/abs/2109.04238} {arXiv:2109.04238 [gr-qc]} \BibitemShut
  {NoStop}%
\bibitem [{\citenamefont {Bieri}\ and\ \citenamefont
  {Garfinkle}(2014)}]{Bieri:2013ada}%
  \BibitemOpen
  \bibfield  {author} {\bibinfo {author} {\bibfnamefont {L.}~\bibnamefont
  {Bieri}}\ and\ \bibinfo {author} {\bibfnamefont {D.}~\bibnamefont
  {Garfinkle}},\ }\bibfield  {title} {\bibinfo {title} {{Perturbative and gauge
  invariant treatment of gravitational wave memory}},\ }\href
  {https://doi.org/10.1103/PhysRevD.89.084039} {\bibfield  {journal} {\bibinfo
  {journal} {Phys. Rev. D}\ }\textbf {\bibinfo {volume} {89}},\ \bibinfo
  {pages} {084039} (\bibinfo {year} {2014})},\ \Eprint
  {https://arxiv.org/abs/1312.6871} {arXiv:1312.6871 [gr-qc]} \BibitemShut
  {NoStop}%
\bibitem [{\citenamefont {Nichols}(2017)}]{Nichols:2017rqr}%
  \BibitemOpen
  \bibfield  {author} {\bibinfo {author} {\bibfnamefont {D.~A.}\ \bibnamefont
  {Nichols}},\ }\bibfield  {title} {\bibinfo {title} {{Spin memory effect for
  compact binaries in the post-Newtonian approximation}},\ }\href
  {https://doi.org/10.1103/PhysRevD.95.084048} {\bibfield  {journal} {\bibinfo
  {journal} {Phys. Rev. D}\ }\textbf {\bibinfo {volume} {95}},\ \bibinfo
  {pages} {084048} (\bibinfo {year} {2017})},\ \Eprint
  {https://arxiv.org/abs/1702.03300} {arXiv:1702.03300 [gr-qc]} \BibitemShut
  {NoStop}%
\bibitem [{\citenamefont {Mitman}\ \emph {et~al.}(2020)\citenamefont {Mitman},
  \citenamefont {Moxon}, \citenamefont {Scheel}, \citenamefont {Teukolsky},
  \citenamefont {Boyle}, \citenamefont {Deppe}, \citenamefont {Kidder},\ and\
  \citenamefont {Throwe}}]{Mitman:2020pbt}%
  \BibitemOpen
  \bibfield  {author} {\bibinfo {author} {\bibfnamefont {K.}~\bibnamefont
  {Mitman}}, \bibinfo {author} {\bibfnamefont {J.}~\bibnamefont {Moxon}},
  \bibinfo {author} {\bibfnamefont {M.~A.}\ \bibnamefont {Scheel}}, \bibinfo
  {author} {\bibfnamefont {S.~A.}\ \bibnamefont {Teukolsky}}, \bibinfo {author}
  {\bibfnamefont {M.}~\bibnamefont {Boyle}}, \bibinfo {author} {\bibfnamefont
  {N.}~\bibnamefont {Deppe}}, \bibinfo {author} {\bibfnamefont {L.~E.}\
  \bibnamefont {Kidder}},\ and\ \bibinfo {author} {\bibfnamefont
  {W.}~\bibnamefont {Throwe}},\ }\bibfield  {title} {\bibinfo {title}
  {{Computation of displacement and spin gravitational memory in numerical
  relativity}},\ }\href {https://doi.org/10.1103/PhysRevD.102.104007}
  {\bibfield  {journal} {\bibinfo  {journal} {Phys. Rev. D}\ }\textbf {\bibinfo
  {volume} {102}},\ \bibinfo {pages} {104007} (\bibinfo {year} {2020})},\
  \Eprint {https://arxiv.org/abs/2007.11562} {arXiv:2007.11562 [gr-qc]}
  \BibitemShut {NoStop}%
\bibitem [{\citenamefont {Nitz}\ \emph {et~al.}(2024)\citenamefont {Nitz},
  \citenamefont {Harry}, \citenamefont {Brown}, \citenamefont {Biwer},
  \citenamefont {Willis}, \citenamefont {Canton}, \citenamefont {Capano},
  \citenamefont {Dent}, \citenamefont {Pekowsky}, \citenamefont {Davies},
  \citenamefont {De}, \citenamefont {Cabero}, \citenamefont {Wu}, \citenamefont
  {Williamson}, \citenamefont {Machenschalk}, \citenamefont {Macleod},
  \citenamefont {Pannarale}, \citenamefont {Kumar}, \citenamefont {Reyes},
  \citenamefont {dfinstad}, \citenamefont {Kumar}, \citenamefont {Tápai},
  \citenamefont {Singer}, \citenamefont {Kumar}, \citenamefont {veronica
  villa}, \citenamefont {maxtrevor}, \citenamefont {Gadre}, \citenamefont
  {Khan}, \citenamefont {Fairhurst},\ and\ \citenamefont
  {Tolley}}]{alex_nitz_2024_10473621}%
  \BibitemOpen
  \bibfield  {author} {\bibinfo {author} {\bibfnamefont {A.}~\bibnamefont
  {Nitz}}, \bibinfo {author} {\bibfnamefont {I.}~\bibnamefont {Harry}},
  \bibinfo {author} {\bibfnamefont {D.}~\bibnamefont {Brown}}, \bibinfo
  {author} {\bibfnamefont {C.~M.}\ \bibnamefont {Biwer}}, \bibinfo {author}
  {\bibfnamefont {J.}~\bibnamefont {Willis}}, \bibinfo {author} {\bibfnamefont
  {T.~D.}\ \bibnamefont {Canton}}, \bibinfo {author} {\bibfnamefont
  {C.}~\bibnamefont {Capano}}, \bibinfo {author} {\bibfnamefont
  {T.}~\bibnamefont {Dent}}, \bibinfo {author} {\bibfnamefont {L.}~\bibnamefont
  {Pekowsky}}, \bibinfo {author} {\bibfnamefont {G.~S.~C.}\ \bibnamefont
  {Davies}}, \bibinfo {author} {\bibfnamefont {S.}~\bibnamefont {De}}, \bibinfo
  {author} {\bibfnamefont {M.}~\bibnamefont {Cabero}}, \bibinfo {author}
  {\bibfnamefont {S.}~\bibnamefont {Wu}}, \bibinfo {author} {\bibfnamefont
  {A.~R.}\ \bibnamefont {Williamson}}, \bibinfo {author} {\bibfnamefont
  {B.}~\bibnamefont {Machenschalk}}, \bibinfo {author} {\bibfnamefont
  {D.}~\bibnamefont {Macleod}}, \bibinfo {author} {\bibfnamefont
  {F.}~\bibnamefont {Pannarale}}, \bibinfo {author} {\bibfnamefont
  {P.}~\bibnamefont {Kumar}}, \bibinfo {author} {\bibfnamefont
  {S.}~\bibnamefont {Reyes}}, \bibinfo {author} {\bibnamefont {dfinstad}},
  \bibinfo {author} {\bibfnamefont {S.}~\bibnamefont {Kumar}}, \bibinfo
  {author} {\bibfnamefont {M.}~\bibnamefont {Tápai}}, \bibinfo {author}
  {\bibfnamefont {L.}~\bibnamefont {Singer}}, \bibinfo {author} {\bibfnamefont
  {P.}~\bibnamefont {Kumar}}, \bibinfo {author} {\bibnamefont {veronica
  villa}}, \bibinfo {author} {\bibnamefont {maxtrevor}}, \bibinfo {author}
  {\bibfnamefont {B.~U.~V.}\ \bibnamefont {Gadre}}, \bibinfo {author}
  {\bibfnamefont {S.}~\bibnamefont {Khan}}, \bibinfo {author} {\bibfnamefont
  {S.}~\bibnamefont {Fairhurst}},\ and\ \bibinfo {author} {\bibfnamefont
  {A.}~\bibnamefont {Tolley}},\ }\href
  {https://doi.org/10.5281/zenodo.10473621} {\bibinfo {title} {gwastro/pycbc:
  v2.3.3 release of pycbc}} (\bibinfo {year} {2024})\BibitemShut {NoStop}%
\bibitem [{\citenamefont {Boyle}\ \emph {et~al.}(2019)\citenamefont {Boyle}
  \emph {et~al.}}]{Boyle:2019kee}%
  \BibitemOpen
  \bibfield  {author} {\bibinfo {author} {\bibfnamefont {M.}~\bibnamefont
  {Boyle}} \emph {et~al.},\ }\bibfield  {title} {\bibinfo {title} {{The SXS
  Collaboration catalog of binary black hole simulations}},\ }\href
  {https://doi.org/10.1088/1361-6382/ab34e2} {\bibfield  {journal} {\bibinfo
  {journal} {Class. Quant. Grav.}\ }\textbf {\bibinfo {volume} {36}},\ \bibinfo
  {pages} {195006} (\bibinfo {year} {2019})},\ \Eprint
  {https://arxiv.org/abs/1904.04831} {arXiv:1904.04831 [gr-qc]} \BibitemShut
  {NoStop}%
\bibitem [{\citenamefont {Seto}\ \emph {et~al.}(2001)\citenamefont {Seto},
  \citenamefont {Kawamura},\ and\ \citenamefont {Nakamura}}]{Seto:2001qf}%
  \BibitemOpen
  \bibfield  {author} {\bibinfo {author} {\bibfnamefont {N.}~\bibnamefont
  {Seto}}, \bibinfo {author} {\bibfnamefont {S.}~\bibnamefont {Kawamura}},\
  and\ \bibinfo {author} {\bibfnamefont {T.}~\bibnamefont {Nakamura}},\
  }\bibfield  {title} {\bibinfo {title} {{Possibility of direct measurement of
  the acceleration of the universe using 0.1-Hz band laser interferometer
  gravitational wave antenna in space}},\ }\href
  {https://doi.org/10.1103/PhysRevLett.87.221103} {\bibfield  {journal}
  {\bibinfo  {journal} {Phys. Rev. Lett.}\ }\textbf {\bibinfo {volume} {87}},\
  \bibinfo {pages} {221103} (\bibinfo {year} {2001})},\ \Eprint
  {https://arxiv.org/abs/astro-ph/0108011} {arXiv:astro-ph/0108011 [astro-ph]}
  \BibitemShut {NoStop}%
\bibitem [{\citenamefont {{Kawamura}}\ \emph {et~al.}(2019)\citenamefont
  {{Kawamura}} \emph {et~al.}}]{decigo2019}%
  \BibitemOpen
  \bibfield  {author} {\bibinfo {author} {\bibfnamefont {S.}~\bibnamefont
  {{Kawamura}}} \emph {et~al.},\ }\bibfield  {title} {\bibinfo {title} {{Space
  gravitational-wave antennas DECIGO and B-DECIGO}},\ }\href
  {https://doi.org/10.1142/S0218271818450013} {\bibfield  {journal} {\bibinfo
  {journal} {Int. J. of Mod. Phys. D}\ }\textbf {\bibinfo {volume} {28}},\
  \bibinfo {eid} {1845001} (\bibinfo {year} {2019})}\BibitemShut {NoStop}%
\bibitem [{\citenamefont {Kawamura}\ \emph {et~al.}(2021)\citenamefont
  {Kawamura} \emph {et~al.}}]{Kawamura:2020pcg}%
  \BibitemOpen
  \bibfield  {author} {\bibinfo {author} {\bibfnamefont {S.}~\bibnamefont
  {Kawamura}} \emph {et~al.},\ }\bibfield  {title} {\bibinfo {title} {{Current
  status of space gravitational wave antenna DECIGO and B-DECIGO}},\ }\href
  {https://doi.org/10.1093/ptep/ptab019} {\bibfield  {journal} {\bibinfo
  {journal} {PTEP}\ }\textbf {\bibinfo {volume} {2021}},\ \bibinfo {pages}
  {05A105} (\bibinfo {year} {2021})},\ \Eprint
  {https://arxiv.org/abs/2006.13545} {arXiv:2006.13545 [gr-qc]} \BibitemShut
  {NoStop}%
\bibitem [{\citenamefont {Abbott}\ \emph {et~al.}(2017)\citenamefont {Abbott}
  \emph {et~al.}}]{Evans:2016mbw}%
  \BibitemOpen
  \bibfield  {author} {\bibinfo {author} {\bibfnamefont {B.~P.}\ \bibnamefont
  {Abbott}} \emph {et~al.} (\bibinfo {collaboration} {LIGO Scientific}),\
  }\bibfield  {title} {\bibinfo {title} {{Exploring the Sensitivity of Next
  Generation Gravitational Wave Detectors}},\ }\href
  {https://doi.org/10.1088/1361-6382/aa51f4} {\bibfield  {journal} {\bibinfo
  {journal} {Class. Quant. Grav.}\ }\textbf {\bibinfo {volume} {34}},\ \bibinfo
  {pages} {044001} (\bibinfo {year} {2017})},\ \Eprint
  {https://arxiv.org/abs/1607.08697} {arXiv:1607.08697 [astro-ph.IM]}
  \BibitemShut {NoStop}%
\bibitem [{\citenamefont {Punturo}\ \emph {et~al.}(2010)\citenamefont {Punturo}
  \emph {et~al.}}]{Punturo:2010zza}%
  \BibitemOpen
  \bibfield  {author} {\bibinfo {author} {\bibfnamefont {M.}~\bibnamefont
  {Punturo}} \emph {et~al.},\ }\bibfield  {title} {\bibinfo {title} {{The third
  generation of gravitational wave observatories and their science reach}},\
  }\bibfield  {booktitle} {\emph {\bibinfo {booktitle} {{Gravitational waves.
  Proceedings, 8th Edoardo Amaldi Conference, Amaldi 8, New York, USA, June
  22-26, 2009}}},\ }\href {https://doi.org/10.1088/0264-9381/27/8/084007}
  {\bibfield  {journal} {\bibinfo  {journal} {Class. Quant. Grav.}\ }\textbf
  {\bibinfo {volume} {27}},\ \bibinfo {pages} {084007} (\bibinfo {year}
  {2010})}\BibitemShut {NoStop}%
\bibitem [{\citenamefont {Babak}\ \emph {et~al.}(2021)\citenamefont {Babak},
  \citenamefont {Petiteau},\ and\ \citenamefont {Hewitson}}]{Babak:2021mhe}%
  \BibitemOpen
  \bibfield  {author} {\bibinfo {author} {\bibfnamefont {S.}~\bibnamefont
  {Babak}}, \bibinfo {author} {\bibfnamefont {A.}~\bibnamefont {Petiteau}},\
  and\ \bibinfo {author} {\bibfnamefont {M.}~\bibnamefont {Hewitson}},\
  }\bibfield  {title} {\bibinfo {title} {{LISA Sensitivity and SNR
  Calculations}},\ }\href@noop {} {\bibfield  {journal} {\bibinfo  {journal}
  {arXiv}\ } (\bibinfo {year} {2021})},\ \Eprint
  {https://arxiv.org/abs/2108.01167} {arXiv:2108.01167 [astro-ph.IM]}
  \BibitemShut {NoStop}%
\bibitem [{\citenamefont {Pratten}\ \emph {et~al.}(2021)\citenamefont {Pratten}
  \emph {et~al.}}]{Pratten:2020ceb}%
  \BibitemOpen
  \bibfield  {author} {\bibinfo {author} {\bibfnamefont {G.}~\bibnamefont
  {Pratten}} \emph {et~al.},\ }\bibfield  {title} {\bibinfo {title}
  {{Computationally efficient models for the dominant and subdominant harmonic
  modes of precessing binary black holes}},\ }\href
  {https://doi.org/10.1103/PhysRevD.103.104056} {\bibfield  {journal} {\bibinfo
   {journal} {Phys. Rev. D}\ }\textbf {\bibinfo {volume} {103}},\ \bibinfo
  {pages} {104056} (\bibinfo {year} {2021})},\ \Eprint
  {https://arxiv.org/abs/2004.06503} {arXiv:2004.06503 [gr-qc]} \BibitemShut
  {NoStop}%
\bibitem [{Note1()}]{Note1}%
  \BibitemOpen
  \bibinfo {note} {According to Eqs.~\protect \textup {\hbox {\mathsurround \z@
  \protect \normalfont (\ignorespaces \ref {eq-phi-lm}\unskip \@@italiccorr
  )}}, \protect \textup {\hbox {\mathsurround \z@ \protect \normalfont
  (\ignorespaces \ref {eq-def-cl}\unskip \@@italiccorr )}} and \protect \textup
  {\hbox {\mathsurround \z@ \protect \normalfont (\ignorespaces \ref
  {eq-mem-1}\unskip \@@italiccorr )}}, for a chosen pair of $h_{\protect \hat
  \ell \protect \hat m}$ and $h_{\protect \tilde \ell \protect \tilde m}$,
  $\protect \mathcal C_{\ell m}(-2,\protect \hat \ell ,\protect \hat
  m;2,\protect \tilde \ell ,\protect \tilde m)$ might be vanishing for a lot of
  $(\ell ,m)$ pairs.}\BibitemShut {Stop}%
\bibitem [{Note2()}]{Note2}%
  \BibitemOpen
  \bibinfo {note} {It might be possible that the approximate step-function like
  behavior of the time-domain memory signal is degenerate with the instrumental
  glitches or non-Gaussian noise. However, we will not discuss this possibility
  in the current work, as the focus is on the lensing effect. As discussed
  below, the step-function like behavior disappears after the memory signal is
  strongly lensed.}\BibitemShut {Stop}%
\bibitem [{\citenamefont {Ezquiaga}\ and\ \citenamefont
  {Zumalac{\'a}rregui}(2020)}]{Ezquiaga:2020dao}%
  \BibitemOpen
  \bibfield  {author} {\bibinfo {author} {\bibfnamefont {J.~M.}\ \bibnamefont
  {Ezquiaga}}\ and\ \bibinfo {author} {\bibfnamefont {M.}~\bibnamefont
  {Zumalac{\'a}rregui}},\ }\bibfield  {title} {\bibinfo {title} {{Gravitational
  wave lensing beyond general relativity: birefringence, echoes and shadows}},\
  }\href {https://doi.org/10.1103/PhysRevD.102.124048} {\bibfield  {journal}
  {\bibinfo  {journal} {Phys. Rev. D}\ }\textbf {\bibinfo {volume} {102}},\
  \bibinfo {pages} {124048} (\bibinfo {year} {2020})},\ \Eprint
  {https://arxiv.org/abs/2009.12187} {arXiv:2009.12187 [gr-qc]} \BibitemShut
  {NoStop}%
\bibitem [{\citenamefont {Xu}\ \emph {et~al.}(2022)\citenamefont {Xu},
  \citenamefont {Ezquiaga},\ and\ \citenamefont {Holz}}]{Xu:2021bfn}%
  \BibitemOpen
  \bibfield  {author} {\bibinfo {author} {\bibfnamefont {F.}~\bibnamefont
  {Xu}}, \bibinfo {author} {\bibfnamefont {J.~M.}\ \bibnamefont {Ezquiaga}},\
  and\ \bibinfo {author} {\bibfnamefont {D.~E.}\ \bibnamefont {Holz}},\
  }\bibfield  {title} {\bibinfo {title} {{Please Repeat: Strong Lensing of
  Gravitational Waves as a Probe of Compact Binary and Galaxy Populations}},\
  }\href {https://doi.org/10.3847/1538-4357/ac58f8} {\bibfield  {journal}
  {\bibinfo  {journal} {Astrophys. J.}\ }\textbf {\bibinfo {volume} {929}},\
  \bibinfo {pages} {9} (\bibinfo {year} {2022})},\ \Eprint
  {https://arxiv.org/abs/2105.14390} {arXiv:2105.14390 [astro-ph.CO]}
  \BibitemShut {NoStop}%
\bibitem [{\citenamefont {Weinberg}(2008)}]{Weinberg:2008zzc}%
  \BibitemOpen
  \bibfield  {author} {\bibinfo {author} {\bibfnamefont {S.}~\bibnamefont
  {Weinberg}},\ }\href {http://www.oup.com/uk/catalogue/?ci=9780198526827}
  {\emph {\bibinfo {title} {{Cosmology}}}}\ (\bibinfo  {publisher} {Oxford, UK:
  Oxford Univ. Pr. (2008) 593 p},\ \bibinfo {year} {2008})\BibitemShut
  {NoStop}%
\bibitem [{\citenamefont {Liao}\ \emph {et~al.}(2019)\citenamefont {Liao},
  \citenamefont {Biesiada},\ and\ \citenamefont {Fan}}]{Liao:2019aqq}%
  \BibitemOpen
  \bibfield  {author} {\bibinfo {author} {\bibfnamefont {K.}~\bibnamefont
  {Liao}}, \bibinfo {author} {\bibfnamefont {M.}~\bibnamefont {Biesiada}},\
  and\ \bibinfo {author} {\bibfnamefont {X.-L.}\ \bibnamefont {Fan}},\
  }\bibfield  {title} {\bibinfo {title} {{The wave nature of continuous
  gravitational waves from microlensing}},\ }\href
  {https://doi.org/10.3847/1538-4357/ab1087} {\bibfield  {journal} {\bibinfo
  {journal} {Astrophys. J.}\ }\textbf {\bibinfo {volume} {875}},\ \bibinfo
  {pages} {139} (\bibinfo {year} {2019})},\ \Eprint
  {https://arxiv.org/abs/1903.06612} {arXiv:1903.06612 [gr-qc]} \BibitemShut
  {NoStop}%
\bibitem [{Note3()}]{Note3}%
  \BibitemOpen
  \bibinfo {note} {For the saddle point, one can compare this $k_a$ with that
  in Ref.~\cite {Takahashi:2003ix}, which is $e^{-i\pi /2}$. The factor
  $e^{-i\pi /2}$ violates the reality condition $F^*(\omega ,\protect \vec
  X)=F(-\omega ,\protect \vec X)$.}\BibitemShut {Stop}%
\bibitem [{\citenamefont {Allen}\ \emph {et~al.}(2003)\citenamefont {Allen},
  \citenamefont {Creighton}, \citenamefont {Flanagan},\ and\ \citenamefont
  {Romano}}]{Allen:2002jw}%
  \BibitemOpen
  \bibfield  {author} {\bibinfo {author} {\bibfnamefont {B.}~\bibnamefont
  {Allen}}, \bibinfo {author} {\bibfnamefont {J.~D.~E.}\ \bibnamefont
  {Creighton}}, \bibinfo {author} {\bibfnamefont {E.~E.}\ \bibnamefont
  {Flanagan}},\ and\ \bibinfo {author} {\bibfnamefont {J.~D.}\ \bibnamefont
  {Romano}},\ }\bibfield  {title} {\bibinfo {title} {{Robust statistics for
  deterministic and stochastic gravitational waves in nonGaussian noise. 2.
  Bayesian analyses}},\ }\href {https://doi.org/10.1103/PhysRevD.67.122002}
  {\bibfield  {journal} {\bibinfo  {journal} {Phys. Rev. D}\ }\textbf {\bibinfo
  {volume} {67}},\ \bibinfo {pages} {122002} (\bibinfo {year} {2003})},\
  \Eprint {https://arxiv.org/abs/gr-qc/0205015} {arXiv:gr-qc/0205015}
  \BibitemShut {NoStop}%
\bibitem [{\citenamefont {Allen}\ \emph {et~al.}(2012)\citenamefont {Allen},
  \citenamefont {Anderson}, \citenamefont {Brady}, \citenamefont {Brown},\ and\
  \citenamefont {Creighton}}]{Allen:2005fk}%
  \BibitemOpen
  \bibfield  {author} {\bibinfo {author} {\bibfnamefont {B.}~\bibnamefont
  {Allen}}, \bibinfo {author} {\bibfnamefont {W.~G.}\ \bibnamefont {Anderson}},
  \bibinfo {author} {\bibfnamefont {P.~R.}\ \bibnamefont {Brady}}, \bibinfo
  {author} {\bibfnamefont {D.~A.}\ \bibnamefont {Brown}},\ and\ \bibinfo
  {author} {\bibfnamefont {J.~D.~E.}\ \bibnamefont {Creighton}},\ }\bibfield
  {title} {\bibinfo {title} {{FINDCHIRP: An Algorithm for detection of
  gravitational waves from inspiraling compact binaries}},\ }\href
  {https://doi.org/10.1103/PhysRevD.85.122006} {\bibfield  {journal} {\bibinfo
  {journal} {Phys. Rev. D}\ }\textbf {\bibinfo {volume} {85}},\ \bibinfo
  {pages} {122006} (\bibinfo {year} {2012})},\ \Eprint
  {https://arxiv.org/abs/gr-qc/0509116} {arXiv:gr-qc/0509116} \BibitemShut
  {NoStop}%
\bibitem [{\citenamefont {Tinto}\ and\ \citenamefont
  {Dhurandhar}(2014)}]{Tinto:2014lxa}%
  \BibitemOpen
  \bibfield  {author} {\bibinfo {author} {\bibfnamefont {M.}~\bibnamefont
  {Tinto}}\ and\ \bibinfo {author} {\bibfnamefont {S.~V.}\ \bibnamefont
  {Dhurandhar}},\ }\bibfield  {title} {\bibinfo {title} {{Time-Delay
  Interferometry}},\ }\href {https://doi.org/10.12942/lrr-2014-6} {\bibfield
  {journal} {\bibinfo  {journal} {Living Rev. Rel.}\ }\textbf {\bibinfo
  {volume} {17}},\ \bibinfo {pages} {6} (\bibinfo {year} {2014})}\BibitemShut
  {NoStop}%
\bibitem [{\citenamefont {Marsat}\ and\ \citenamefont
  {Baker}(2018)}]{Marsat:2018oam}%
  \BibitemOpen
  \bibfield  {author} {\bibinfo {author} {\bibfnamefont {S.}~\bibnamefont
  {Marsat}}\ and\ \bibinfo {author} {\bibfnamefont {J.~G.}\ \bibnamefont
  {Baker}},\ }\bibfield  {title} {\bibinfo {title} {{Fourier-domain modulations
  and delays of gravitational-wave signals}},\ }\href@noop {} {\bibfield
  {journal} {\bibinfo  {journal} {arXiv}\ } (\bibinfo {year} {2018})},\ \Eprint
  {https://arxiv.org/abs/1806.10734} {arXiv:1806.10734 [gr-qc]} \BibitemShut
  {NoStop}%
\bibitem [{\citenamefont {Marsat}\ \emph {et~al.}(2021)\citenamefont {Marsat},
  \citenamefont {Baker},\ and\ \citenamefont {Dal~Canton}}]{Marsat:2020rtl}%
  \BibitemOpen
  \bibfield  {author} {\bibinfo {author} {\bibfnamefont {S.}~\bibnamefont
  {Marsat}}, \bibinfo {author} {\bibfnamefont {J.~G.}\ \bibnamefont {Baker}},\
  and\ \bibinfo {author} {\bibfnamefont {T.}~\bibnamefont {Dal~Canton}},\
  }\bibfield  {title} {\bibinfo {title} {{Exploring the Bayesian parameter
  estimation of binary black holes with LISA}},\ }\href
  {https://doi.org/10.1103/PhysRevD.103.083011} {\bibfield  {journal} {\bibinfo
   {journal} {Phys. Rev. D}\ }\textbf {\bibinfo {volume} {103}},\ \bibinfo
  {pages} {083011} (\bibinfo {year} {2021})},\ \Eprint
  {https://arxiv.org/abs/2003.00357} {arXiv:2003.00357 [gr-qc]} \BibitemShut
  {NoStop}%
\bibitem [{\citenamefont {Garc{\'\i}a-Quir{\'o}s}\ \emph
  {et~al.}(2025)\citenamefont {Garc{\'\i}a-Quir{\'o}s}, \citenamefont
  {Tiwari},\ and\ \citenamefont {Babak}}]{Garcia-Quiros:2025usi}%
  \BibitemOpen
  \bibfield  {author} {\bibinfo {author} {\bibfnamefont {C.}~\bibnamefont
  {Garc{\'\i}a-Quir{\'o}s}}, \bibinfo {author} {\bibfnamefont {S.}~\bibnamefont
  {Tiwari}},\ and\ \bibinfo {author} {\bibfnamefont {S.}~\bibnamefont
  {Babak}},\ }\bibfield  {title} {\bibinfo {title} {{GPU-accelerated LISA
  parameter estimation with full time-domain response}},\ }\href
  {https://doi.org/10.1103/79kn-53nt} {\bibfield  {journal} {\bibinfo
  {journal} {Phys. Rev. D}\ }\textbf {\bibinfo {volume} {112}},\ \bibinfo
  {pages} {064017} (\bibinfo {year} {2025})},\ \Eprint
  {https://arxiv.org/abs/2501.08261} {arXiv:2501.08261 [gr-qc]} \BibitemShut
  {NoStop}%
\bibitem [{Note4()}]{Note4}%
  \BibitemOpen
  \bibinfo {note} {Please refer to \protect \href {https://dlmf.nist.gov/}{NIST
  Digital Library of Mathematical Functions}.}\BibitemShut {Stop}%
\bibitem [{Note5()}]{Note5}%
  \BibitemOpen
  \bibinfo {note} {Here, by ``same'', we mean the same functional form for the
  antenna pattern function, but not the same value, if the space-borne
  interferometer is used to observe the lensed memory signal.}\BibitemShut
  {Stop}%
\bibitem [{\citenamefont {Zhang}\ \emph {et~al.}(2020)\citenamefont {Zhang},
  \citenamefont {Gao}, \citenamefont {Gong}, \citenamefont {Wang},
  \citenamefont {Weinstein},\ and\ \citenamefont {Zhang}}]{Zhang:2020khm}%
  \BibitemOpen
  \bibfield  {author} {\bibinfo {author} {\bibfnamefont {C.}~\bibnamefont
  {Zhang}}, \bibinfo {author} {\bibfnamefont {Q.}~\bibnamefont {Gao}}, \bibinfo
  {author} {\bibfnamefont {Y.}~\bibnamefont {Gong}}, \bibinfo {author}
  {\bibfnamefont {B.}~\bibnamefont {Wang}}, \bibinfo {author} {\bibfnamefont
  {A.~J.}\ \bibnamefont {Weinstein}},\ and\ \bibinfo {author} {\bibfnamefont
  {C.}~\bibnamefont {Zhang}},\ }\bibfield  {title} {\bibinfo {title} {{Full
  analytical formulas for frequency response of space-based gravitational wave
  detectors}},\ }\href {https://doi.org/10.1103/PhysRevD.101.124027} {\bibfield
   {journal} {\bibinfo  {journal} {Phys. Rev. D}\ }\textbf {\bibinfo {volume}
  {101}},\ \bibinfo {pages} {124027} (\bibinfo {year} {2020})},\ \Eprint
  {https://arxiv.org/abs/2003.01441} {arXiv:2003.01441 [gr-qc]} \BibitemShut
  {NoStop}%
\bibitem [{\citenamefont {Apostolatos}(1995)}]{Apostolatos:1995pj}%
  \BibitemOpen
  \bibfield  {author} {\bibinfo {author} {\bibfnamefont {T.~A.}\ \bibnamefont
  {Apostolatos}},\ }\bibfield  {title} {\bibinfo {title} {{Search templates for
  gravitational waves from precessing, inspiraling binaries}},\ }\href
  {https://doi.org/10.1103/PhysRevD.52.605} {\bibfield  {journal} {\bibinfo
  {journal} {Phys. Rev. D}\ }\textbf {\bibinfo {volume} {52}},\ \bibinfo
  {pages} {605} (\bibinfo {year} {1995})}\BibitemShut {NoStop}%
\bibitem [{\citenamefont {Barausse}\ \emph {et~al.}(2023)\citenamefont
  {Barausse}, \citenamefont {Dey}, \citenamefont {Crisostomi}, \citenamefont
  {Panayada}, \citenamefont {Marsat},\ and\ \citenamefont
  {Basak}}]{Barausse:2023yrx}%
  \BibitemOpen
  \bibfield  {author} {\bibinfo {author} {\bibfnamefont {E.}~\bibnamefont
  {Barausse}}, \bibinfo {author} {\bibfnamefont {K.}~\bibnamefont {Dey}},
  \bibinfo {author} {\bibfnamefont {M.}~\bibnamefont {Crisostomi}}, \bibinfo
  {author} {\bibfnamefont {A.}~\bibnamefont {Panayada}}, \bibinfo {author}
  {\bibfnamefont {S.}~\bibnamefont {Marsat}},\ and\ \bibinfo {author}
  {\bibfnamefont {S.}~\bibnamefont {Basak}},\ }\bibfield  {title} {\bibinfo
  {title} {{Implications of the pulsar timing array detections for massive
  black hole mergers in the LISA band}},\ }\href
  {https://doi.org/10.1103/PhysRevD.108.103034} {\bibfield  {journal} {\bibinfo
   {journal} {Phys. Rev. D}\ }\textbf {\bibinfo {volume} {108}},\ \bibinfo
  {pages} {103034} (\bibinfo {year} {2023})},\ \Eprint
  {https://arxiv.org/abs/2307.12245} {arXiv:2307.12245 [astro-ph.GA]}
  \BibitemShut {NoStop}%
\bibitem [{\citenamefont {Tarafdar}\ \emph {et~al.}(2022)\citenamefont
  {Tarafdar} \emph {et~al.}}]{Tarafdar:2022toa}%
  \BibitemOpen
  \bibfield  {author} {\bibinfo {author} {\bibfnamefont {P.}~\bibnamefont
  {Tarafdar}} \emph {et~al.},\ }\bibfield  {title} {\bibinfo {title} {{The
  Indian Pulsar Timing Array: First data release}},\ }\href
  {https://doi.org/10.1017/pasa.2022.46} {\bibfield  {journal} {\bibinfo
  {journal} {Publ. Astron. Soc. Austral.}\ }\textbf {\bibinfo {volume} {39}},\
  \bibinfo {pages} {e053} (\bibinfo {year} {2022})},\ \Eprint
  {https://arxiv.org/abs/2206.09289} {arXiv:2206.09289 [astro-ph.IM]}
  \BibitemShut {NoStop}%
\bibitem [{\citenamefont {Antoniadis}\ \emph {et~al.}(2023)\citenamefont
  {Antoniadis} \emph {et~al.}}]{EPTA:2023fyk}%
  \BibitemOpen
  \bibfield  {author} {\bibinfo {author} {\bibfnamefont {J.}~\bibnamefont
  {Antoniadis}} \emph {et~al.} (\bibinfo {collaboration} {EPTA, InPTA:}),\
  }\bibfield  {title} {\bibinfo {title} {{The second data release from the
  European Pulsar Timing Array - III. Search for gravitational wave signals}},\
  }\href {https://doi.org/10.1051/0004-6361/202346844} {\bibfield  {journal}
  {\bibinfo  {journal} {Astron. Astrophys.}\ }\textbf {\bibinfo {volume}
  {678}},\ \bibinfo {pages} {A50} (\bibinfo {year} {2023})},\ \Eprint
  {https://arxiv.org/abs/2306.16214} {arXiv:2306.16214 [astro-ph.HE]}
  \BibitemShut {NoStop}%
\bibitem [{\citenamefont {Agazie}\ \emph {et~al.}(2023)\citenamefont {Agazie}
  \emph {et~al.}}]{NANOGrav:2023gor}%
  \BibitemOpen
  \bibfield  {author} {\bibinfo {author} {\bibfnamefont {G.}~\bibnamefont
  {Agazie}} \emph {et~al.} (\bibinfo {collaboration} {NANOGrav}),\ }\bibfield
  {title} {\bibinfo {title} {{The NANOGrav 15 yr Data Set: Evidence for a
  Gravitational-wave Background}},\ }\href
  {https://doi.org/10.3847/2041-8213/acdac6} {\bibfield  {journal} {\bibinfo
  {journal} {Astrophys. J. Lett.}\ }\textbf {\bibinfo {volume} {951}},\
  \bibinfo {pages} {L8} (\bibinfo {year} {2023})},\ \Eprint
  {https://arxiv.org/abs/2306.16213} {arXiv:2306.16213 [astro-ph.HE]}
  \BibitemShut {NoStop}%
\bibitem [{\citenamefont {Reardon}\ \emph {et~al.}(2023)\citenamefont {Reardon}
  \emph {et~al.}}]{Reardon:2023gzh}%
  \BibitemOpen
  \bibfield  {author} {\bibinfo {author} {\bibfnamefont {D.~J.}\ \bibnamefont
  {Reardon}} \emph {et~al.},\ }\bibfield  {title} {\bibinfo {title} {{Search
  for an Isotropic Gravitational-wave Background with the Parkes Pulsar Timing
  Array}},\ }\href {https://doi.org/10.3847/2041-8213/acdd02} {\bibfield
  {journal} {\bibinfo  {journal} {Astrophys. J. Lett.}\ }\textbf {\bibinfo
  {volume} {951}},\ \bibinfo {pages} {L6} (\bibinfo {year} {2023})},\ \Eprint
  {https://arxiv.org/abs/2306.16215} {arXiv:2306.16215 [astro-ph.HE]}
  \BibitemShut {NoStop}%
\bibitem [{\citenamefont {Xu}\ \emph {et~al.}(2023)\citenamefont {Xu} \emph
  {et~al.}}]{Xu:2023wog}%
  \BibitemOpen
  \bibfield  {author} {\bibinfo {author} {\bibfnamefont {H.}~\bibnamefont {Xu}}
  \emph {et~al.},\ }\bibfield  {title} {\bibinfo {title} {{Searching for the
  Nano-Hertz Stochastic Gravitational Wave Background with the Chinese Pulsar
  Timing Array Data Release I}},\ }\href
  {https://doi.org/10.1088/1674-4527/acdfa5} {\bibfield  {journal} {\bibinfo
  {journal} {Res. Astron. Astrophys.}\ }\textbf {\bibinfo {volume} {23}},\
  \bibinfo {pages} {075024} (\bibinfo {year} {2023})},\ \Eprint
  {https://arxiv.org/abs/2306.16216} {arXiv:2306.16216 [astro-ph.HE]}
  \BibitemShut {NoStop}%
\bibitem [{\citenamefont {Hannuksela}\ \emph {et~al.}(2019)\citenamefont
  {Hannuksela}, \citenamefont {Haris}, \citenamefont {Ng}, \citenamefont
  {Kumar}, \citenamefont {Mehta}, \citenamefont {Keitel}, \citenamefont {Li},\
  and\ \citenamefont {Ajith}}]{Hannuksela:2019kle}%
  \BibitemOpen
  \bibfield  {author} {\bibinfo {author} {\bibfnamefont {O.~A.}\ \bibnamefont
  {Hannuksela}}, \bibinfo {author} {\bibfnamefont {K.}~\bibnamefont {Haris}},
  \bibinfo {author} {\bibfnamefont {K.~K.~Y.}\ \bibnamefont {Ng}}, \bibinfo
  {author} {\bibfnamefont {S.}~\bibnamefont {Kumar}}, \bibinfo {author}
  {\bibfnamefont {A.~K.}\ \bibnamefont {Mehta}}, \bibinfo {author}
  {\bibfnamefont {D.}~\bibnamefont {Keitel}}, \bibinfo {author} {\bibfnamefont
  {T.~G.~F.}\ \bibnamefont {Li}},\ and\ \bibinfo {author} {\bibfnamefont
  {P.}~\bibnamefont {Ajith}},\ }\bibfield  {title} {\bibinfo {title} {{Search
  for gravitational lensing signatures in LIGO-Virgo binary black hole
  events}},\ }\href {https://doi.org/10.3847/2041-8213/ab0c0f} {\bibfield
  {journal} {\bibinfo  {journal} {Astrophys. J.}\ }\textbf {\bibinfo {volume}
  {874}},\ \bibinfo {pages} {L2} (\bibinfo {year} {2019})},\ \Eprint
  {https://arxiv.org/abs/1901.02674} {arXiv:1901.02674 [gr-qc]} \BibitemShut
  {NoStop}%
\bibitem [{\citenamefont {McIsaac}\ \emph {et~al.}(2020)\citenamefont
  {McIsaac}, \citenamefont {Keitel}, \citenamefont {Collett}, \citenamefont
  {Harry}, \citenamefont {Mozzon}, \citenamefont {Edy},\ and\ \citenamefont
  {Bacon}}]{McIsaac:2019use}%
  \BibitemOpen
  \bibfield  {author} {\bibinfo {author} {\bibfnamefont {C.}~\bibnamefont
  {McIsaac}}, \bibinfo {author} {\bibfnamefont {D.}~\bibnamefont {Keitel}},
  \bibinfo {author} {\bibfnamefont {T.}~\bibnamefont {Collett}}, \bibinfo
  {author} {\bibfnamefont {I.}~\bibnamefont {Harry}}, \bibinfo {author}
  {\bibfnamefont {S.}~\bibnamefont {Mozzon}}, \bibinfo {author} {\bibfnamefont
  {O.}~\bibnamefont {Edy}},\ and\ \bibinfo {author} {\bibfnamefont
  {D.}~\bibnamefont {Bacon}},\ }\bibfield  {title} {\bibinfo {title} {{Search
  for strongly lensed counterpart images of binary black hole mergers in the
  first two LIGO observing runs}},\ }\href
  {https://doi.org/10.1103/PhysRevD.102.084031} {\bibfield  {journal} {\bibinfo
   {journal} {Phys. Rev. D}\ }\textbf {\bibinfo {volume} {102}},\ \bibinfo
  {pages} {084031} (\bibinfo {year} {2020})},\ \Eprint
  {https://arxiv.org/abs/1912.05389} {arXiv:1912.05389 [gr-qc]} \BibitemShut
  {NoStop}%
\bibitem [{\citenamefont {Li}\ \emph {et~al.}(2023)\citenamefont {Li},
  \citenamefont {Lo}, \citenamefont {Sachdev}, \citenamefont {Chan},
  \citenamefont {Lin}, \citenamefont {Li},\ and\ \citenamefont
  {Weinstein}}]{Li:2019osa}%
  \BibitemOpen
  \bibfield  {author} {\bibinfo {author} {\bibfnamefont {A.~K.~Y.}\
  \bibnamefont {Li}}, \bibinfo {author} {\bibfnamefont {R.~K.~L.}\ \bibnamefont
  {Lo}}, \bibinfo {author} {\bibfnamefont {S.}~\bibnamefont {Sachdev}},
  \bibinfo {author} {\bibfnamefont {J.~C.~L.}\ \bibnamefont {Chan}}, \bibinfo
  {author} {\bibfnamefont {E.~T.}\ \bibnamefont {Lin}}, \bibinfo {author}
  {\bibfnamefont {T.~G.~F.}\ \bibnamefont {Li}},\ and\ \bibinfo {author}
  {\bibfnamefont {A.~J.}\ \bibnamefont {Weinstein}} (\bibinfo {collaboration}
  {LIGO Scientific, Virgo}),\ }\bibfield  {title} {\bibinfo {title} {{Targeted
  subthreshold search for strongly lensed gravitational-wave events}},\ }\href
  {https://doi.org/10.1103/PhysRevD.107.123014} {\bibfield  {journal} {\bibinfo
   {journal} {Phys. Rev. D}\ }\textbf {\bibinfo {volume} {107}},\ \bibinfo
  {pages} {123014} (\bibinfo {year} {2023})},\ \Eprint
  {https://arxiv.org/abs/1904.06020} {arXiv:1904.06020 [gr-qc]} \BibitemShut
  {NoStop}%
\bibitem [{\citenamefont {Ezquiaga}\ \emph {et~al.}(2023)\citenamefont
  {Ezquiaga}, \citenamefont {Hu},\ and\ \citenamefont {Lo}}]{Ezquiaga:2023xfe}%
  \BibitemOpen
  \bibfield  {author} {\bibinfo {author} {\bibfnamefont {J.~M.}\ \bibnamefont
  {Ezquiaga}}, \bibinfo {author} {\bibfnamefont {W.}~\bibnamefont {Hu}},\ and\
  \bibinfo {author} {\bibfnamefont {R.~K.~L.}\ \bibnamefont {Lo}},\ }\bibfield
  {title} {\bibinfo {title} {{Identifying strongly lensed gravitational waves
  through their phase consistency}},\ }\href
  {https://doi.org/10.1103/PhysRevD.108.103520} {\bibfield  {journal} {\bibinfo
   {journal} {Phys. Rev. D}\ }\textbf {\bibinfo {volume} {108}},\ \bibinfo
  {pages} {103520} (\bibinfo {year} {2023})},\ \Eprint
  {https://arxiv.org/abs/2308.06616} {arXiv:2308.06616 [astro-ph.CO]}
  \BibitemShut {NoStop}%
\bibitem [{\citenamefont {Janquart}\ \emph {et~al.}(2023)\citenamefont
  {Janquart} \emph {et~al.}}]{Janquart:2023mvf}%
  \BibitemOpen
  \bibfield  {author} {\bibinfo {author} {\bibfnamefont {J.}~\bibnamefont
  {Janquart}} \emph {et~al.},\ }\bibfield  {title} {\bibinfo {title}
  {{Follow-up analyses to the O3 LIGO{\textendash}Virgo{\textendash}KAGRA
  lensing searches}},\ }\href {https://doi.org/10.1093/mnras/stad2909}
  {\bibfield  {journal} {\bibinfo  {journal} {Mon. Not. Roy. Astron. Soc.}\
  }\textbf {\bibinfo {volume} {526}},\ \bibinfo {pages} {3832} (\bibinfo {year}
  {2023})},\ \Eprint {https://arxiv.org/abs/2306.03827} {arXiv:2306.03827
  [gr-qc]} \BibitemShut {NoStop}%
\bibitem [{\citenamefont {Cutler}\ and\ \citenamefont
  {Holz}(2009)}]{Cutler:2009qv}%
  \BibitemOpen
  \bibfield  {author} {\bibinfo {author} {\bibfnamefont {C.}~\bibnamefont
  {Cutler}}\ and\ \bibinfo {author} {\bibfnamefont {D.~E.}\ \bibnamefont
  {Holz}},\ }\bibfield  {title} {\bibinfo {title} {{Ultra-high precision
  cosmology from gravitational waves}},\ }\href
  {https://doi.org/10.1103/PhysRevD.80.104009} {\bibfield  {journal} {\bibinfo
  {journal} {Phys. Rev. D}\ }\textbf {\bibinfo {volume} {80}},\ \bibinfo
  {pages} {104009} (\bibinfo {year} {2009})},\ \Eprint
  {https://arxiv.org/abs/0906.3752} {arXiv:0906.3752 [astro-ph.CO]}
  \BibitemShut {NoStop}%
\bibitem [{\citenamefont {Camera}\ and\ \citenamefont
  {Nishizawa}(2013)}]{Camera:2013xfa}%
  \BibitemOpen
  \bibfield  {author} {\bibinfo {author} {\bibfnamefont {S.}~\bibnamefont
  {Camera}}\ and\ \bibinfo {author} {\bibfnamefont {A.}~\bibnamefont
  {Nishizawa}},\ }\bibfield  {title} {\bibinfo {title} {{Beyond Concordance
  Cosmology with Magnification of Gravitational-Wave Standard Sirens}},\ }\href
  {https://doi.org/10.1103/PhysRevLett.110.151103} {\bibfield  {journal}
  {\bibinfo  {journal} {Phys. Rev. Lett.}\ }\textbf {\bibinfo {volume} {110}},\
  \bibinfo {pages} {151103} (\bibinfo {year} {2013})},\ \Eprint
  {https://arxiv.org/abs/1303.5446} {arXiv:1303.5446 [astro-ph.CO]}
  \BibitemShut {NoStop}%
\bibitem [{\citenamefont {Congedo}\ and\ \citenamefont
  {Taylor}(2019)}]{Congedo:2018wfn}%
  \BibitemOpen
  \bibfield  {author} {\bibinfo {author} {\bibfnamefont {G.}~\bibnamefont
  {Congedo}}\ and\ \bibinfo {author} {\bibfnamefont {A.}~\bibnamefont
  {Taylor}},\ }\bibfield  {title} {\bibinfo {title} {{Joint cosmological
  inference of standard sirens and gravitational wave weak lensing}},\ }\href
  {https://doi.org/10.1103/PhysRevD.99.083526} {\bibfield  {journal} {\bibinfo
  {journal} {Phys. Rev. D}\ }\textbf {\bibinfo {volume} {99}},\ \bibinfo
  {pages} {083526} (\bibinfo {year} {2019})},\ \Eprint
  {https://arxiv.org/abs/1812.02730} {arXiv:1812.02730 [astro-ph.CO]}
  \BibitemShut {NoStop}%
\bibitem [{\citenamefont {Jung}\ and\ \citenamefont
  {Shin}(2019)}]{Jung:2017flg}%
  \BibitemOpen
  \bibfield  {author} {\bibinfo {author} {\bibfnamefont {S.}~\bibnamefont
  {Jung}}\ and\ \bibinfo {author} {\bibfnamefont {C.~S.}\ \bibnamefont
  {Shin}},\ }\bibfield  {title} {\bibinfo {title} {{Gravitational-Wave Fringes
  at LIGO: Detecting Compact Dark Matter by Gravitational Lensing}},\ }\href
  {https://doi.org/10.1103/PhysRevLett.122.041103} {\bibfield  {journal}
  {\bibinfo  {journal} {Phys. Rev. Lett.}\ }\textbf {\bibinfo {volume} {122}},\
  \bibinfo {pages} {041103} (\bibinfo {year} {2019})},\ \Eprint
  {https://arxiv.org/abs/1712.01396} {arXiv:1712.01396 [astro-ph.CO]}
  \BibitemShut {NoStop}%
\bibitem [{\citenamefont {Tambalo}\ \emph {et~al.}(2023)\citenamefont
  {Tambalo}, \citenamefont {Zumalac{\'a}rregui}, \citenamefont {Dai},\ and\
  \citenamefont {Cheung}}]{Tambalo:2022wlm}%
  \BibitemOpen
  \bibfield  {author} {\bibinfo {author} {\bibfnamefont {G.}~\bibnamefont
  {Tambalo}}, \bibinfo {author} {\bibfnamefont {M.}~\bibnamefont
  {Zumalac{\'a}rregui}}, \bibinfo {author} {\bibfnamefont {L.}~\bibnamefont
  {Dai}},\ and\ \bibinfo {author} {\bibfnamefont {M.~H.-Y.}\ \bibnamefont
  {Cheung}},\ }\bibfield  {title} {\bibinfo {title} {{Gravitational wave
  lensing as a probe of halo properties and dark matter}},\ }\href
  {https://doi.org/10.1103/PhysRevD.108.103529} {\bibfield  {journal} {\bibinfo
   {journal} {Phys. Rev. D}\ }\textbf {\bibinfo {volume} {108}},\ \bibinfo
  {pages} {103529} (\bibinfo {year} {2023})},\ \Eprint
  {https://arxiv.org/abs/2212.11960} {arXiv:2212.11960 [astro-ph.CO]}
  \BibitemShut {NoStop}%
\bibitem [{\citenamefont {Sereno}\ \emph {et~al.}(2011)\citenamefont {Sereno},
  \citenamefont {Jetzer}, \citenamefont {Sesana},\ and\ \citenamefont
  {Volonteri}}]{Sereno:2011ty}%
  \BibitemOpen
  \bibfield  {author} {\bibinfo {author} {\bibfnamefont {M.}~\bibnamefont
  {Sereno}}, \bibinfo {author} {\bibfnamefont {P.}~\bibnamefont {Jetzer}},
  \bibinfo {author} {\bibfnamefont {A.}~\bibnamefont {Sesana}},\ and\ \bibinfo
  {author} {\bibfnamefont {M.}~\bibnamefont {Volonteri}},\ }\bibfield  {title}
  {\bibinfo {title} {{Cosmography with strong lensing of LISA gravitational
  wave sources}},\ }\href {https://doi.org/10.1111/j.1365-2966.2011.18895.x}
  {\bibfield  {journal} {\bibinfo  {journal} {Mon. Not. Roy. Astron. Soc.}\
  }\textbf {\bibinfo {volume} {415}},\ \bibinfo {pages} {2773} (\bibinfo {year}
  {2011})},\ \Eprint {https://arxiv.org/abs/1104.1977} {arXiv:1104.1977
  [astro-ph.CO]} \BibitemShut {NoStop}%
\bibitem [{\citenamefont {Liao}\ \emph {et~al.}(2017)\citenamefont {Liao},
  \citenamefont {Fan}, \citenamefont {Ding}, \citenamefont {Biesiada},\ and\
  \citenamefont {Zhu}}]{Liao:2017ioi}%
  \BibitemOpen
  \bibfield  {author} {\bibinfo {author} {\bibfnamefont {K.}~\bibnamefont
  {Liao}}, \bibinfo {author} {\bibfnamefont {X.-L.}\ \bibnamefont {Fan}},
  \bibinfo {author} {\bibfnamefont {X.-H.}\ \bibnamefont {Ding}}, \bibinfo
  {author} {\bibfnamefont {M.}~\bibnamefont {Biesiada}},\ and\ \bibinfo
  {author} {\bibfnamefont {Z.-H.}\ \bibnamefont {Zhu}},\ }\bibfield  {title}
  {\bibinfo {title} {{Precision cosmology from future lensed gravitational wave
  and electromagnetic signals}},\ }\href
  {https://doi.org/10.1038/s41467-017-01152-9, 10.1038/s41467-017-02135-6}
  {\bibfield  {journal} {\bibinfo  {journal} {Nature Commun.}\ }\textbf
  {\bibinfo {volume} {8}},\ \bibinfo {pages} {1148} (\bibinfo {year} {2017})},\
  \bibinfo {note} {[Erratum: Nature Commun.8,no.1,2136(2017)]},\ \Eprint
  {https://arxiv.org/abs/1703.04151} {arXiv:1703.04151 [astro-ph.CO]}
  \BibitemShut {NoStop}%
\bibitem [{\citenamefont {Choi}\ \emph {et~al.}(2007)\citenamefont {Choi},
  \citenamefont {Park},\ and\ \citenamefont {Vogeley}}]{Choi:2006qg}%
  \BibitemOpen
  \bibfield  {author} {\bibinfo {author} {\bibfnamefont {Y.-Y.}\ \bibnamefont
  {Choi}}, \bibinfo {author} {\bibfnamefont {C.}~\bibnamefont {Park}},\ and\
  \bibinfo {author} {\bibfnamefont {M.~S.}\ \bibnamefont {Vogeley}},\
  }\bibfield  {title} {\bibinfo {title} {{Internal and Collective Properties of
  Galaxies in the Sloan Digital Sky Survey}},\ }\href
  {https://doi.org/10.1086/511060} {\bibfield  {journal} {\bibinfo  {journal}
  {Astrophys. J.}\ }\textbf {\bibinfo {volume} {658}},\ \bibinfo {pages} {884}
  (\bibinfo {year} {2007})},\ \Eprint {https://arxiv.org/abs/astro-ph/0611607}
  {arXiv:astro-ph/0611607 [astro-ph]} \BibitemShut {NoStop}%
\bibitem [{\citenamefont {Biesiada}\ \emph {et~al.}(2014)\citenamefont
  {Biesiada}, \citenamefont {Ding}, \citenamefont {Piorkowska},\ and\
  \citenamefont {Zhu}}]{Biesiada:2014kwa}%
  \BibitemOpen
  \bibfield  {author} {\bibinfo {author} {\bibfnamefont {M.}~\bibnamefont
  {Biesiada}}, \bibinfo {author} {\bibfnamefont {X.}~\bibnamefont {Ding}},
  \bibinfo {author} {\bibfnamefont {A.}~\bibnamefont {Piorkowska}},\ and\
  \bibinfo {author} {\bibfnamefont {Z.-H.}\ \bibnamefont {Zhu}},\ }\bibfield
  {title} {\bibinfo {title} {{Strong gravitational lensing of gravitational
  waves from double compact binaries - perspectives for the Einstein
  Telescope}},\ }\href {https://doi.org/10.1088/1475-7516/2014/10/080}
  {\bibfield  {journal} {\bibinfo  {journal} {JCAP}\ }\textbf {\bibinfo
  {volume} {10}},\ \bibinfo {pages} {080}},\ \Eprint
  {https://arxiv.org/abs/1409.8360} {arXiv:1409.8360 [astro-ph.HE]}
  \BibitemShut {NoStop}%
\end{thebibliography}%

\end{document}